\documentclass{aa}  
\bibpunct{(}{)}{;}{a}{}{,} 

\usepackage{graphicx}
\usepackage{caption}
\usepackage{subcaption}
\usepackage{subcaption}

\usepackage{cleveref}

\renewcommand\thesubfigure{(\alph{subfigure})}
\usepackage{alphalph}
\renewcommand*{\thesubfigure}{\thefigure.
\arabic{subfigure}}
\usepackage{hyperref}
\usepackage[toc,page]{appendix}
\usepackage{chngcntr}

\hypersetup{colorlinks=true,linkcolor=blue,citecolor=blue,filecolor=blue,urlcolor=blue}
\usepackage{txfonts}
\usepackage[utf8]{inputenc}
\let\ACMmaketitle=\maketitle
\renewcommand{\maketitle}{\begingroup\let\footnote=\thanks \ACMmaketitle\endgroup}
\begin{document}

   \title{LoTSS DR1: Double-double radio galaxies in the HETDEX field\footnote{The VLA images (FITS files) are available in electronic form via anonymous ftp to cdsarc.u-strasbg.fr (130.79.128.5) or via \texttt{http://cdsarc.u-strasbg.fr/viz-bin/qcat?J/A+A/vol/page}}}

   \author{V. H. Mahatma\inst{1}\thanks{Email: v.mahatma2@herts.ac.uk},
          M. J. Hardcastle\inst{1},
          W. L. Williams\inst{1},
          P.N.Best\inst{2},
          J. H. Croston\inst{3},
          K. Duncan\inst{4},
          B. Mingo\inst{3},
          R. Morganti\inst{5,6},
          M. Brienza\inst{7},
          R. K. Cochrane\inst{2},
          G. G\"urkan\inst{8},
          J. J. Harwood\inst{1},
          M. J. Jarvis\inst{9,10},
          M. Jamrozy\inst{11},
          N. Jurlin\inst{5,6},
          L. K. Morabito\inst{9},
          H. J. A. R\"ottgering\inst{4},
          J. Sabater\inst{2},
          T. W. Shimwell\inst{5},
          D. J. B. Smith \inst{1},
          A. Shulevski\inst{12} and
          C. Tasse\inst{13,14}
          }

   \institute{Centre for Astrophysics Research, School of Physics, Astronomy and Mathematics, University of Hertfordshire, College Lane, Hatfield AL10 9AB, UK
         \and
         SUPA, Institute for Astronomy, Royal Observatory, Blackford Hill, Edinburgh, EH9 3HJ, UK
         \and
         School of Physical Sciences, The Open University, Walton Hall, Milton Keynes, MK7 6AA, UK
         \and 
         Leiden Observatory, Leiden University, PO Box 9513, NL-2300 RA Leiden, the Netherlands
         \and
         ASTRON, the Netherlands Institute for Radio Astronomy, Postbus 2,7990 AA, Dwingeloo, The Netherlands
         \and 
                 Kapteyn Astronomical Institute, University of Groningen, P.O. Box 800, 9700 AV Groningen, The Netherlands
         \and 
         INAF – Istituto di Radioastronomia, via Gobetti 101, 40129, Bologna, Italy
         \and
         CSIRO Astronomy and Space Science, PO Box 1130, Bentley WA 6102, Australia
         \and 
         Astrophysics, University of Oxford, Denys Wilkinson Building, Keble Road, Oxford, OX1 3RH, UK
         \and
         Physics and Astronomy Department, University of the Western Cape, Bellville 7535, South Africa
         \and 
         Astronomical Observatory, Jagiellonian University, ul. Orla 171, 30-244 Krakow, Poland
         \and 
         Anton Pannekoek Institute for Astronomy, University of Amsterdam, Postbus 94249, 1090 GE Amsterdam, The Netherlands
         \and 
         GEPI \& USN, Observatoire de Paris, Université PSL, CNRS, 5 Place Jules Janssen, 92190 Meudon, France
         \and 
         Department of Physics \& Electronics, Rhodes University, PO Box 94, Grahamstown, 6140, South Africa
            }

   \date{Received 27/07/2018; accepted 21/09/2018}

 
  \abstract
   {Double-double radio galaxies (DDRGs) represent a short but unique phase in the life-cycle of some of the most powerful radio-loud active galactic nuclei (RLAGN). These galaxies display large-scale remnant radio plasma in the intergalactic medium left behind by a past episode of active galactic nuclei (AGN) activity, and meanwhile, the radio jets have restarted in a new episode. The knowledge of what causes the jets to switch off and restart is crucial to our understanding of galaxy evolution, while it is important to know if DDRGs form a host galaxy dichotomy relative to RLAGN.}
   {The sensitivity and field of view of LOFAR enables the observation of DDRGs on a population basis rather than single-source observations. Using statistical comparisons with a control sample of RLAGN, we may obtain insights into the nature of DDRGs in the context of their host galaxies, where physical differences in their hosts compared to RLAGN as a population may allow us to infer the conditions that drive restarting jets.}
   {We utilised the LOFAR Two-Metre Sky Survey (LoTSS) DR1, using a visual identification method to compile a sample of morphologically selected candidate DDRGs, showing two pairs of radio lobes. To confirm the restarted nature in each of the candidate sources, we obtained follow-up observations with the Karl. G. Jansky Very Large Array (VLA) at higher resolution to observe the inner lobes or restarted jets, the confirmation of which created a robust sample of 33 DDRGs. We created a comparison sample of 777 RLAGN, matching the luminosity distribution of the DDRG sample, and compared the optical and infrared magnitudes and colours of their host galaxies.} 
   {We find that there is no statistically significant difference in the brightness of the host galaxies between double-doubles and single-cycle RLAGN. The DDRG and RLAGN samples also have similar distributions in WISE mid-infrared colours, indicating similar ages of stellar populations and dust levels in the hosts of DDRGs. We conclude that DDRGs and `normal' RLAGN are hosted by galaxies of the same type, and that DDRG activity is simply a normal part of the life cycle of RLAGN. Restarted jets, particularly for the class of low-excitation radio galaxies, rather than being a product of a particular event in the life of a host galaxy, must instead be caused by smaller scale changes, such as in the accretion system surrounding the black hole.}
   {}

\authorrunning{Mahatma et al.}
\titlerunning{DDRGs in the HETDEX field}
   \maketitle
%

\section{Introduction}
Radio galaxies can display large-scale and powerful jets that are
associated with active galactic nuclei (AGN) in the centres of the
most massive galaxies \citep{mclu99,best05}. These powerful outflows
are suggested to play a significant role in offsetting gas cooling and
the consequent suppression of star formation \citep{fabi12} while,
conversely, AGN-triggered star formation is also thought to be possible \citep[e.g.][]{brue85,zinn13}. The link between jet energetics and gas reservoirs present in the interstellar or intergalactic medium over cosmic time may lead to the observed decline of the galaxy mass function at high stellar masses \citep{bald08}. The jets of radio galaxies, or radio-loud AGN (RLAGN), form an essential component of evolutionary models for massive galaxies, as well as of feedback processes that might affect the hot thermal medium present in the centres of galaxy clusters \citep[see review by][]{fabi12}. 

In determining the interplay between the radio jets and host galaxy dynamics, a key ingredient is the duty cycle, i.e the fraction of time spent by a RLAGN in its active phase relative to its total lifetime (active and inactive). Simulations predict that the jets may be disrupted after $\sim 10^8$ yr of activity \citep[e.g.][]{tuck97,omma04}, while spectral ageing models yield ages of a few $\times 10^7$ yr \citep{al-le87}, after which the jets are no longer active. In this so-called remnant phase the radio emission from the jets is expected to fade quickly; the light-travel time of a 300-kpc scale relativistic jet is $\sim 1$ Myr. However the radio lobes inflated by the jets may radiate via synchrotron emission for a longer period. It is unclear for how long the remnant phase lasts, although both modelling and the small remnant fractions found in recent studies \citep{godf17,brie17,hard18,maha18} imply short radiative lifetimes\footnote{It should be borne in mind that small remnant fractions are simply observable quantities and do not necessarily translate to a short `off-phase' for AGN activity -- these large-sample studies only infer information about the timescales of the radio emitting plasma.}. These studies, including those of \cite{shul12}, \cite{murg11}, and \cite{turn18}, have provided some information about the dynamics and energetics of RLAGN in their remnant phase.

The jets are expected eventually to restart with a new episode of AGN
activity\footnote{The relaunch of jets, and the timescales between
  episodes may depend on the type of RLAGN. Sources operating in the
  `radio' or `jet mode' feedback role, often associated with
  low-excitation radio galaxies \citep[LERGs:][]{evan07} are expected
  to have a long duty cycle \citep{turn15} and some may have
  intermittent life-cycles, while for `quasar mode' or high-excitation
  radio galaxies (HERGs) the re-triggering of jets may be related to a
  particular episode of fuel being made available, for example through the
  infall of gas via mergers \citep{best12}.}. In either case, if a
jet starts up again soon after the earlier episode of activity, the
newly restarted jets drive a fresh pair of radio lobes into the
pre-existing remnant plasma known as a restarted radio-loud AGN (RRLAGN);
therefore RRLAGN represent a brief, but interesting, phase in the
life cycle of RLAGN.

Physical jet properties as well as observational selection effects
mean that observing RRLAGN is difficult. Given the short light travel
time for powerful sources, the reborn jets may quickly merge with the
remnant lobe plasma left behind by the previous activity, essentially
removing any history of past radio jets from radio observations. In
the instance of capturing a luminous restarted radio galaxy before
this happens, we may expect to observe a `double-double radio
galaxy' \citep[DDRG;][]{scho00}\footnote{\cite{broc07} discovered a
  \textit{third} pair of radio lobes in B0925+420 -- a `triple-double
  radio galaxy', suggesting that the inner double associated with
  DDRGs are indeed restarting jets, as opposed to bright knots in the
  underlying jet.} -- a pair of bright inner radio lobes together with
(and often embedded in) an outer pair of faint remnant lobes. We make a crucial
point on nomenclature, for clarity. Double-double radio galaxies are a class of restarted
radio galaxies that are exclusive to classical double
\citep[Fanaroff-Riley type II or FR-II;][]{FR74} radio galaxies
\citep{scho00}, whereas other classes of RRLAGN do not produce two
pairs of edge-brightened radio lobes. Examples of objects that are
RRLAGN but not DDRG include the observations of \citealt{jamr07}, showing extended diffuse emission
  around a compact inner double for the radio source 4C29.30, or the
  multiple episodes of activity in the Fanaroff-Riley class I object
  4C32.26 seen by \cite{jeth08}.

The aims of this paper are not to understand the exact triggering mechanisms of DDRGs, but rather to understand their global properties as a population. Nevertheless, we briefly mention the possible scenarios that may cause the jets of RLAGN to be disrupted and/or restart, referencing these ideas in the context of our results in Section \ref{sect-results}.
\begin{itemize}
\item The large-scale infall of gas driven by a galaxy merger or a black
  hole merger as a disruption event and subsequent re-triggering of
  the AGN. Observations of DDRGs show that the restarted jets are
  usually driven along the same spatial axis as the old
  jets \footnote{This may be a selection effect in which samples are
    chosen on a purely morphological basis. Detached outer lobes are
    trivial to associate with a DDRG if they lie on the same spatial
    axis as the inner jets and misaligned DDRGs may be
    missed in such samples where the misalignment of the outer lobes
    is significantly large so that they cannot visually be associated with the same source containing the inner jets.}
  \citep{scho00,kais00}, providing constraints on the accretion system
  producing the jets: the spin vector of the black hole very plausibly
  controls the direction of the jets, and hence a black
  hole merger that may significantly change the black hole spin does
  not seem plausible as the origin of DDRGs. Moreover, \cite{nata98} suggested that the jet direction is instead controlled by the angular momentum of the accretion material, but this would require the infalling material from a galaxy merger to have the same angular momentum as the previously accreting gas, if the jet direction is unchanged.
\item Variations in the accretion system disrupting jet production on short timescales. If intermittent jet activity is solely governed by the accretion or black hole parameters rather than the environment of the galaxy, then it may be plausible to suggest a change in magnetic flux dragged by a spinning black hole \citep{blan77} or the accretion rate may cause a disruption in jet activity. While this may seem credible, there is no \textit{a priori} reason for such a change, other than a galaxy merger, or an entirely random process. Moreover a different black hole spin magnitude between the old and restarted jets, the driver of which might control intermittent activity, is likely to cause a difference in jet power between the old and restarting jets (for a fixed accretion rate). However, \cite{kona13} found a striking resemblance between the observed radio properties of the inner and outer hotspots of DDRGs, meaning that we might tend to observe DDRGs with old and new jets of similar power, although it is possible that this is only a selection effect. 

\end{itemize}

\cite{kais00} used an analytic model constrained by observations to
predict the properties of DDRGs. They deduced that the low densities
in the outer lobes created by the old jets are insufficient to explain
the observed properties of the restarted jets -- jets interacting with
a denser environment produce stronger shocks and hence bright
hotspots leading to the detection of an inner source structure. To
account for the observations they proposed that the remnant lobes
mixed with warm clouds in the interstellar medium. An alternative, but not exclusive, bow-shock model was introduced by \cite{clar91}, which also describes the inner jets propagating in a low-density environment, but which drive bow shocks into the remnant lobes and re-energise the rapidly ageing particles. It is expected that a combination of these physical processes causes the restarted jets of DDRGs to become observable in radio surveys.

Compared to other types of RRLAGN, DDRG are easier to identify based
on their morphology. Therefore it is relatively easy to construct samples
of DDRGs, although such samples would be biased towards the more luminous (FR-II-type) sources. As such, DDRGs have been known for many years
\citep[e.g.][]{scho00,kais00}, but large robust samples are limited. \cite{nand12} presented a search for DDRGs in the Faint Images of the Radio-Sky at Twenty
centimetres \citep[FIRST;][]{beck95} survey, constraining 23 sources as only candidate DDRGs. Later \cite{kuzm17} presented a larger sample that included 74 radio
sources with evidence of recurrent activity ($85$ per cent of which
are DDRGs) using both FIRST and the NRAO VLA Sky Survey
\citep[NVSS;][]{cond98}. 

Such radio surveys are not sensitive
to faint or low-luminosity sources, or the most compact structures that exist for the inner edge-brightened double of a DDRG. The well-studied Third Cambridge
Catalogue of Radio Sources \citep[the 3CRR survey;][]{3crr}, for example, with a
detection sensitivity of $10.9$ Jy at $178$ MHz,  preferentially
selects for the brightest radio sources. The fraction of remnant
(switched-off) RLAGN in this survey is only
around $1-3$ per cent \citep{giov88}. Since remnant outer lobes are a
precondition for the detection of DDRG, the numbers of DDRGs
visible in such surveys would be expected to be limited; however it is important to note that not all remnants necessarily become DDRGs. It is plausible to suggest that for some of the
classical doubles detected in these surveys that have had repeated
activity in the past, we may only view their inner double, while the
outer double has faded beyond the sensitivity limits. This might
suggest that DDRGs, or restarting jets, are simply a normal but brief
phase in the life cycle of RLAGN \citep{broc11}. In order to test this with a large sample, sensitivity is crucial. Studies of remnants  with the LOw Frequency ARray \citep[LOFAR;][]{vanh13} in the Herschel-ATLAS field \citep{maha18}, for sources $>80$ mJy at $150$ MHz, and in the similar LOFAR study by \cite{brie17}, found upper-limit remnant fractions of $\sim 10$ per cent; these are potentially a larger percentage of remnants than the percentage found in 3CRR, although this is clearly a non-systematic comparison.

While the existence of previous remnant activity may not be true of
all classical doubles observed in radio observations, it is necessary
to understand the physical source properties of DDRGs relative to
normal, or single-cycle, RLAGN. A detailed investigation of a large
sample of these objects in a statistical sense could result in a
deeper understanding of the conditions that drive restarting AGN.
Observations with LOFAR can
provide the large samples and high sensitivity needed to
capture a larger population of DDRGs with clear evidence for outer
remnant radio lobes associated with restarted sources. The combination of long and short baselines of LOFAR at a resolution comparable to
FIRST ($\sim 6$ arcsec) enables observations of both the inner and
more diffuse outer double of DDRGs.

To understand whether small- or large-scale galaxy processes determine
the life cycles of jets, it is important to understand if any
fundamental difference exists between the host galaxies of DDRGs and
single-cycle RLAGN. It might be expected that if large-scale processes
in the host galaxy disrupt or trigger jets, then DDRGs would be hosted
by galaxies of a certain type, relative to single-cycle RLAGN. Large
samples of DDRGs with good host galaxy measurements are needed to test this model. Moreover, the implications of such information on the AGN duty cycle and how variable AGN activity is related to host galaxy properties is crucial in our current understanding of galaxy evolution. \cite{kuzm17} presented a study comparing the host galaxies of DDRGs and FR-IIs, finding that the host stellar masses of DDRGs are lower and also suggesting that the hosts of restarting sources have had a history of merger events. However, their DDRG and FR-II comparison samples are inhomogeneously selected from multiple surveys at different observing frequencies and varying levels of completeness (see Section \ref{sect-results} for a further discussion). A systematic host galaxy comparison between the population of RLAGN and a robust sample of DDRGs using a single, sensitive radio survey will improve our understanding of the nature of DDRGs.

In this paper, we utilise the first data release of the LOFAR
Two-Metre Sky Survey \citep[LoTSS DR1;][and submitted.]{Shimwell+17} to
create a sample of candidate DDRGs. We confirm their restarted nature
with follow-up observations with the VLA, leading to a robust sample
of DDRGs. The main scope of this paper consists of analyses of host
galaxy properties between DDRGs and a control sample of RLAGN obtained
from LoTSS DR1. In Section \ref{sect-observations}, we briefly
describe LoTSS DR1, the selection of DDRGs, and present our follow-up
VLA observations of the DDRGs. In Section \ref{sect-analysis} we
describe our analysis of the host galaxy properties of DDRGs and
RLAGN. In Section \ref{sect-results} we summarise our main findings
and conclude with our results in Section \ref{sect-conclusions}.
Observed magnitudes are in the AB system \citep{oke83}. Throughout this paper we use a cosmology in which $H_0=70$ km s$^{-1}$, $\Omega_m=0.3$ and $\Omega_{\Lambda}=0.7$.
\section{Observations} \label{sect-observations}
\subsection{LoTSS DR1} \label{subsect-lotss}
The LoTSS is an ongoing low-frequency radio survey of the northern sky \citep[][]{Shimwell+17}. The current release (DR1; Shimwell et al., submitted) covers the area of the Hobble-Eberly Telescope Dark Energy eXperiment (HETDEX: \citealt{hetdex}) Spring field; over 420 square degrees on the sky within $161<$RA$<231$ degrees and $45.5<$DEC$<57$ degrees, observed at 6 arcsec resolution with a median sensitivity of $\sim 70  \mu$Jy beam$^{-1}$. Recently developed procedures for direction-dependent calibration \citep{tass18} were applied to the pre-processed data after the standard direction-independent calibration pipeline \citep[prefactor;][]{Shimwell+17}. The survey detected $318,542$  individual radio sources. Host galaxy identification using optical Pan-STARRS \citep{panstarrs} or mid-infrared Wide-field Infrared Survey Explorer \citep[\textit{WISE;}][]{wise} images, and association with other radio components nearby in the field, was primarily performed using a likelihood ratio analysis, while sources that were more complex or extended had their different radio components associated and host galaxies identified through visual inspection (Williams et al. submitted). Photometric redshifts, in cases in which spectroscopic redshifts were unavailable, were estimated using the methods of Duncan et al. (submitted). Full details on processing strategies, imaging methods, and all other information on the first data release, are given in the aforementioned papers.   
\subsection{Double-double radio galaxy selection}\label{subsect-rrlagn_selection}
\subsubsection{Preliminary selection}
For a preliminary selection of candidate DDRGs from LoTSS DR1, we utilised the visual
inspection strategy of Williams et al. (submitted) based on the
Zooniverse framework. Volunteers selected candidate DDRGs or
restarting sources based on the LOFAR morphology combined with cross-matched images from the FIRST survey. Typically the LOFAR images displayed the extended
diffuse emission surrounding the central optical ID or outer
double of a DDRG, while the higher-frequency and similar resolution
FIRST images usually, but not exclusively, displayed emission from
more compact structures such as the radio core. We identified 91 candidate
DDRGs and restarting sources by this visual inspection process.

Targets in this candidate list were then visually inspected again by
some of the authors (VHM, MJH, WLW) to select the most obvious
DDRG systems out of the 91 candidate restarting sources. We required an optical ID to select a DDRG for follow-up,
which removed a number of potential high-$z$ objects where no ID was
present. We then rejected other objects on morphological grounds. The rejected objects were largely either faint, and
therefore had ambiguous morphology (although some may have been other
classes of RRLAGN), or sources in which a possible inner
double showed signs of extended downstream emission in the LOFAR
images, which would be more characteristic of FRI or wide-angle tailed
(WAT) sources.

After this visual process, we were left with a sample of the 40 most credible candidate 
DDRGs from our initial pool. This sample however, does not include a robust indication of
restarted jets. Typically the FIRST emission where the inner double
was assumed to be located (using the LOFAR morphology) was either unresolved or the individual inner lobes were unresolved. This raised the question of whether we actually see the edge-brightened restarting jets, which are an exclusive property for DDRGs, for the majority of the sources in this sample.
Moreover, any observed bright inner structures could also be
interpreted as the bases of WAT-type jets, or the core-brightened jets
of classical FR-I radio galaxies. For a DDRG, the jets are
edge-brightened and end in compact hotspots, which may also be missed
by the resolution of FIRST. \cite{nand12} were only able to confirm 23 out of 242 of their candidate objects as DDRGs using FIRST alone, while 63 required higher resolution follow-up observations. To clarify the nature of our candidate
DDRG sample, we obtained follow-up VLA observations at higher
resolution to determine whether compact hotspots exist within the
inner double. These observations are described below.

  \subsubsection{VLA observations} \label{subsect-vlaobs}
We obtained snapshot VLA
observations of our 40 candidate DDRGs at 1.4 GHz in the A array. In this configuration and observing frequency, the VLA has a synthesised beam size of $1.3$ arcsec, giving a
substantial improvement in angular resolution over FIRST ($5$ arcsec). Recent VLA observations of candidate remnant radio galaxies
\citep{maha18} have demonstrated the ability of the VLA to detect compact
cores where missed by the sensitivity and resolution of FIRST. With
these observations we are able to check which of our candidate DDRGs contain
compact inner hotspots or an edge brightened jet associated with an
inner double, which are the clear signatures of DDRGs. 

Two sets of observations were conducted on the 27 and 28 March 2018, both consisting of 5-minute snapshot observations of each target source (as detailed in Table \ref{source_table}) with the same hardware setup. Scans of target sources that were spatially distributed on the sky within approximately 15 degrees of right ascension and 5 degrees in declination were bracketed by two $\sim$ 1-minute scans of a nearby phase calibrator. Owing to the large area of sky covered by our sources, five phase calibrators were used in total to correct for ionospheric variations throughout the observation time of four hours. 3C286 was observed as the primary flux and bandpass calibrator. 

The two epochs of observations were reduced separately. Prior to data reduction, the AOFlagger algorithm \citep{offr12} was applied to the data sets to flag for obvious radio-frequency interference (RFI). The measurement sets (MSs) containing the observations were then reduced using the Common Astronomy Software Applications \citep[CASA:][]{mcmu07} VLA pipeline version 1.3.11. Various calibration tables were inspected to check for the quality of calibration, and baselines displaying residual RFI or erratic phase variations were flagged manually in CASA. The CASA \texttt{rflag} algorithm was subsequently applied to flag further residual RFI. Images were produced by combining both epochs of observations in the $uv$-plane using the CASA image reconstruction technique \texttt{CLEAN} (making use of the \texttt{clarke} algorithm; \citealt{clar80}). Different values for the Briggs robust weighting \citep{brig95} parameter between $-1.0<$ \texttt{ROBUST} $<1.5$ were used for the imaging of targets, depending on the visibility of a compact core or inner hotspots. Images are shown in Figure \ref{overlays}. For presentation purposes, we scaled the VLA images logarithmically and convolved with a Gaussian function with a full width at half maximum (FWHM) of three times the beam. We also overlay contours of the LOFAR source to view the outer extended remnant emission.

After visually checking the VLA images for the inner hotspots or restarted
jets of a DDRG (Figure \ref{overlays}), we removed six sources from our sample: ILTJ105955.01+492615.4 has a very bright outer northern hotspot, but does not have a significant detection of an inner northern hotspot along the jet axis, and may be a classical FR-II. ILTJ113201.82+472829.9 does not display clear edge-brightened inner jets and is plausibly an FR-I radio galaxy.
ILTJ124240.48+483706.8 has features that are poorly resolved with the VLA, i.e. natural weighting was required to see structure in the image, and it is unclear whether these features are emission from hotspots or from diffuse emission related to a young FR-I source. ILTJ131115.53+534356.8 and ILTJ133135.09+455957.0 both display jets typical of an FR-I based on the VLA data. ILTJ133252.97+544103.2 also displays FR-I-type jets based on the VLA data, while the LOFAR morphology is typical of a double source. The remaining 34 objects in our sample have clear evidence for edge-brightened restarting jets in a DDRG; hereafter, we call these objects our DDRG sample. We note that some objects only have a single inner hotspot detected with the VLA. We interpret this as an (or a combination of) effect(s) due to the very compact nature of hotspots, relativistic beaming often seen in X-ray observations of RLAGN (e.g \textit{Chandra} observations of 3C303; \citealt{kata03}), and an asymmetric environment rendering the counter-hotspot undetectable, while the observation of an inner jet implies the existence a counter-jet. We visually cross-matched the position of the hotspot in these sources with Pan-STARRS and \textit{WISE} to ascertain that no optical host lies in their locations and that hotspots are not misidentified as background quasars or foreground stars.

\subsubsection{Optical ID mis-identifications} \label{subsubsect-misids}
While our sample includes host galaxy identifications, it is important to ensure that the DDRGs each have the correct optical identification as our analysis predominantly compares host galaxy properties. As discussed in Section \ref{subsect-lotss} above, host galaxy identifications were based on a visual method of cross-matching the LOFAR source with FIRST core emission at the position of a Pan-STARSS or \textit{WISE}-detected galaxy, if any. While \cite{maha18} showed that in a similar LOFAR study 10 per cent of the largest sources may be misidentified in the absence of FIRST core emission, it is still possible that a small fraction may still be misidentified even with FIRST emission in the central regions. Our VLA observations, at higher resolution and sensitivity, are more sensitive to the flat-spectrum cores of RLAGN.

Where compact radio cores were detected with our VLA observations (Figure \ref{overlays}), we performed a  positional cross-match with the nearest Pan-STARSS and \textit{WISE} galaxies in the vicinity of the radio source. Source ILT140255.12+512726.28 has a compact core detection spatially along the jet axis and between the two outer lobes (see Figure \ref{overlays}), which lies directly at the position of a different host ID than that made in LoTSS DR1. Since no other compact structures or hotspots are detected either side of the new host ID other than the hotspots in the outer double, we cannot confirm this as a DDRG. Hence, we removed this source from our sample, reducing our DDRG sample to 33 sources. Source ILTJ111033.09+555310.8 has a VLA detection at its optical ID, but also has bright compact emission further south towards the centroid of the source and is perhaps more likely to be emission from the radio core. A positional cross-match with Pan-STARRS and \textit{WISE} at this location shows no other possible host galaxy, and hence we retained the ID for this source. We cannot rule out compact emission being associated with bright jet knots if not associated with an optical host galaxy. For this source and the remaining DDRGs in our sample with core detections, we confirm that these sources have the best possible host ID, giving confidence to the number of correct host IDs chosen for the bulk of extended RLAGN in LoTSS DR1.

 \begin{table*}
\centering
\begin{tabular}{ccccc}
Source & RA & DEC & ID & $z^{\dagger}$ \\
\hline 
ILTJ105133.89+514451.1 & 10:51:33.89 & +51:44:51.18 & AllWISE J105134.42+514455.4 & -- \\
ILTJ105742.50+510558.5 & 10:57:42.50 & +51:05:58.59 & PSO J105743.090+510557.747 & 0.463$^s$ \\
ILTJ105955.01+492615.4$^{+}$ & 10:59:55.01 & +49:26:15.48 & AllWISE J105955.51+492607.4 & -- \\
ILTJ111033.09+555310.8 & 11:10:33.09 & +55:53:10.86 & AllWISE J111033.19+555313.8 & -- \\
ILTJ111417.63+461058.9 & 11:14:17.63 & +46:10:58.90 & AllWISE J111417.56+461102.0 & -- \\
ILTJ111449.99+485640.2 & 11:14:49.99 & +48:56:40.25 & AllWISE J111450.75+485640.5 & -- \\
ILTJ112218.41+555047.7 & 11:22:18.41 & +55:50:47.70 & PSO J112218.514+555033.651 & 0.910$^s$ \\
ILTJ112425.85+554607.6 & 11:24:25.85 & +55:46:07.61 & PSO J112425.079+554615.740 & 0.809$^s$ \\
ILTJ113201.82+472829.9$^{+}$ & 11:32:01.82 & +47:28:29.93 & PSO J113202.310+472824.218 & 0.264$^s$ \\
ILTJ115527.32+485039.0 & 11:55:27.32 & +48:50:39.05 & PSO J115528.238+485044.446 & 0.788$^p$ \\
ILTJ120459.87+475825.4 & 12:04:59.87 & +47:58:25.45 & PSO J120459.941+475827.470 & 0.585$^p$ \\
ILTJ120808.48+462940.6 & 12:08:08.48 & +46:29:40.65 & PSO J120808.882+462941.772 & 0.546$^p$ \\
ILTJ121136.54+505537.5 & 12:11:36.54 & +50:55:37.50 & PSO J121136.398+505537.743 & 0.487$^s$ \\
ILTJ121502.39+474641.1 & 12:15:02.39 & +47:46:41.10 & PSO J121502.262+474641.710 & 0.597$^s$ \\
ILTJ121541.21+502517.9 & 12:15:41.21 & +50:25:17.92 & AllWISE J121541.20+502517.3 & -- \\
ILTJ122544.63+515951.7 & 12:25:44.63 & +51:59:51.75 & AllWISE J122544.41+515953.0 & -- \\
ILTJ123005.72+491516.8 & 12:30:05.72 & +49:15:16.87 & AllWISE J123005.44+491515.9 & -- \\
ILTJ123857.80+483823.5 & 12:38:57.80 & +48:38:23.50 & PSO J123857.795+483818.428 & 0.458$^{p}$ \\
ILTJ124240.48+483706.8$^{+}$ & 12:42:40.48 & +48:37:06.85 & AllWISE J124240.92+483708.9 & -- \\
ILTJ124411.02+500922.1 & 12:44:11.02 & +50:09:22.17 & PSO J124410.502+500921.925 & 0.232$^s$ \\
ILTJ124548.75+563109.7 & 12:45:48.75 & +56:31:09.70 & PSO J124548.730+563111.869 & 0.702$^p$ \\
ILTJ130357.58+464250.4 & 13:03:57.58 & +46:42:50.49 & PSO J130357.872+464250.488 & 0.584$^s$ \\
ILTJ131115.53+534356.8$^{+}$ & 13:11:15.53 & +53:43:56.84 & PSO J131115.649+534353.418 & 0.491$^s$ \\
ILTJ131158.61+475847.5 & 13:11:58.61 & +47:58:47.54 & PSO J131158.419+475848.393 & 0.914$^p$ \\
ILTJ131403.17+543939.6 & 13:14:03.17 & +54:39:39.64 & PSO J131404.616+543937.998 & 0.347$^s$ \\
ILTJ131941.97+555345.3 & 13:19:41.97 & +55:53:45.37 & PSO J131941.787+555328.909 & 0.136$^s$ \\
ILTJ132049.67+480445.6 & 13:20:49.67 & +48:04:45.65 & AllWISE J132049.70+480442.7 & -- \\
ILTJ133135.09+455957.0$^{+}$ & 13:31:35.09 & +45:59:57.01 & PSO J133135.279+455955.454 & 0.385$^s$ \\
ILTJ133252.97+544103.2$^{+}$ & 13:32:52.97 & +54:41:03.21 & PSO J133252.957+544107.657 & 0.143$^s$ \\
ILTJ134727.92+545233.7 & 13:47:27.92 & +54:52:33.79 & PSO J134727.819+545233.142 & 0.841$^p$ \\
ILTJ140255.12+512726.8 & 14:02:55.12 & +51:27:26.87 & PSO J140256.329+512730.053 $^{*}$ & -- \\
ILTJ143735.74+514434.3 & 14:37:35.74 & +51:44:34.31 & PSO J143737.636+514446.316 & 0.963$^p$ \\
ILTJ144049.79+480444.0 & 14:40:49.79 & +48:04:44.04 & AllWISE J144050.07+480445.3 & -- \\
ILTJ145147.28+484123.5 & 14:51:47.28 & +48:41:23.54 & PSO J145145.215+484127.668 & 0.231$^s$ \\
ILTJ145447.14+542232.2 & 14:54:47.14 & +54:22:32.28 & PSO J145447.069+542232.933 & 0.102$^s$ \\
ILTJ145610.69+481923.0 & 14:56:10.69 & +48:19:23.06 & PSO J145611.291+481927.866 & 0.774$^p$ \\
ILTJ145641.07+484940.5 & 14:56:41.07 & +48:49:40.50 & PSO J145640.671+484942.791 & 0.782$^p$ \\
ILTJ151216.35+514731.8 & 15:12:16.35 & +51:47:31.86 & PSO J151216.252+514725.545 & 0.584$^p$ \\
ILTJ151933.09+500706.2 & 15:19:33.09 & +50:07:06.20 & PSO J151933.756+500724.858 & 0.830$^{s}$ \\
ILTJ152105.64+521442.0 & 15:21:05.64 & +52:14:42.02 & PSO J152105.891+521439.872 & 0.731$^p$ \\
\hline \\
\end{tabular}
\caption{Our candidate DDRGs. $\dagger$ Superscripts `$p$' and `$s$' denote photometric and spectroscopic redshifts, and where both are available for a source, we quote the spectroscopic redshift. $^{+}$Sources removed as having a lack of evidence of being classed as a DDRG (see Section \ref{subsect-rrlagn_selection}). $^{*}$Misidentified source; see Section \ref{subsubsect-misids}}
 \label{source_table}
\end{table*}
\subsection{Comparison radio-loud AGN selection} \label{subsect-rlagn_selection}
To form a control sample for host galaxy comparisons with the DDRG sample,
we used the sample of Hardcastle et al. (submitted), which is a RLAGN-selected sample from LoTSS DR1 (hereafter the RLAGN sample).
The details of the selection of RLAGN from the LoTSS DR1
catalogue are given by Hardcastle et al. (submitted), but we briefly describe the formation of this sample. Starting with the LoTSS DR1 catalogue of 318,542 radio sources, a flux density cut of >0.5 mJy was imposed to produce a flux-complete sample; Shimwell et al. (submitted) showed that the catalogue is close to complete at this level at 145 MHz. Further, sources were selected as having an optical ID (either Pan-STARRs or WISE as our DDRG sample) and either a spectroscopic redshift or a photometric redshift with a fractional error < 10\%. From this sample of 71,955 sources, a set of criteria were applied to select AGN based on a mixture of their radio luminosities and their host galaxy $K_s$-band absolute magnitudes.

After applying these criteria, 23,344 sources were left. The caveats to these methods have been outlined by Hardcastle et al. (submitted), but it is also important to mention them in this work. Owing to the nature of the selection criteria applied, it is likely that some RLAGN have been missed, particularly from sources close to the boundary containing star-forming galaxies (SFG) in the WISE colour-colour diagram (discussed in Section \ref{sect-analysis}). Moreover, the selection does not include potentially strong SFG that host RLAGN unless their $L_{144}>10^{25}$ W Hz$^{-1}$. For the purposes of our study the current sample sufficiently describes the population of RLAGN detected in LoTSS DR1. 

This RLAGN sample was selected using a combination of host galaxy
properties and extended radio properties. Using it directly as a comparison
sample with the DDRG sample has a clear drawback; there is a
relationship between the hosts of RLAGN and their radio
luminosities. It is well known that HERGs, which are the more radio-luminous
class of RLAGN, tend to have lower stellar masses and bluer host
colours than those of LERGs \citep{best12}. While this is a generic trend between the
hosts of HERGs and LERGs rather than a one-to-one relationship
\citep{hard18_nature}, such a bias in host galaxy properties may be
manifested in our analysis if the two samples have different
distributions in radio luminosity. Moreover, \cite{best12} and
\cite{jans12} showed a dependency of radio luminosity on the fraction
of galaxies classed as radio-loud, as a function of stellar mass. It
is plausible to suggest that such trends may themselves affect relationships between RLAGN samples. Thus, a comparison of host galaxy properties between samples of RLAGN \textit{must} be matched in radio luminosity. This further ensures that we sampled both populations with similar intrinsic brightness, but also with a similar group of evolutionary states. \cite{hard18} showed that the modelled radio luminosity of a RLAGN of a given jet power varies substantially with source age and size. Radio luminosity is still jet power and environment-dependent, and these physical parameters are difficult to determine with the existing data, but this selection represents the best-matched sample we can produce with current techniques. Another aspect of producing a luminosity-matched sample is that it removes any contamination from extended radio-bright SFG, if they existed in the sample. 

Since the RLAGN sample of Hardcastle et al. (submitted) have redshift estimates, we removed DDRGs from our sample that do not have redshifts, and hence do not have luminosity estimates (9/33), leaving our final sample of 24 DDRGs that can be used to construct a luminosity-matched sample of RLAGN. Our DDRGs were selected to be, in conjunction, the brightest and largest sources in angular size in the DR1 catalogue, and hence we applied the following observational criteria to the original RLAGN sample of 23,302 sources:
\begin{itemize}
\item Sources with 144-MHz total flux density $< 35$ mJy, which is the minimum flux density in the DDRG sample, were removed.
\item Sources with an angular size $< 70$ arcsec, which is the minimum angular size in the DDRG sample, were removed. This criterion was used to remove all compact RLAGN present in the sample, which may represent
a different class of AGN (i.e. compact steep-spectrum and gigahertz
steep-spectrum sources).
\end{itemize}
While this filtering allowed a more representative
  sample of RLAGN relative to DDRGs, our samples were still unmatched
  in $L_{144}$. A two-sided Kolmogorov-Smirnov (KS) test
  \citep{kolm33} returned a $p-$value of < 5 per cent, meaning that we
  can reject the null hypothesis that the two samples are drawn from
  the same distribution at the 95 per cent confidence level. Our RLAGN
  span a broader range of $L_{144}$, and in particular have more
  low-luminosity sources. In order to generate a better-matched sample, we
  restricted the range of $L_{144}$ to that spanned by our DDRGs
  ($10^{24.50}\leqslant L_{144}$ (W Hz$^{-1}$)$\leqslant 10^{27.14}$).
  We further restricted the range in physical sizes of RLAGN to that spanned by our large DDRGs, which improved the match by removing more compact sources. From this pool of 1185 sources, which were still not matched with our RLAGN at the 95 per cent confidence level, we used a sampling technique to construct an $L_{144}$-matched sample:
\begin{itemize}
\item From the sample of 1185 sources, we created ten subsamples that have 10 per cent of the original sample of sources removed, at random.
\item For these ten subsamples, we performed a KS test with the control DDRG sample, comparing their $L_{144}$, using the resulting $p$-value as a test statistic. 
\item For the highest $p$-value out of the ten subsamples, if that $p-$value is $\geqslant$ 0.05, then we used this subsample as our $L_{144}$-matched sample. If the $p-$value is $\leqslant$ 0.05, we repeated step 1, using this reduced subsample as the initial sample.
\end{itemize}
Finally, in order to ensure as much as possible that we sampled only the single-cycle RLAGN, we removed the 91 visually
identified candidate DDRGs from the sample. While only our sample of 24 DDRGs have robust indications of restarted activity, it might be possible that some of the discarded sources from the original sample of 91 contain restarted activity at some level; it is important that this contamination is removed if it exists, although it is likely that the preliminary sample of 91 does not contain \textit{all} of the restarting sources in DR1 and some sources in the RLAGN sample may contain restarting sources. 

\begin{figure}[!h]
\minipage{0.34\textwidth}
        \includegraphics[width=9cm]{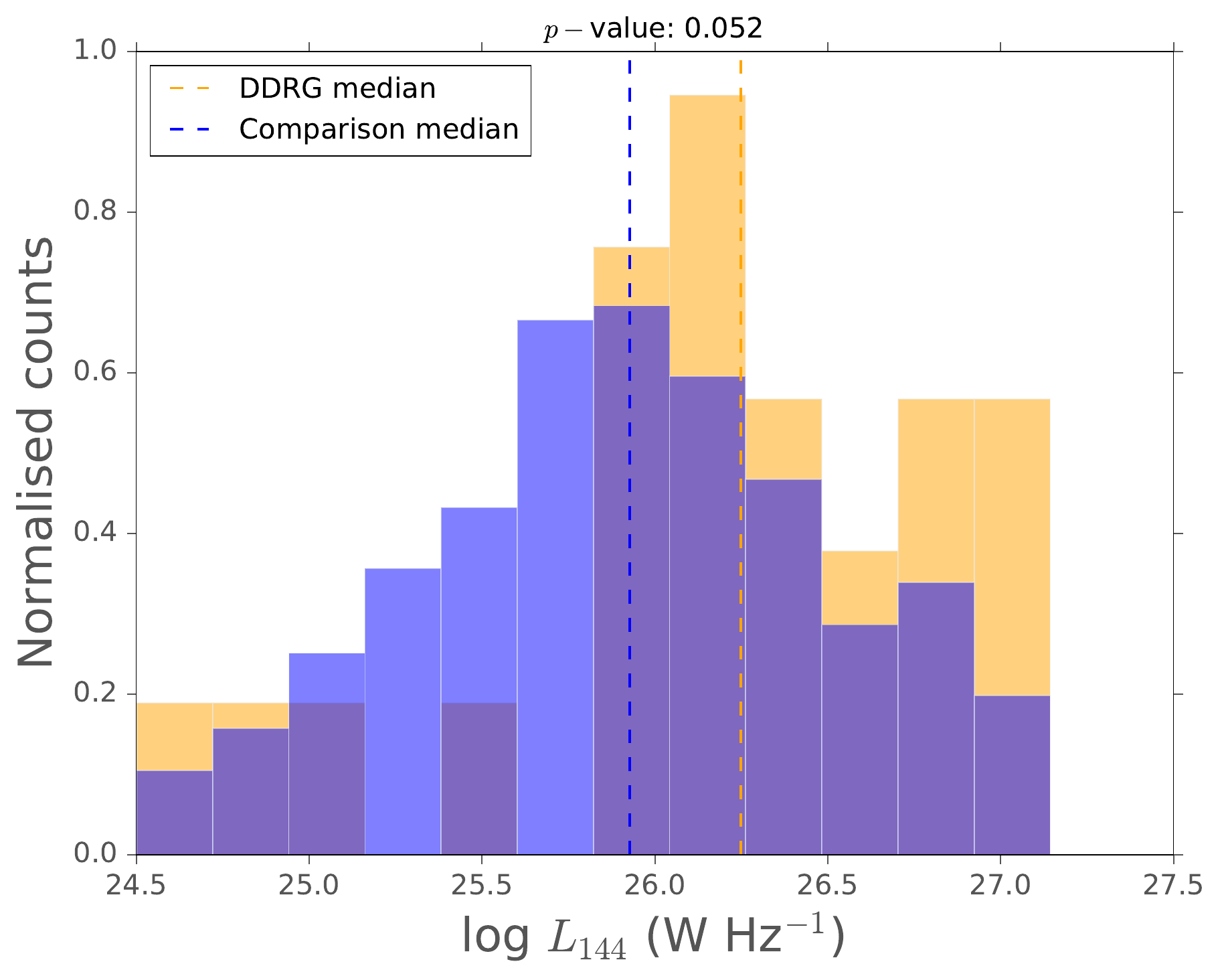}
    \centering
    
\endminipage
\vspace{0.5em}
\minipage{0.34\textwidth}
        \includegraphics[width=9cm]{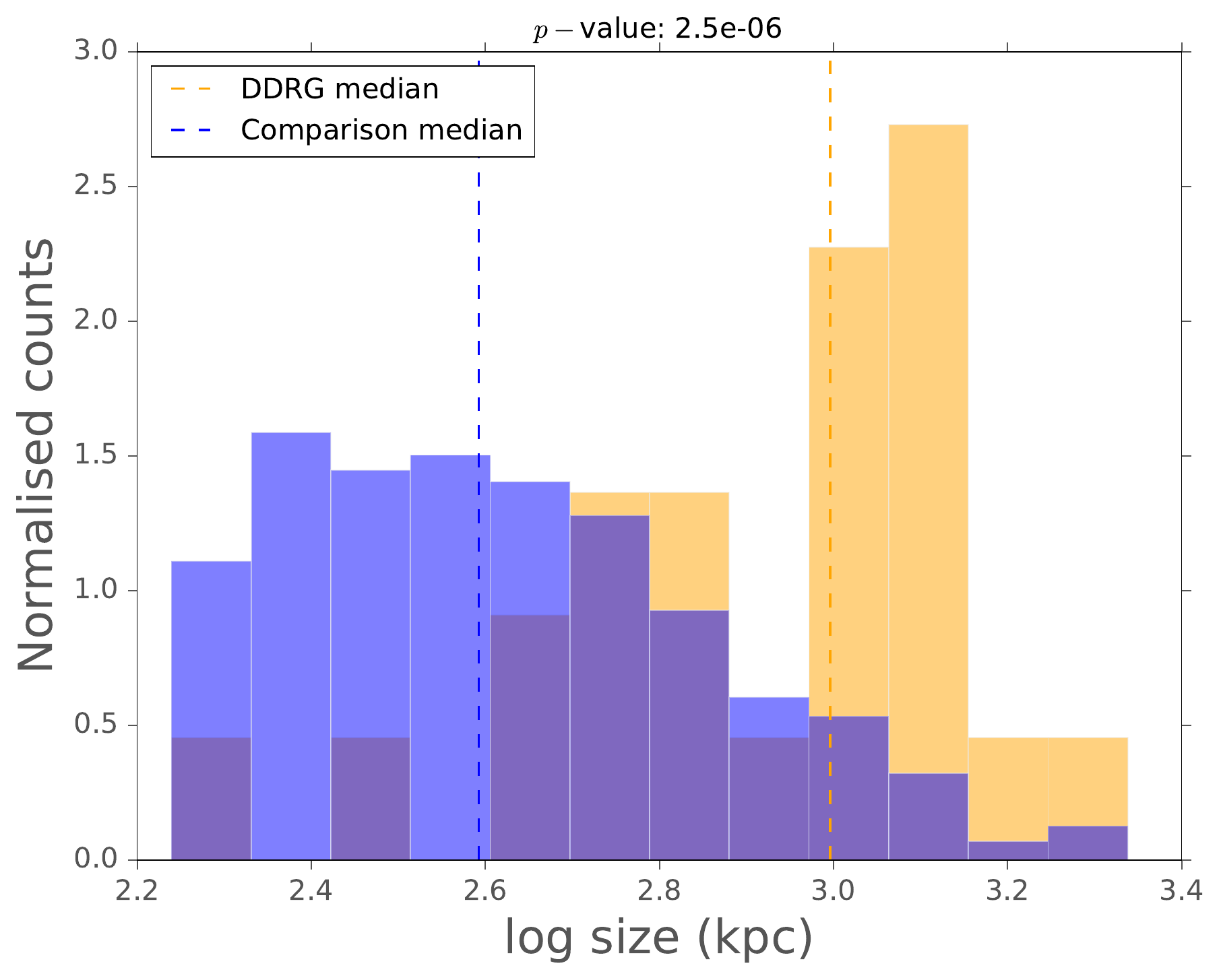}
\endminipage

    \caption{Normalised distributions of total 144-MHz radio luminosity (top panel) and projected total physical sizes in kpc (bottom panel) for our comparison RLAGN (blue; 777 sources) and DDRG (orange; 24 sources) samples. The $p-$value from a KS test between the two samples is given in the figure heading. The dashed lines show the median values from each sample.}
    \label{rlagn-rrlagn_sizes}
\end{figure}

Our final
RLAGN sample consists of 777 sources. Figure \ref{rlagn-rrlagn_sizes}
shows the distribution of $L_{144}$ and physical sizes for our
RLAGN and DDRG samples, based on the total flux densities and sizes
using their combined \textsc{PyBDSF} components (Williams et al. submitted). The $p-$values from their KS tests are stated in the figure heading. While both samples are clearly matched in radio luminosity, it
can be seen that the bulk of the DDRGs have larger physical sizes than
the RLAGN sample. This is likely to be a selection effect due to our
visual inspection method. The DDRGs are most easily identified where the
outer lobe emission is well extended such that the restarted jet has
not reached the end of the outer lobe, causing our selection to be
biased towards both the brightest and largest radio sources. As a check, we created a subsample from our final RLAGN sample, selecting, for each physical size estimate of our DDRGs, five RLAGN with the nearest physical size estimate. This returned a well-matched sample, both in $L_{144}$ and physical size, albeit with a much reduced sample size (120 sources). We confirm that the results of this paper are unchanged with this subsample, and hence use our $L_{144}$-matched-only sample (777 sources). The significantly larger sizes of the DDRGs should be borne in mind for the results presented in Section \ref{sect-results}.

\begin{table}
\centering
\caption{Summary of VLA observations.}
\setlength{\tabcolsep}{4pt}
\begin{tabular}{ccccc}
\textbf{Project} & \textbf{Date} & \textbf{Array} & \textbf{Frequency} & \textbf{Target exposure} \\
\hline
18A-202 & 27/03/2018 & A & 1.4 GHz & ~5 min \\
18A-202 & 28/03/2018 & A & 1.4 GHz & ~5 min\\
\hline
\end{tabular}
\end{table}

\section{Analysis} \label{sect-analysis}
Our analysis is primarily based on comparisons of host galaxy properties between our robustly identified DDRGs and RLAGN sample, on which information is available in the LoTSS DR1 catalogue. This includes observed fluxes, apparent magnitudes, and rest-frame absolute magnitudes (where redshifts are available) in the optical $grizy$ bands and the near-infrared bands including the $K_s$ and WISE bands, as given by the Pan-STARRS 3$\pi$ survey \citep{panstarrs}, the 2MASS extended source catalogue \citep[2MASX;][]{2masx}, and the AllWISE catalogue \citep{allwise}, respectively. With significant differences or similarities between the samples, we may infer the nature of the hosts of DDRGs as a population, and if possible, understand the host galaxy conditions that may drive restarted AGN activity. Where required, we performed two-sample KS tests for each set of distributions, and quote the $p-$value (labelled in our figures), where we use a $95$ per cent confidence level throughout.

\begin{figure}[!h] 
\centering
\minipage{0.45\textwidth}
  \includegraphics[width=\linewidth]{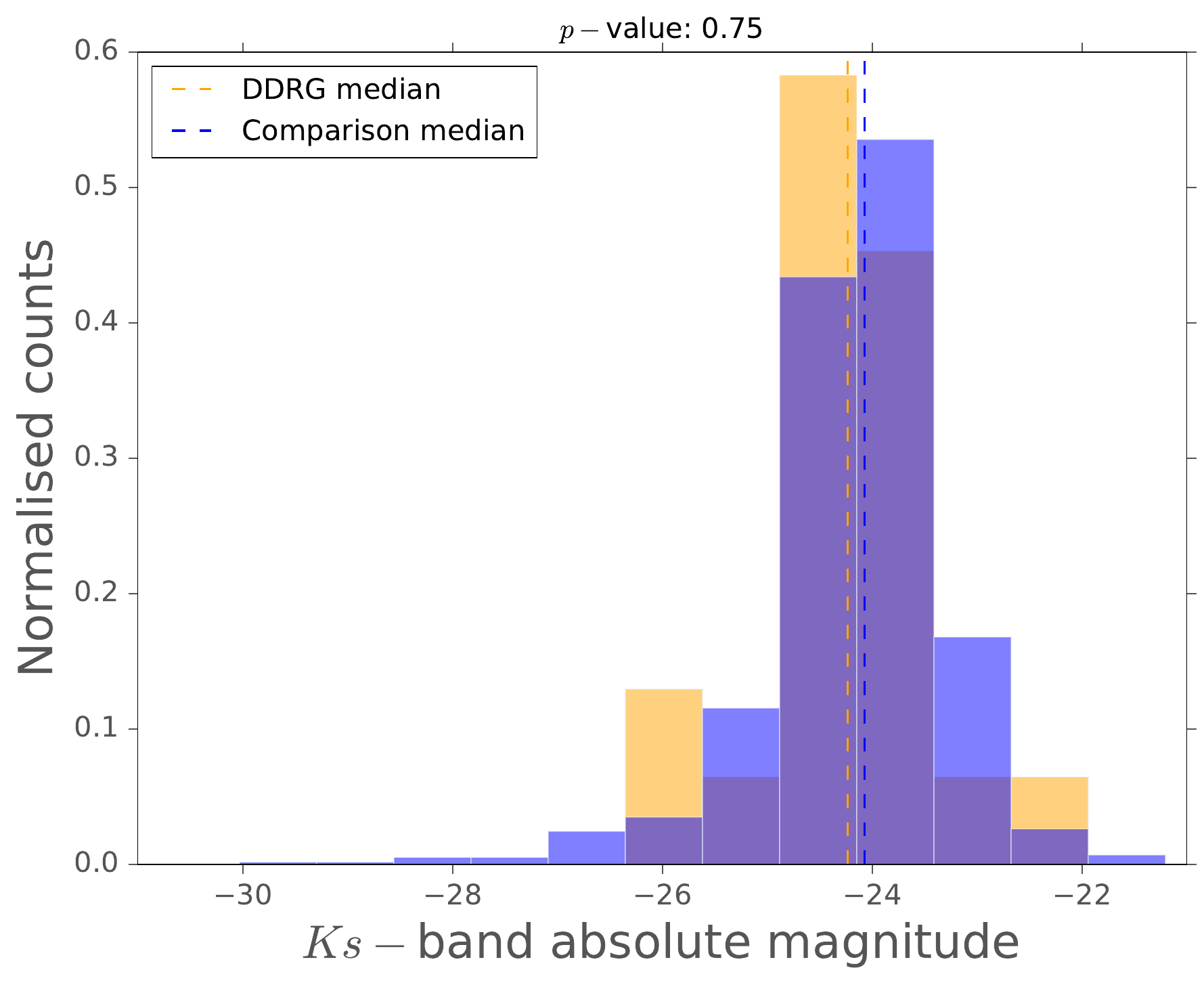}
\endminipage
\hspace{0.01em}
\minipage{0.45\textwidth}
  \includegraphics[width=\linewidth]{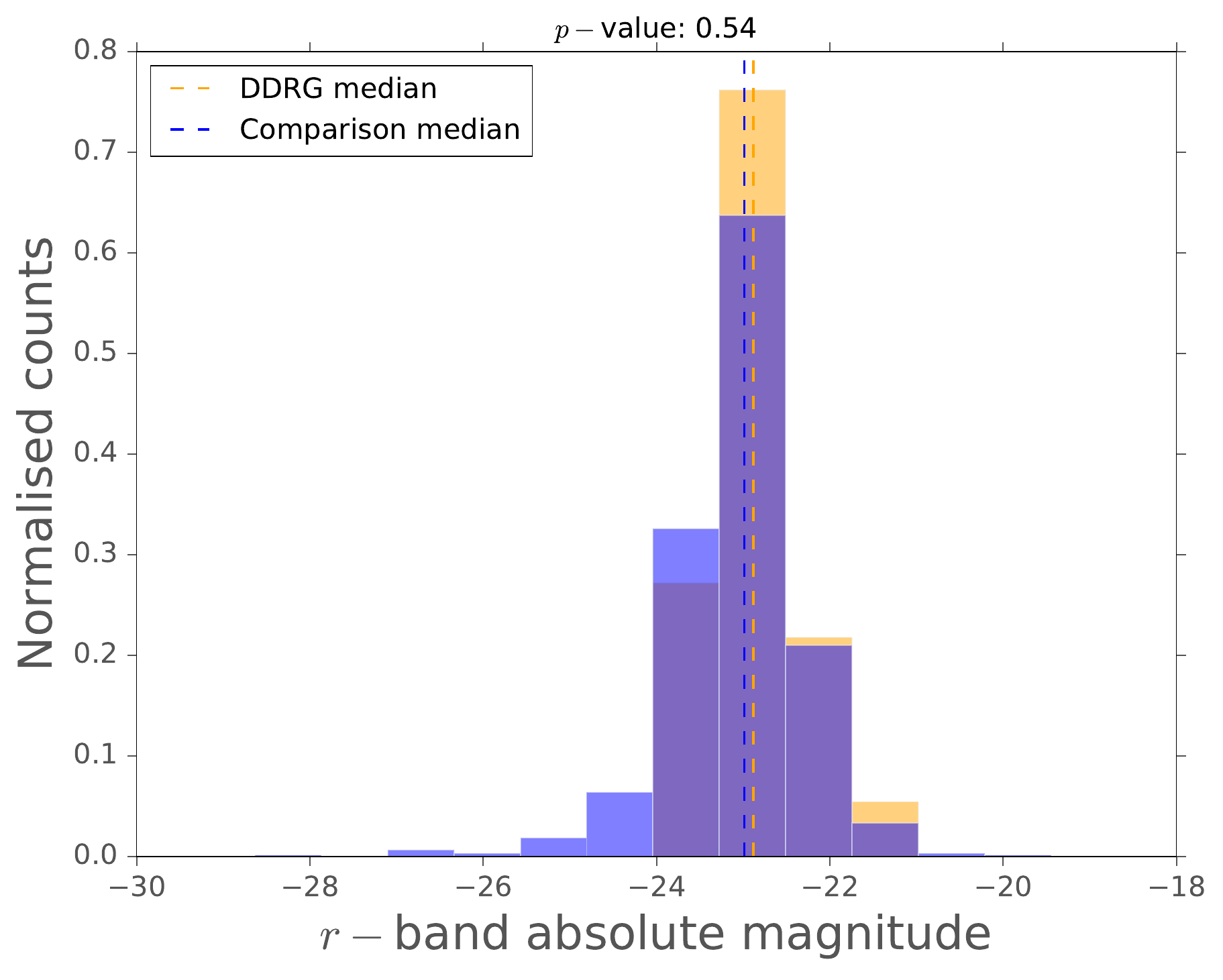}
\endminipage
\hspace{0.01em}
\minipage{0.45\textwidth}%
  \includegraphics[width=\linewidth]{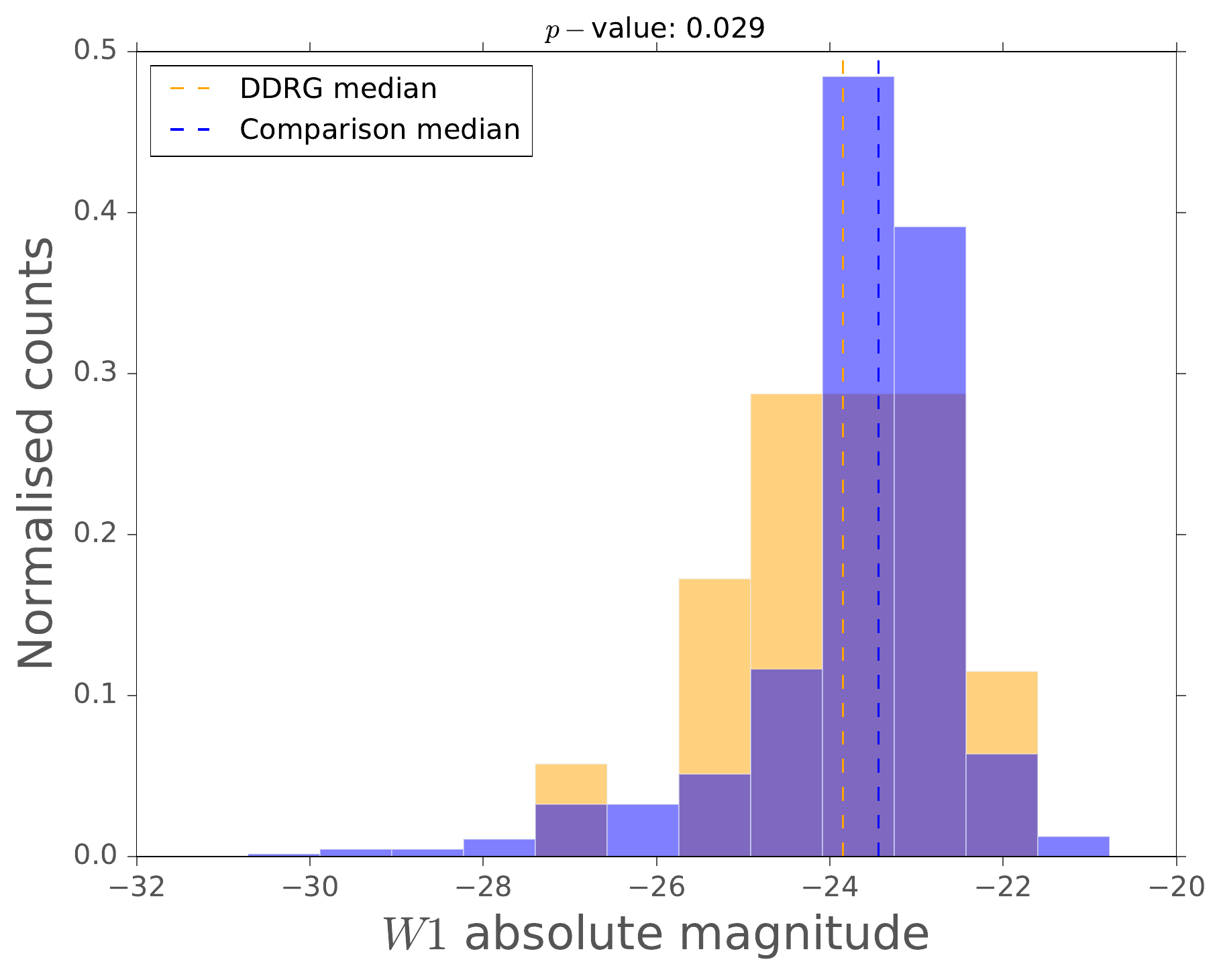}
\endminipage
\caption{Host galaxy rest-frame absolute magnitudes for RLAGN and DDRGs. From left to right: $K_s$-band magnitude, $r$-band magnitude, and $W1$-band magnitude. The $p-$value from a KS test between the two samples is given in the figure heading.}
\label{restframe_hists}
\end{figure}

In Figure \ref{restframe_hists} we plot the normalised distributions (such that the area under the histogram sums to one) of rest-frame absolute magnitudes of DDRGs and RLAGN, in the $K_s$, $r$ and $WISE$ $3.4\mu$m ($W1$) bands. It can be clearly seen that both DDRGs and RLAGN follow the same distributions of host galaxy absolute magnitudes in all three bands with similar median values. The $p$-values from a KS test are $>0.05$ for the distributions in $M_{K_s}$ and $M_r$, as shown in Figure \ref{restframe_hists}, meaning that we cannot reject the hypothesis that both samples can be drawn by the same distribution at the 95 per cent confidence level. The KS test for the distribution in $M_{W1}$, however, gives a $p-$value $\leqslant 0.05$. We attribute this slightly lower $p-$value to the small tail of extremely bright galaxies (likely quasars) with $M_{W1}\leqslant -25$ (see the bottom panel of Figure \ref{restframe_hists}). We computed a Wilcoxon-Mann-Whitney (WMW) test \citep{mann47}, which is similar to the KS test, but is more sensitive to a discrepancy between the peaks of our two samples. The test returned a $p-$value of $0.1$, and hence we cannot reject the null hypothesis that the two samples can be drawn from the same distribution, at the $95$ per cent confidence level. For consistency, for the proceeding analysis we performed both the KS and WMW tests and we confirmed the $p-$values between the tests give the same result at the $95$ per cent confidence level. Henceforth we state only the $p-$values from the KS tests. 

\begin{figure}
        \includegraphics[width=9cm]{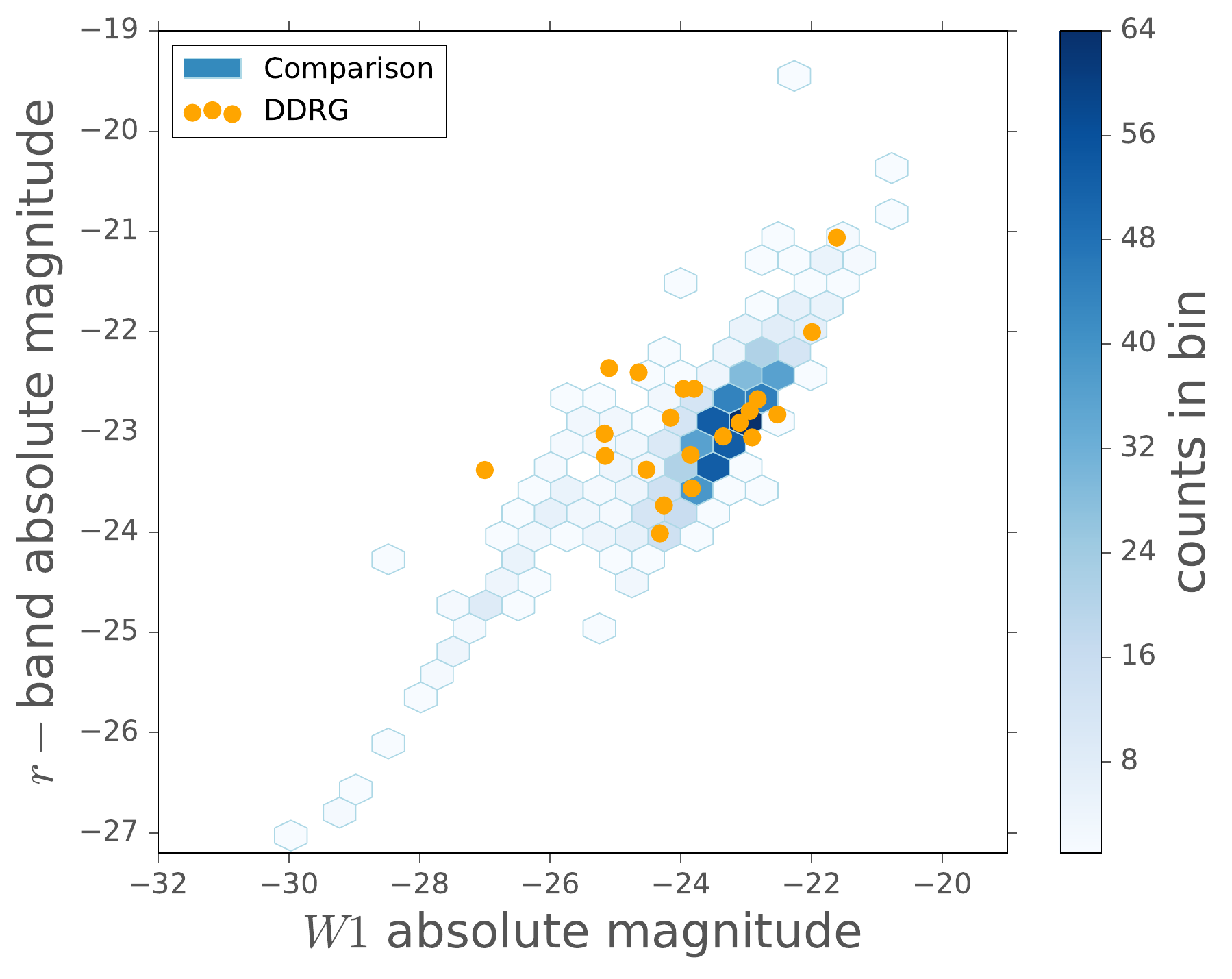}
    \caption{RLAGN and DDRGs plotted for $W1$ rest-frame magnitude against their $r$-band rest-frame magnitude for their hosts.}
    \label{w1_r}
\end{figure}

In Figure \ref{w1_r} we plot $W1$ against the $r-$band rest-frame absolute magnitudes. While there is a clear and expected relationship between the optical and near-infrared host galaxy brightness, both DDRGs and RLAGN lie along the same correlation. The immediate inference is that the population of DDRGs and RLAGN are not hosted by galaxies of significantly different brightness, mass (which we infer from the similar $M_{K_s}$), and emission from stellar populations (traced by $W1$ -- see below). Our data therefore suggests that DDRGs and single-cycle RLAGN are driven by the same type of galaxy, in a statistical sense.

\begin{figure*}[!h]
\centering
\minipage{0.45\textwidth}
  \includegraphics[width=\linewidth]{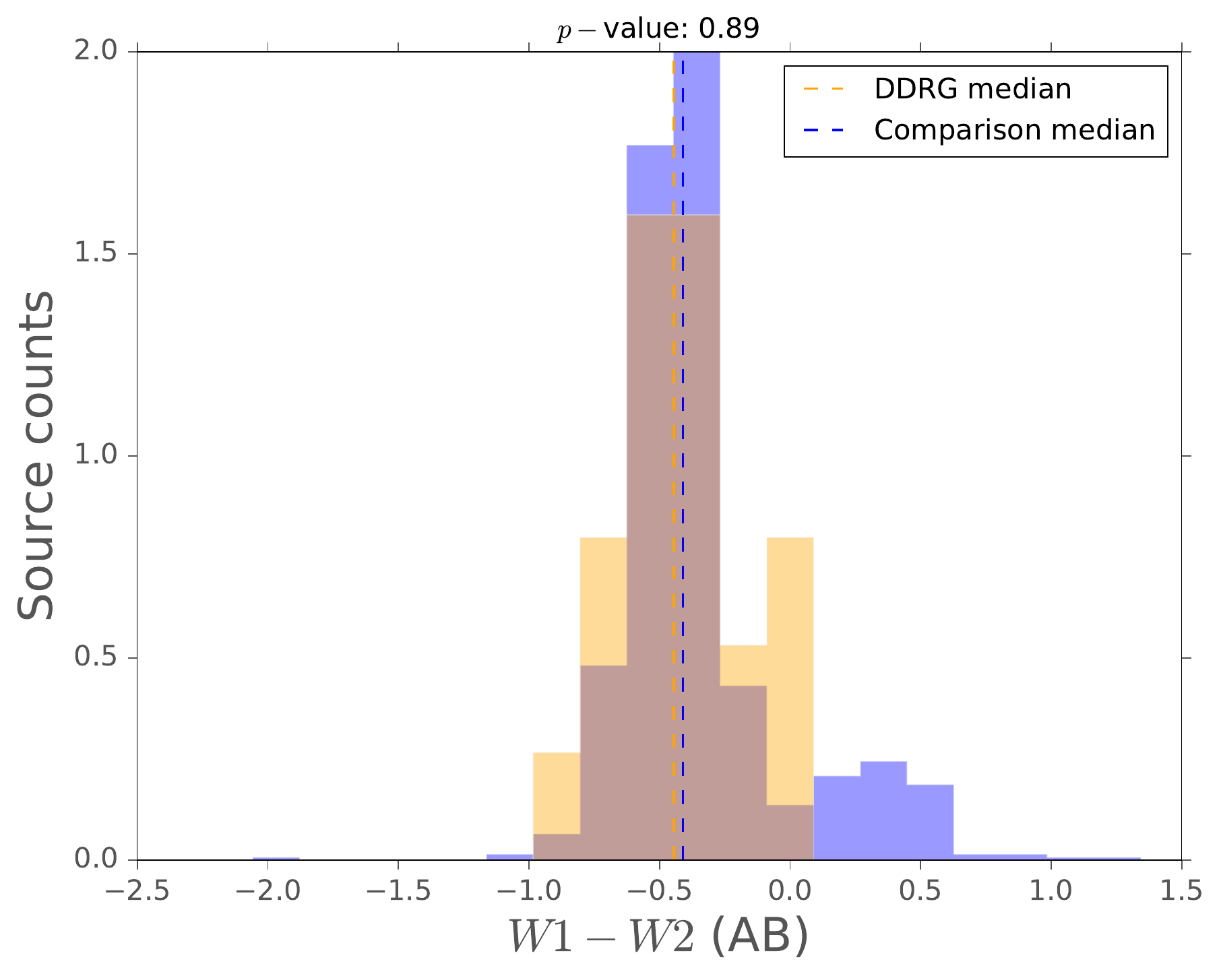}
\endminipage
\hspace{0.1em}
\minipage{0.45\textwidth}
  \includegraphics[width=\linewidth]{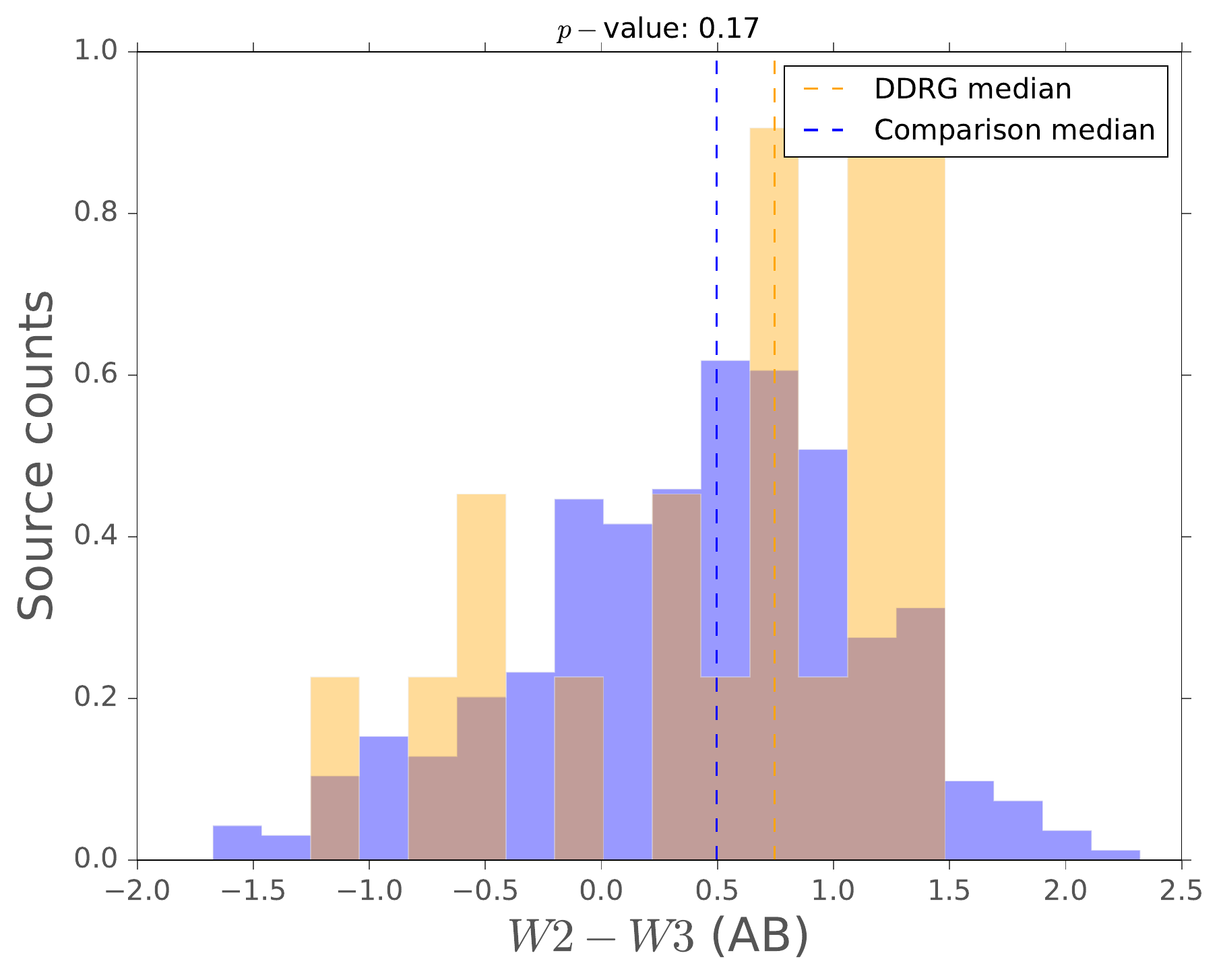}
\endminipage
\hspace{0.1em}
\caption{Host galaxy WISE apparent colours (in AB system) for RLAGN and DDRGs. From left to right: $W1-W2$, $W2-W3$. The $p-$value from a KS test between the two samples is given in the figure heading.}
\label{wise_colours}
\end{figure*}

When comparing the host galaxies of various classes of AGN, it is important to understand the magnitude and effects of dust either due to large amounts of star formation or the formation of a dusty torus around the central AGN. Significant differences in these physical parameters between life cycles of AGN activity can have important implications for the nature of RLAGN and the driving mechanisms for their restarting phase. At low redshifts, the WISE 3.4 ($W1$) and 4.6 ($W2$) $\mu$m bands primarily sample continuum emission from stellar photospheres,
whereas at longer wavelengths the 12 ($W3$) and 22 $\mu$m ($W4$) bands are more sensitive to warm dust emission heated by stars or the dusty torus surrounding some
accreting black holes \citep{wise}. Therefore, a higher $W1-W2$ colour (redder in near-infrared) indicates dustier and/or increasing star-forming objects, while a lower value indicates old stellar populations. The $W2-W3$ colours scale in a similar way, although the $W2$ and $W3$ bands are more sensitive to re-radiated emission from dust rather than direct heat sources (stars). We plot the distributions of WISE apparent colours in Figure \ref{wise_colours}. The KS tests return a $p$-value of more than 5 per cent for both distributions, meaning the $W1-W2$ and $W2-W3$ colour between DDRGs and RLAGN can be drawn from the same distribution, agreeing with the distribution of rest-frame optical magnitudes.

\begin{figure*}[ht!]
        \centering
        \includegraphics[width=15cm]{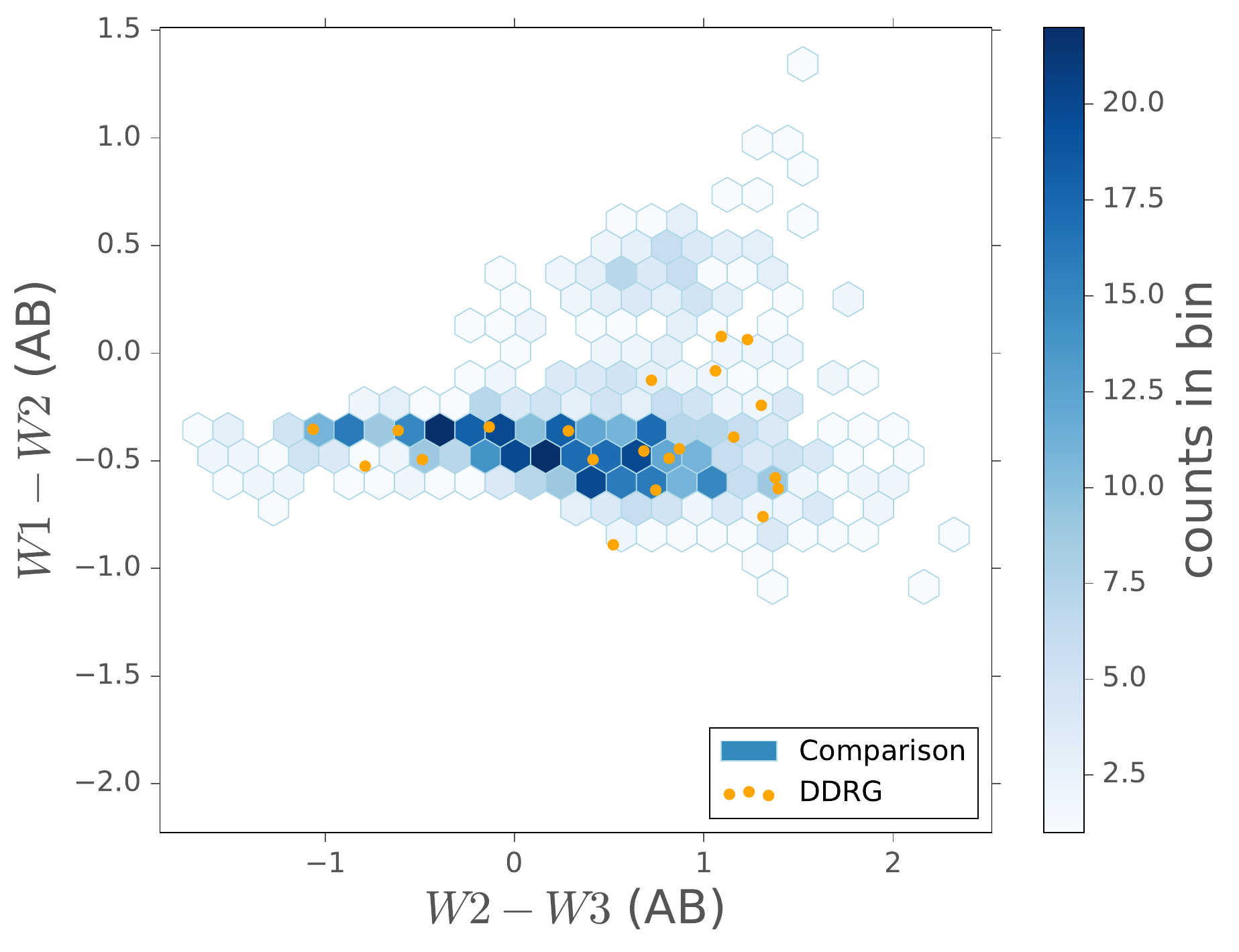}
    \caption{RLAGN and DDRGs in WISE colour-colour space. The hexagonal bins represent the density of RLAGN sources, while the orange scatter points represent the DDRGs}
    \label{wise_colour-colour}
\end{figure*}

The WISE apparent colour-colour diagram \citep{wrig10,yan13} is  known
to effectively separate SFG and AGN in galaxy samples at low redshift.
Significantly higher $W1-W2$ and $W2-W3$ colours than the population
of galaxies hosting AGN tend to select bright quasars presumably with
nuclear obscuring tori, while lower and bluer $W1-W2$ colours, which
primarily sample stellar photospheres,tend to be populated by galaxies
with old stellar populations. Apparent colour-colour diagrams from WISE can
therefore give information on the nature of the hosts of DDRGs
relative to those of RLAGN. Figure \ref{wise_colour-colour} shows the
colour-colour diagram of our DDRGs (orange points) and RLAGN (blue
density bins) samples, which is essentially a two-dimensional
representation of Figure \ref{wise_colours}. It can be seen
immediately that both DDRGs and RLAGN seem to reside in similar host
galaxies. The hosts of DDRGs and RLAGN have the same range and
distribution in levels of dust and emission from stellar populations.
There is indeed a bimodality in colour-colour space, as shown by a
similar figure by Hardcastle et al. (submitted), which shows the entire
RLAGN sample from LoTSS DR1. The smaller population of RLAGN in Figure
\ref{wise_colour-colour} towards higher $W1-W2$ are quasars or have
quasar-like hosts (HERGs), whereas the bulk of the RLAGN have lower
$W1-W2$. These host galaxy colours are indicative of LERGs, which are the dominant population at low
redshift. It is interesting to see that DDRGs also tend to lie along the
parameter space dominated by LERGs, although our selection bias
towards large angular size sources possibly causes us to neglect higher redshift DDRGs, which may have quasar-like colours. 

If galaxy mergers or enhanced rates of star formation were to initiate
the process of disruption and restarting of the jets, we might expect
to observe significantly bluer hosts than those of single-cycle
RLAGN. Although the restarted radio activity may manifest itself sooner than signatures of subsequently enhanced star formation becoming observable following a merger\footnote{\cite{emon06} derived a significant ($\sim$ 0.3 Gyr) time delay between a merger and the onset of a starburst event for the radio galaxy B2 0648+27, a timescale during which many cycles of RLAGN activity may persist.}, restarted jets rapidly drive into the remnant activity on short timescales and effectively become single cycle RLAGN, as observed in radio observations. Hence, if mergers were the significant driver for the formation of DDRGs, enhanced star formation should also be naturally correlated with the hosts of single-cycle RLAGN. The WISE colours of this low-redshift ($z\leqslant 1$) RLAGN sample suggest otherwise; see Figure 4 of Hardcastle et al. (submitted) for a comparison of RLAGN, quasars, and star-forming objects in WISE colour-colour space. 

It is also plausible to suggest a scenario in which a major
merger between an elliptical galaxy originally hosting the RLAGN and a
gas-rich spiral leads to a significant and periodic infall of gas
towards the central AGN. While the merger itself may disrupt the jet
activity, causing a switch to the remnant phase, the subsequent infall
of gas may re-fuel the AGN, causing a restarting jet or a DDRG. The
short timescales of radio-loud activity can support this scenario: if
we assume that the small remnant fraction of $\leqslant 10$ per cent
found by \cite{godf17}, \cite{brie17}, and \cite{maha18} directly
relate to the synchrotron timescales\footnote{In reality, adiabatic
  losses also contribute to the rapid energy losses of remnants and
  hence also to the remnant fractions in these studies.}, then the
remnant (and subsequent restarting shortly after) phase for sources
with an assumed active lifetime of $50$ Myr is $5$ Myr, which may
relate to the timescales of quasi-periodic infall. Within the $\sim 1$
Gyr timescales of a merger, repeated outbursts of AGN activity, or
double-double phases, might take place. However, our finding that DDRG
galaxy magnitudes and colours are similar to those of RLAGN in general does not support this scenario for the population of DDRGs. 

To check for consistency with a single class of RLAGN, we used the
FR-II radio galaxy sample of Mingo et al. (in prep) from LoTSS DR1.
The FR-II sample was obtained via an automated classification
algorithm (Mingo et al. in prep), which applies the traditional
Fanaroff-Riley separation based on whether the peaks in brightness are
closer to the centre or outer edges of the emission. The
algorithm was applied to all resolved sources in the RLAGN sample of
Hardcastle et al. (submitted) and was found to have a reliability of
>96\% (when compared with visual classification) for objects with
S$_{\rm 144MHz} > 10$ mJy and angular size greater than 50 arcsec. The
sample used in this work consists of all sources meeting these criteria with a
classification of FR-II. We further cut the sample in total flux and
angular size, as for the RLAGN sample, and removed any DDRGs contained
in the FR-II sample, leaving a sample size of 323. We first mention
some caveats for the use of this sample. This sample is clearly biased
towards the brighter and more luminous class of RLAGN, as for our DDRG
sample. The main value of this comparison is that the morphologies of
the FR-II sample closely resemble those of our DDRGs, whereas
our RLAGN sample includes a range of morphologies (FR-I and
FR-II). Moreover, we may directly compare our results with those of \cite{kuzm17}, who have used a sample of FR-IIs as a comparison sample against DDRGs; see Section \ref{sect-results}. 

\begin{figure}[ht!]
        \includegraphics[width=9cm]{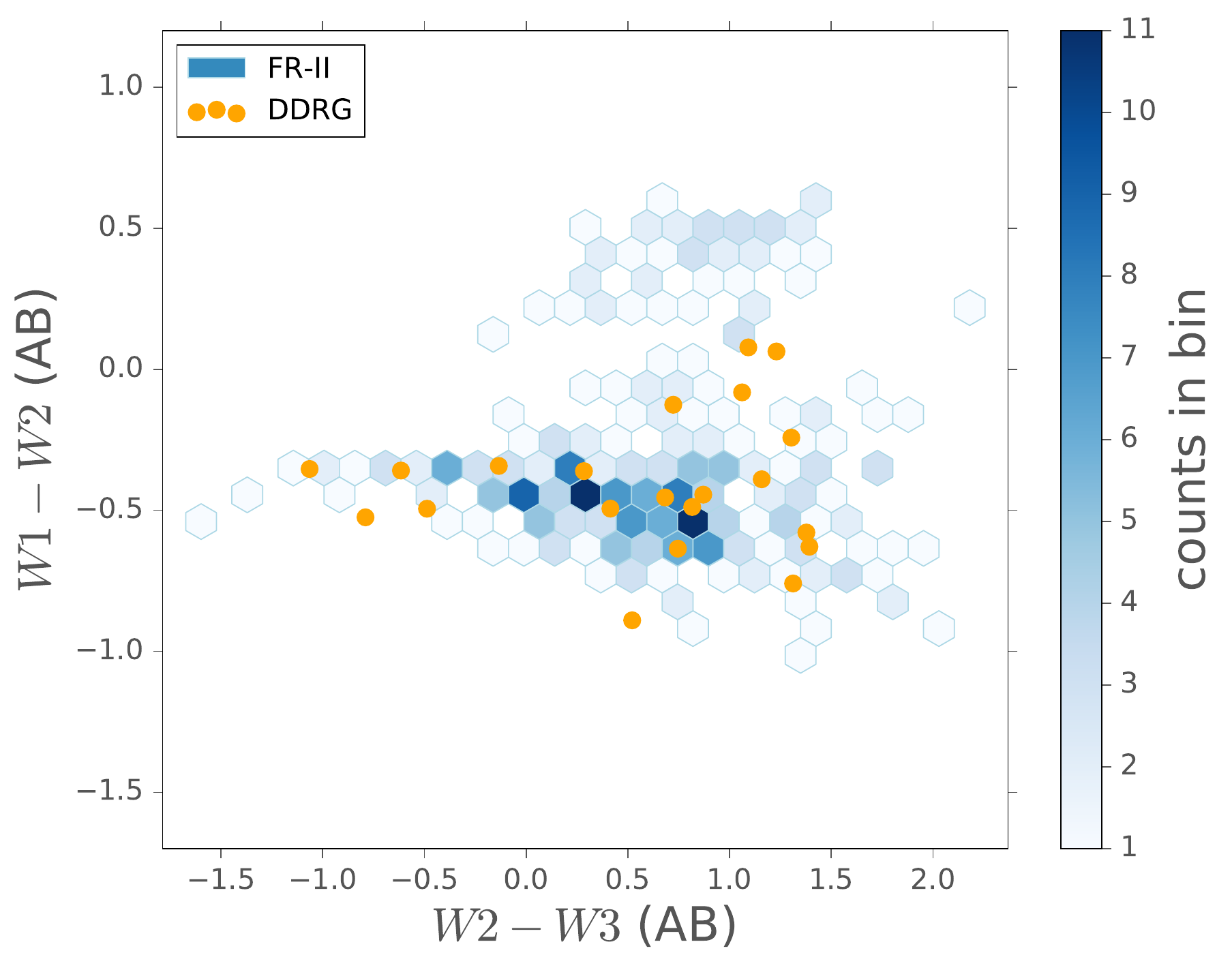}
    \caption{FR-II radio galaxies (blue) and DDRGs (orange) in WISE colour-colour space.}
    \label{wise_colour-colour_fr2}
\end{figure}

We again plot the WISE colour-colour
diagram in Figure
\ref{wise_colour-colour_fr2}, now for the DDRG and FR-II samples, and see a familiar trend as in Figure
\ref{wise_colour-colour}; the FR-II sources seem to trace a similar
range of parameter space in WISE colour as for the population of
RLAGN. Our results confirm that the integrated stellar properties of
galaxies hosted by DDRGs and RLAGN are indistinguishable with our data, both with all classes of RLAGN and with RLAGN of similar morphology as DDRGs.  

\begin{figure}[!h]
        \centering
        \minipage{0.45\textwidth}
                \centering
                \includegraphics[width=\linewidth]{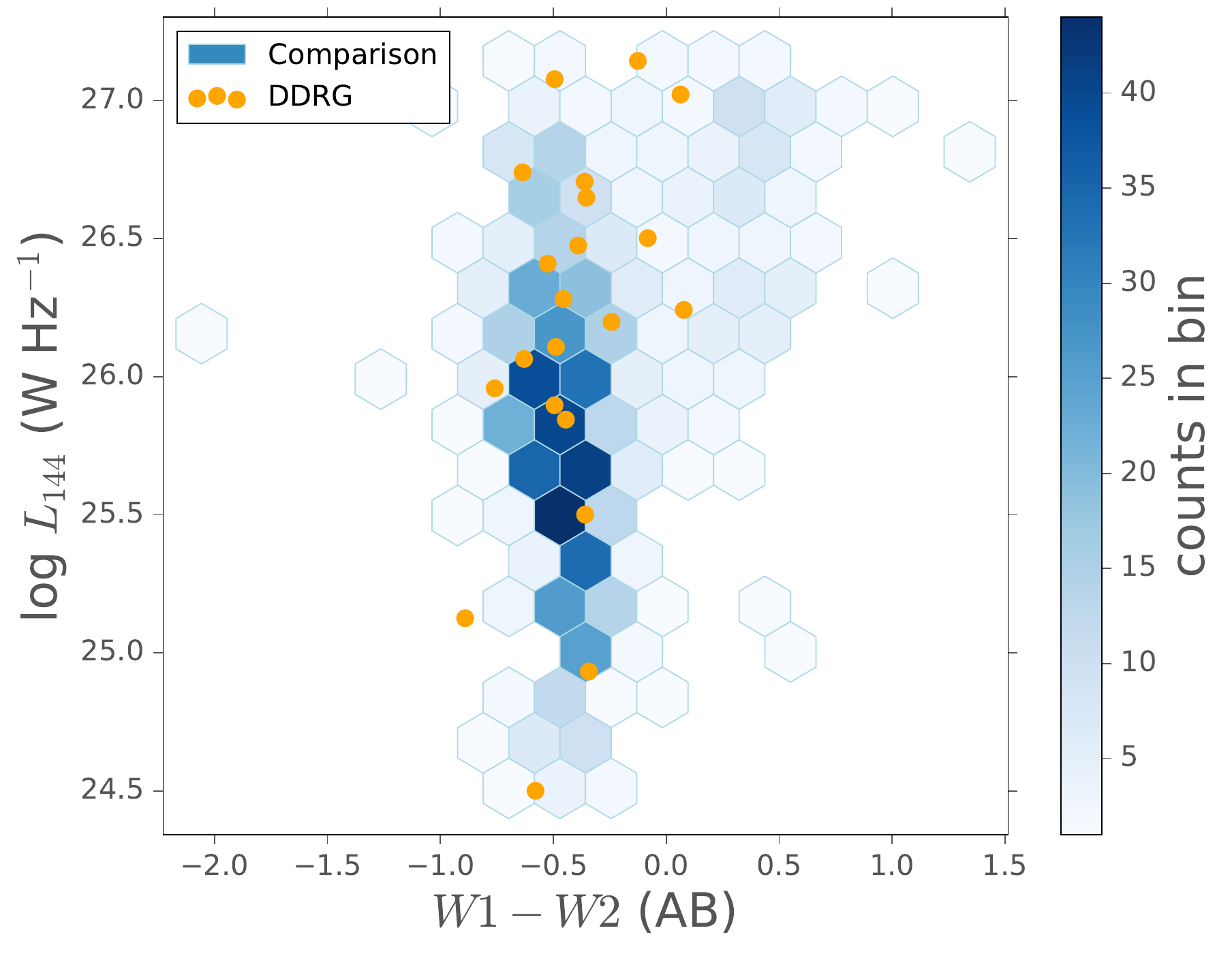}
        \endminipage
\hspace{1em}
        \minipage{0.45\textwidth}
                \centering
                \includegraphics[width=\linewidth]{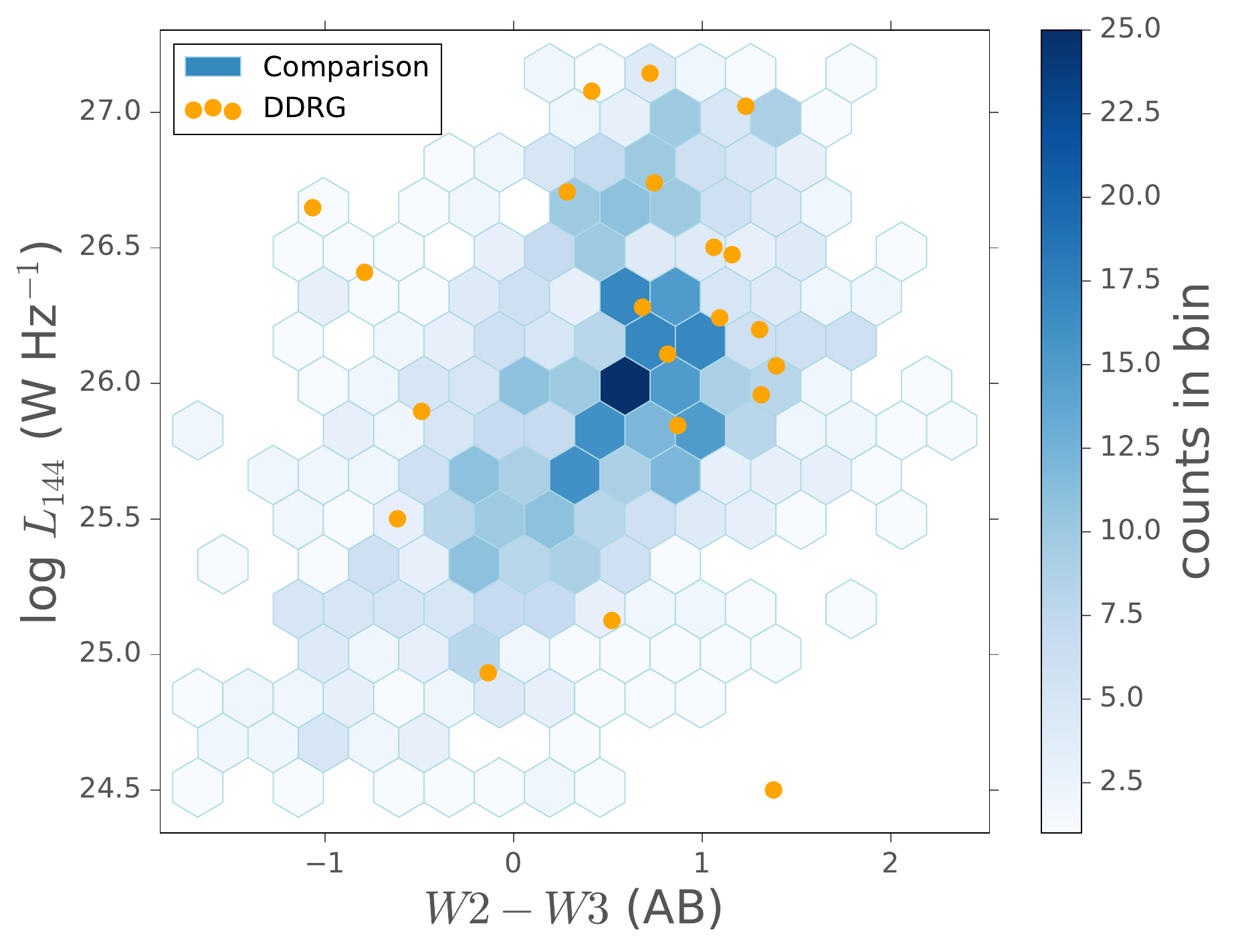}
        \endminipage
\hspace{1em}
\caption{Luminosity distribution of RLAGN (blue density hexagons) and DDRGs (orange points) with $W1-W2$ (top) and $W2-W3$ (bottom).}
\label{wise_colour-l150}
\end{figure}

\begin{figure}
        \includegraphics[width=9cm]{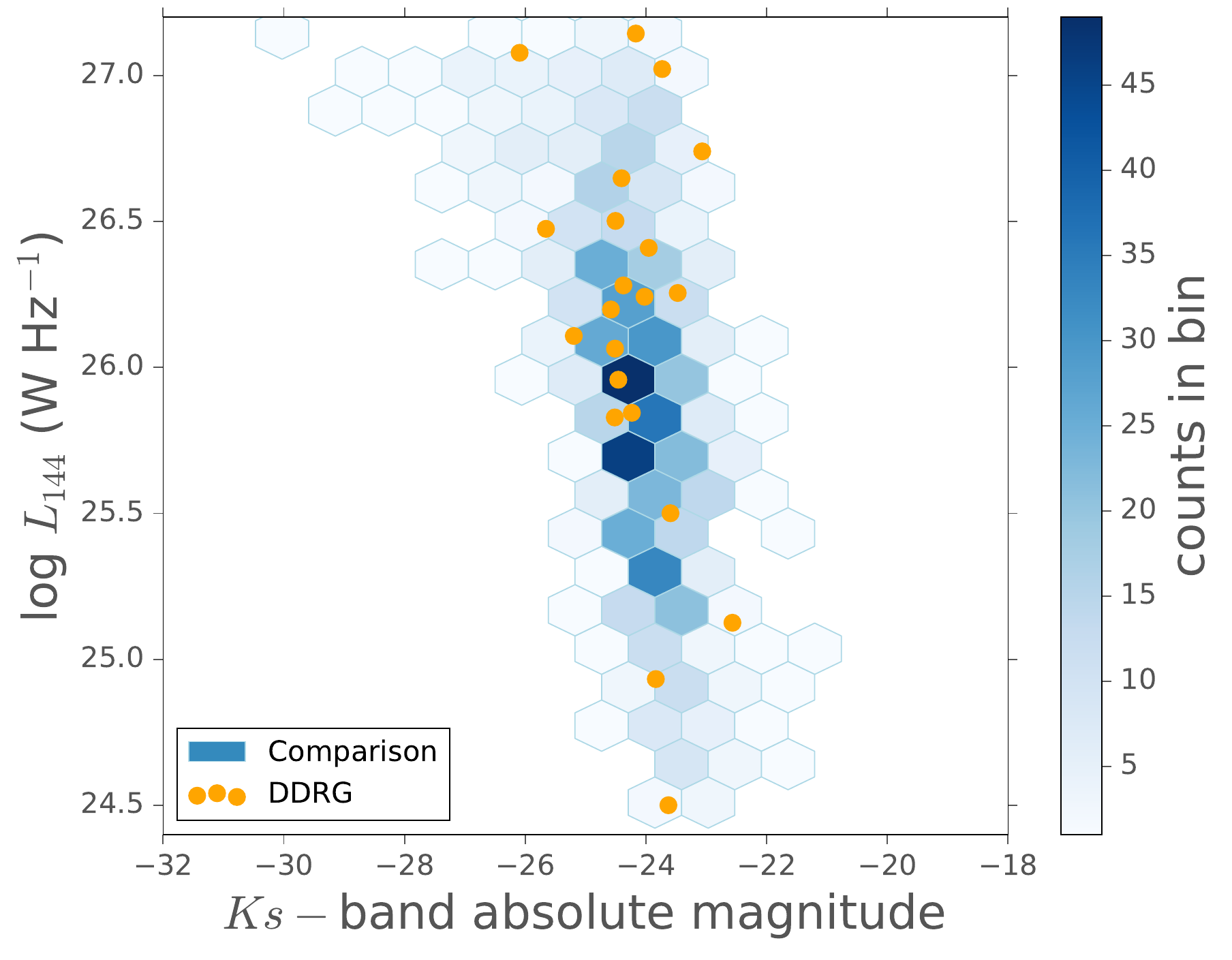}
    \caption{RLAGN and DDRGs plotted for $K_s$ rest-frame magnitude against their $L_{144}$.}
    \label{ks_l150}
\end{figure}

In Figures \ref{wise_colour-l150} and \ref{ks_l150} we plot WISE colours and $M_{K_s}$ against $L_{144}$ of the extended radio emission for our DDRG and RLAGN samples, respectively. It can be clearly seen that the distributions in $M_{K_s}$ and WISE colour, or host galaxy brightness, is independent of radio luminosity, between DDRGs and RLAGN. Therefore, it can be inferred that both DDRGs and RLAGN are hosted by galaxies of similar mass, but also as a function of their radio luminosity (i.e. radio properties). Trends between radio luminosity and host galaxy properties therefore do not affect our results.

\begin{figure}[ht!]
        \centering
        \includegraphics[width=9.0cm]{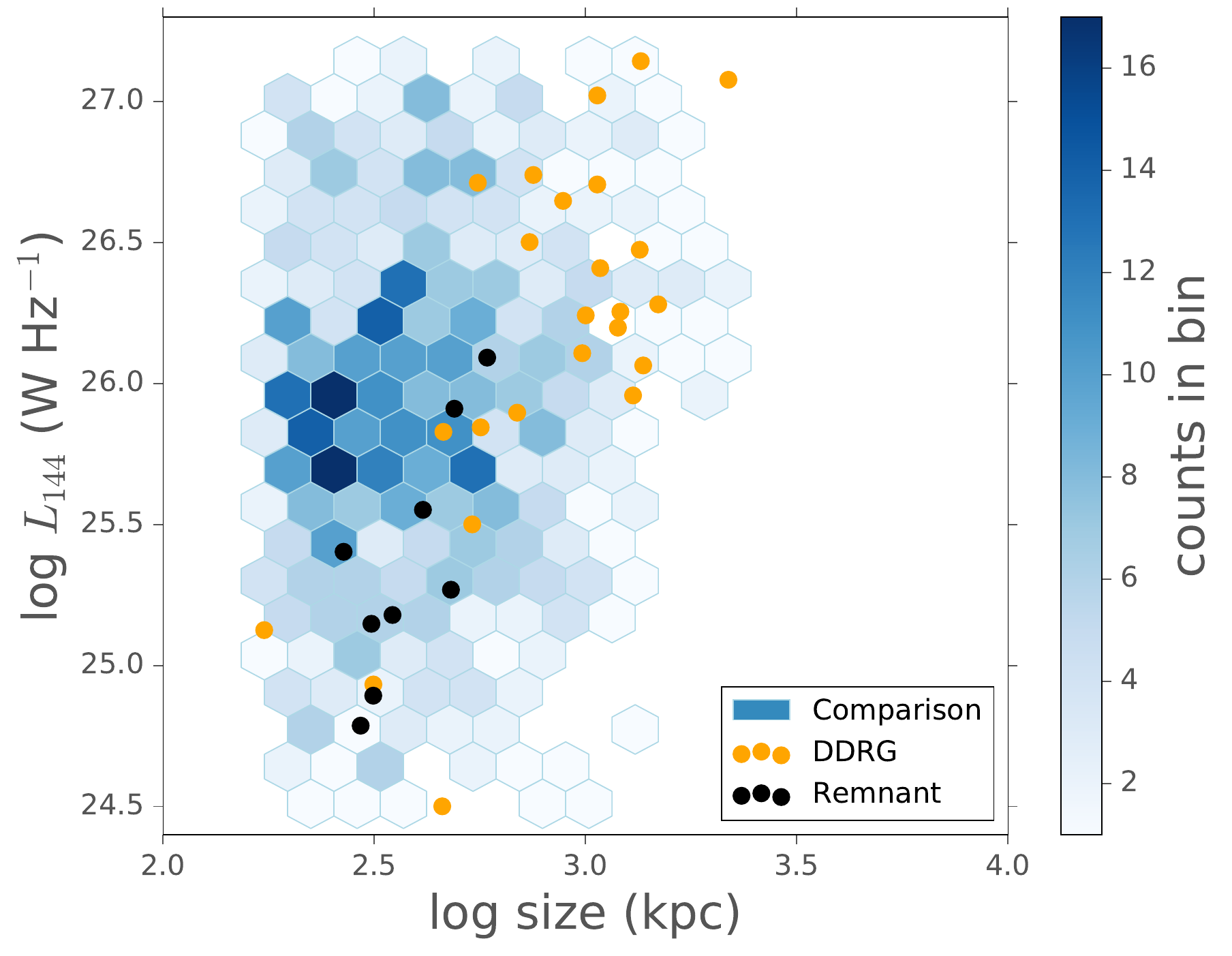}
    \caption{Power-size diagram of DDRGs (orange), RLAGN (blue density hexagons), and candidate remnants (black) from \cite{maha18}.}
    \label{power_size}
\end{figure}

Figure \ref{power_size} shows the total source length against
$144$-MHz luminosity (the so-called power-size diagram, often used to
trace the evolution of RLAGN in their radio properties) of our DDRG
and RLAGN samples. We also overlay the candidate remnant sample of
\cite{maha18}, making use of similar LOFAR observations in the
\textit{Herschel}-ATLAS field \citep{hard_lofar16}. Remnant RLAGN are expected to be similar in linear size with DDRGs, since
restarted activity is expected to occur soon after the original
switch-off, such that the buoyantly rising remnant lobes have not
significantly increased in size in such a short timescale. This sample
of DDRGs, however, are clearly physically larger and more luminous.
This is a selection effect. Larger angular sizes contribute greater
to measured flux densities than smaller sources, but crucially, the
total flux density and angular size cut imposed on the DDRGs are
approximately double in value to those used for the candidate remnants. More
evolved, or larger, remnants that are fainter, rapidly escape
detection and hence may not appear in such non-systematic comparisons.
The LoTSS DR1 observations, probing $\sim 30$ mJy/beam deeper than the
deepest part of the H-ATLAS observations, also
have a higher sensitivity, which may partly explain the
difference in size distribution with the candidate remnants of \cite{maha18}. The RLAGN in blue are
also clearly smaller in physical size, as shown in Figure
\ref{rlagn-rrlagn_sizes}, but the crucial point is that some DDRGs also occupy this space. Thus selection of DDRGs or remnants
  using the power/linear size plot is likely to be difficult without
  follow-up visual inspection of the radio images.

\section{Discussion} \label{sect-results}
Following our analysis in Section \ref{sect-analysis}, our key findings are that
\begin{itemize}
\item DDRGs and normal RLAGN have the same distributions of host galaxy brightness in the optical $r$-band, near-infrared $K_s$-band, and mid-infrared WISE bands.
\item The hosts are also similar in WISE colour, indicating that the global galaxy stellar populations \text{and} the relative levels of cold and warm gas present, on average, are similar. The similarity in WISE colour-colour distribution is repeated when the control sample is a sample of bright nearby FR-II sources. The colour distributions are not radio luminosity-dependent.
\item The bulk of the DDRGs have similar WISE-1 absolute rest-frame magnitude for a given optical rest-frame absolute $r$-band magnitude, providing further evidence for similar stellar populations in the hosts between RLAGN and DDRGs. 
\item In our sample, DDRGs are significantly more luminous and larger in physical size than a small sample of candidate remnants, although this is likely driven by selection effects.
\end{itemize}
Our results on host galaxy property comparisons lead us to the conclusion that the restarting phase of DDRGs are not a consequence of significant changes in their host galaxy -- the galaxies that drive RLAGN also drive DDRGs. Although follow-up optical imaging or studies may give information concerning any signatures of mergers surrounding the hosts of DDRGs, the similar distributions in optical and near-infrared rest-frame magnitudes between DDRGs and RLAGN give evidence to suggest otherwise. The insignificance of host galaxy properties is further supported by the WISE colour-colour diagrams of Figure \ref{wise_colour-colour} and Figure \ref{wise_colour-colour_fr2}.

It is interesting to compare our results with those of \cite{kuzm17},
who have performed a similar study in the comparison of host galaxy
properties between DDRGs and FR-II radio galaxies. Contrary to our
results, \cite{kuzm17} have found a significant difference in host galaxy
properties. These authours have found that hosts of DDRGs have younger stellar populations relative to the
FR-II hosts. They have also found  the hosts of DDRGs tend to have lower stellar masses, and smaller $r-$band
concentration indices, indicating disturbed galaxy morphologies or
non-ellipticals. It is plausible to suggest that most of their sources
are HERGs; none
of their sources occupy the parameter space of the diagnostic Baldwin,
Phillips, \& Terlevich \citep[BPT;][]{bald81} diagram where galaxies
with old stellar populations (i.e. `red and dead' ellipticals) tend to
be located. This is consistent with the WISE colour-colour diagram of
\cite{kuzm17}, which displays a large fraction of both DDRG and
FR-II hosts with $W1-W2 > 0.5$ and/or $W2-W3 > 1.5$, in the region
where HERGs, or dusty/SFG may be expected to be present. Our DDRGs display a different behaviour, predominantly consistent with LERGs, which tend to have redder hosts \citep{best12}.

In terms of selection, their comparison sample consists of the class
of FR-II radio galaxies, whereas our RLAGN sample is drawn from the general
population of RLAGN. It might be expected therefore, based on the
results of \cite{kuzm17}, that host galaxy differences exist between
DDRGs and FR-IIs, but our WISE colour-colour diagram with FR-IIs (Figure
\ref{wise_colour-colour_fr2}) suggests otherwise. As a consistency
check, we repeated the plots shown in Figures \ref{restframe_hists}
--- \ref{power_size}, using the FR-II sample of Mingo et al. (in
prep), as was used for Figure \ref{wise_colour-colour_fr2}. No
significant differences were found between the hosts of DDRGs and
FR-IIs in our samples, similar to the comparisons with the RLAGN
sample. It is possible that selection effects are causing the
discrepancy between this work and that of \cite{kuzm17}. Their sample
of DDRGs is based on a collection of sources from the literature, while also including other types of restarting sources. Moreover, their comparison sample of FR-IIs is not
homogeneously selected from the same population, but is derived from
many different catalogues generated at different frequences. Our
samples are obtained from a single survey at a single observing
frequency, albeit over a much smaller area of sky coverage relative to
theirs, and are matched in radio luminosity. It is possible that
high-luminosity DDRGs (which may be dominated by HERGs)\ relative to their FR-IIs affect their results;  67 per cent of
their sample have hosts with WISE colours typical for spiral galaxies
or SFG, hosts of which are not uncommon for high power sources.

It is possible therefore that the difference seen in these similar studies are due to a population dichotomy between HERGs and LERGs. LERGs are suggested to have a fast duty cycle that is galaxy mass-dependent (\citealt{best05}), in which the highest mass galaxies are thought to have a more recurrent activity; this activity is fuelled by the cooling of their hot gas haloes and thought to be driven by chaotic cold accretion \citep[e.g.][]{gasp12}. On the other hand, HERGs are thought to be driven by the accretion of cold gas, plausibly through the infall of material during a gas-rich merger. In this scenario, assuming our DDRGs and RLAGN are predominantly LERGs (by their WISE colours), it is not surprising that we do not see any significant host galaxy differences and that the HERG DDRGs of \cite{kuzm17} have an expected difference in host galaxies with their presumably HERG FR-II sources. This might indicate that the hosts of DDRG HERGs tend to be driven by different merger-related host galaxies than single-cycle HERG RLAGN. We conclude that
the differences in results seen between this work and that of
\cite{kuzm17} can be explained by a population selection effect
between RLAGN samples, where host galaxy dichotomies do exist. A systematic study with clear associations of HERGs and LERGs between DDRGs and RLAGN will support this further.
Larger samples, such as those that will be provided by the full LoTSS survey, will enable such studies.

It should be borne in mind that observable DDRGs necessarily show
particularly young jets; a significant amount of time would not have
passed since the last episode of AGN activity
\citep{kona13}. Moreover, since the jets are relativistic
on smaller scales, the restarted jets should quickly merge with the
pre-existing remnant plasma on the larger scales, becoming normal
RLAGN. Thus, the general conclusion that the host galaxies
are similar between DDRGs and RLAGN does not directly translate to the
idea that the radio jets do not affect interstellar gas that they
drive through. The effect of jet heating on stellar populations of the
host galaxy is likely to be visible only on much longer timescales. We
know observationally that restarting radio galaxies can affect their
hosts, but these effects would not be detectable in optical photometry. We note the strong interaction between the interstellar medium and
fast outflows of jet-driven neutral hydrogen in the RRLAGN 4C\,12.50 reported by \cite{morg13}, or the shocks
driven by the inner lobes of Centaurus A as seen by \cite{cros09}.

While our data do not allow us to probe the cause of restarted jet
activity, we rule out significantly different galaxies driving DDRGs for the LERG population.
We may then speculate that the restarted or disruption of jet activity
is caused by smaller scale changes. The jet duty cycle may be
governed by changes in the accretion system, independent of the amount
of fuel available for accretion from cold or hot gas reservoirs
present in the most massive galaxies. According to the
Blandford-Znajek process \citep{blan77}, jet activity is governed by
the strength of magnetic flux surrounding the black hole, black
hole spin, and black hole mass itself. Since we do not expect the
black hole mass or spin to change significantly given the timescales
of remnant and restarted activity and the results of this paper, it is
plausible that intrinsic effects causing the magnetic flux to vary
substantially in the accretion system cause the jets to switch off and
quickly restart with a similar jet power. Although it is possible that the nature of chaotic cold accretion, which is thought to be the main driver of jets for LERGs, causes significant accretion variability that in turn drives intermittent activity or a rapid duty cycle, the DDRGs observed in this work and in other aforementioned works could simply be recently restarted objects.

Other accretion-related scenarios have been studied extensively in the context of the intermittent nature of AGN (e.g. radiation pressure instability; \citealt{czer09} and the ionisation instability; \citealt{clar88,jani04}), but it is unclear whether and how these short timescale perturbations, and their effects on accretion rate, directly couple to the jet power and its activity timescales for the population of restarting AGN. Alternatively, \cite{ciel17}
have presented simulations of backflows of powerful jets that channel back
into the accretion system causing a periodic (3-5 Myr) evolution in
mass accretion rate. Although these simulations predicted an overall
increase of jet power rather than intermittent or restarting jet activity and some
version of this model, in which backflows may disrupt the central
accretion system on short timescales, may operate in the RLAGN population.

\section{Summary and conclusions}\label{sect-conclusions}
Our findings suggest that DDRGs and normal RLAGN are hosted by the same type of host galaxy, and that the restarted phase is a natural phenomenon that exists particularly for the class of LERGs. We summarise our results and conclusions below:
\begin{itemize}
\item The host galaxies of DDRGs are similar in brightness and colour to those of normal RLAGN matched in radio luminosity.
\item DDRG do not occupy a special region in WISE colour-colour space relative to the bulk of the normal RLAGN population, indicating that both systems are driven by the same types of host galaxies.
\item Selection effects mean that visually identified samples of DDRGs tend to be significantly larger and more luminous than the dominant population of RLAGN and remnant RLAGN.
\item Restarting jets are essentially an intrinsic property of RLAGN, rather than a cause, or a driver of, bulk changes in their host galaxies.
\item If restarted activity is not directly correlated with changes in the host galaxy,
  then it is likely caused by changes in the accretion system only.
  Accreted magnetic flux variation or variations in the mass accretion rate on short timescales may drive restarted activity.
\end{itemize}
This study has confirmed that DDRGs and single-cycle RLAGN can be drawn from the same population of host galaxies, while supporting the idea that mergers alone do not control restarted activity for classical double objects, although this is likely to only be the case for the population of LERGs. In the future, a more morphologically complete selection of restarting
objects will be presented by Jurlin et al. (in prep) based on the
LoTSS DR1 catalogue. Furthermore, understanding how many single-cycle RLAGN have had previous activity, resulting in radio lobes that are undetectable given the sensitivity limits of current instruments, will be beneficial as the LoTSS survey is completed and as further deep radio surveys become available in the future. Both the study of DDRGs and of RRLAGN in general
will be greatly advanced by the vastly increased sky area of the full
LoTSS survey, which will become
available over the next few years. Moreover, optical spectroscopy will become available for these objects; eventually, the bulk of the LOFAR-detected sources in LoTSS will also become available using the WEAVE-LOFAR spectroscopy survey \citep{smit16}, allowing more detailed studies of the hosts of RLAGN in their various life cycles.

\begin{acknowledgements}
We would like to thank the anonymous referee for useful comments on this paper. We would also like to thank Elias Brinks, Kimberly Emig, and Catherine Hale for useful comments on earlier drafts of this paper.

This research has made use of data analysed using the University of
Hertfordshire high-performance computing facility
(\url{http://uhhpc.herts.ac.uk/}) and the LOFAR-UK computing facility
located at the University of Hertfordshire and supported by STFC
[ST/P000096/1]. 

VHM thanks the University of Hertfordshire for a research studentship
[ST/N504105/1]. MJH and WLW acknowledge support from the UK Science
and Technology Facilities Council [ST/M001008/1]. PNB and JS are grateful for
support from the UK STFC via grant ST/M001229/1. JHC acknowledges support from the Science and Technology Facilities Council (STFC) under grants ST/R00109X/1 and ST/R000794/1. KJD acknowledges support from the ERC Advanced Investigator programme NewClusters 321271. RM gratefully acknowledges support from the European Research Council under the European Union's Seventh Framework Programme (FP/2007-2013) /ERC Advanced Grant RADIOLIFE-320745. MB acknowledges support from INAF under PRIN SKA/CTA ‘FORECaST’ RKC is grateful for support from the UK STFC. GG acknowledges the CSIRO OCE Postdoctoral Fellowship. MJJ and LKM acknowledge support from Oxford Hintze Centre for Astrophysical Surveys, which is funded through generous support from the Hintze Family Charitable Foundation. HJAR acknowledges support from the European Research Council under the European Unions Seventh Framework Programme (FP/2007- 2013) /ERC Advanced Grant NEWCLUSTERS-321271. This publication arises from research partly funded by the John Fell Oxford University Press (OUP) Research Fund.

This paper is based (in part) on data obtained with the International LOFAR Telescope (ILT). LOFAR (van Haarlem et al. 2013) is the LOw Frequency ARray designed and constructed by ASTRON. It has observing, data processing, and data storage facilities in several countries, which are owned by various parties (each with their own funding sources), and are collectively operated by the ILT foundation under a joint scientific policy. The ILT resources have benefitted from the following recent major funding sources: CNRS-INSU, Observatoire de Paris and Université d'Orléans, France; BMBF, MIWF-NRW, MPG, Germany; Science Foundation Ireland (SFI), Department of Business, Enterprise and Innovation (DBEI), Ireland; NWO, The Netherlands; The Science and Technology Facilities Council, UK; Ministry of Science and Higher Education, Poland.\\

Part of this work was carried out on the Dutch national e-infrastructure with the support of the SURF Cooperative through grant e-infra 160022 \& 160152. The LOFAR software and dedicated reduction packages on \url{https://github.com/apmechev/GRID\_LRT} were deployed on the e-infrastructure by the LOFAR e-infragroup, consisting of J. B. R. Oonk (ASTRON \& Leiden Observatory), A. P. Mechev (Leiden Observatory) and T. Shimwell (ASTRON) with support from N. Danezi (SURFsara) and C. Schrijvers (SURFsara).
\end{acknowledgements}

\bibliographystyle{aa}
\bibliography{references}
\appendix
\counterwithin{figure}{section}
\section{VLA 1.4-GHz maps}
\begin{figure*}[h!]
\begin{subfigure}{.5\textwidth}
  \centering
  \includegraphics[width=.75\linewidth]{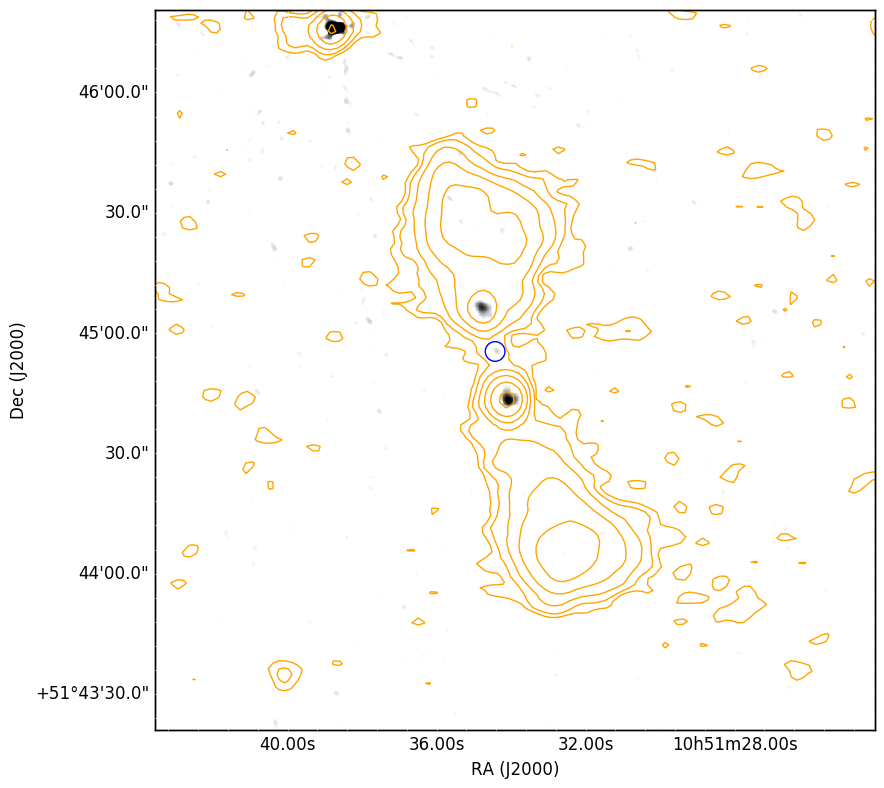}
  \caption{ILTJ105133.89+514451.1 \\
  \texttt{ROBUST: -0.5}}
  \label{fig:sfig1}
\end{subfigure}%
\begin{subfigure}{.5\textwidth}
  \centering
  \includegraphics[width=.75\linewidth]{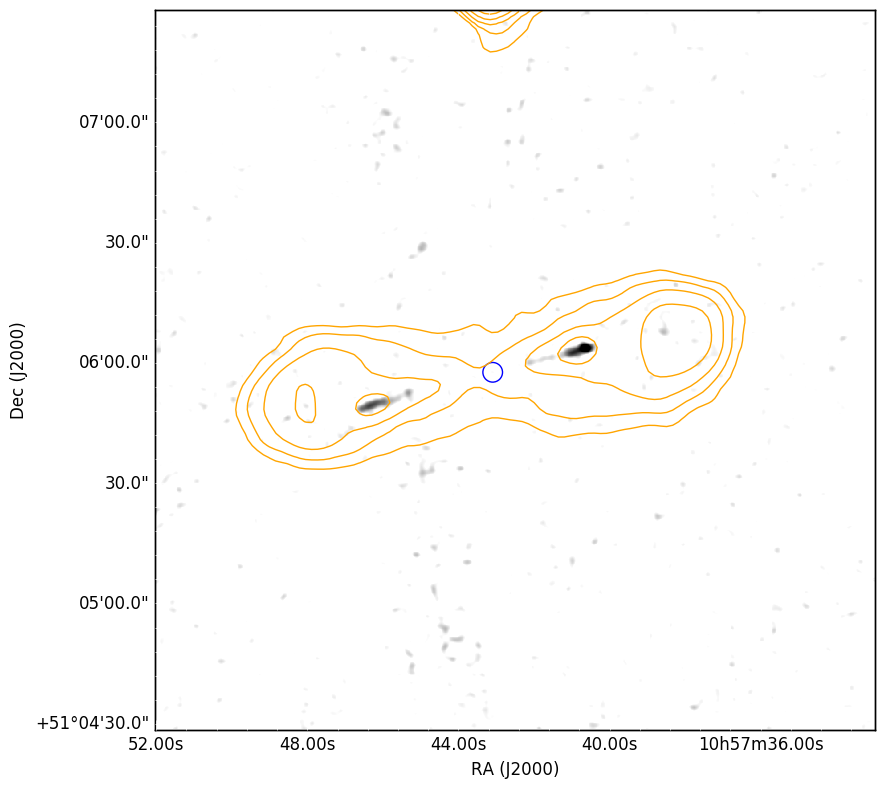}
  \caption{ILTJ105742.50+510558.5$^{\dagger}$ \\
  \texttt{ROBUST: -0.5}}
  \label{fig:sfig2}
\end{subfigure}
\vskip\baselineskip
\begin{subfigure}{.5\textwidth}
  \centering
  \includegraphics[width=.75\linewidth]{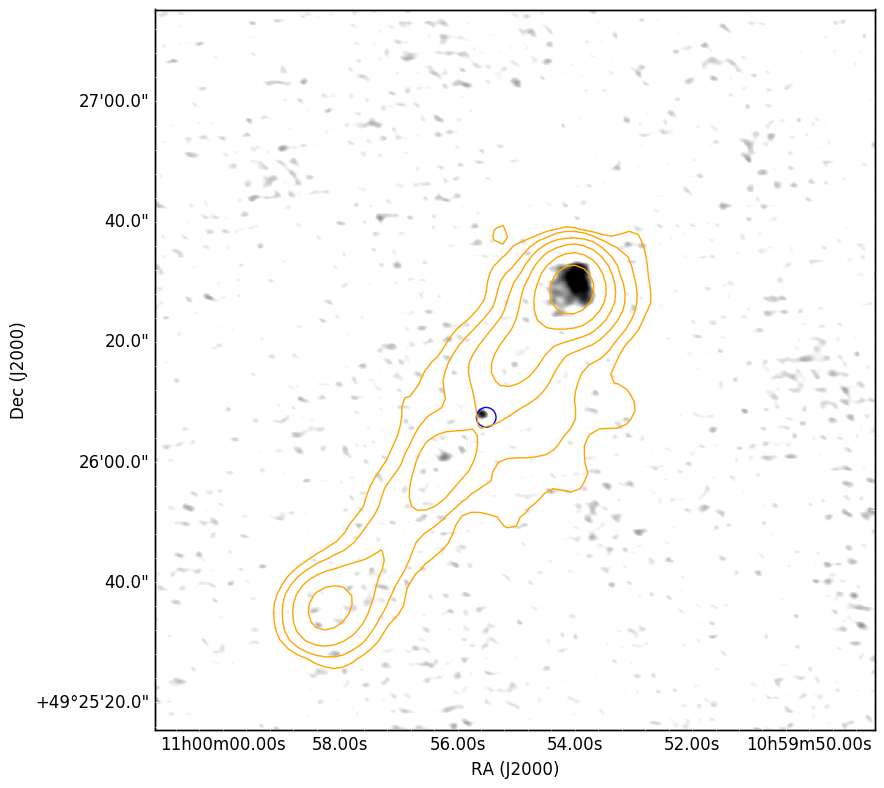}
  \caption{ILTJ105955.01+492615.4$^{\dagger}$ (removed from sample) \\
  \texttt{ROBUST: -1.0}}
  \label{fig:sfig2}
\end{subfigure}
\begin{subfigure}{.5\textwidth}
  \centering
  \includegraphics[width=.75\linewidth]{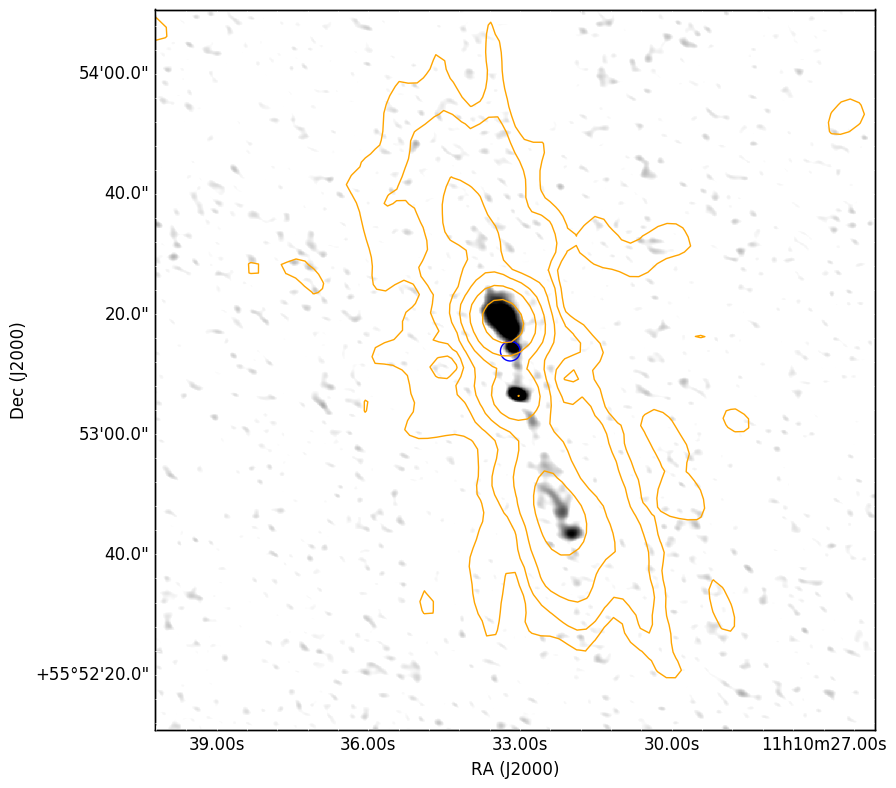}
  \caption{ILTJ111033.09+555310.8$^{\dagger}$ \\
  \texttt{ROBUST: 0.0}}
  \label{fig:sfig2}
\end{subfigure}
\vskip\baselineskip
\begin{subfigure}{.5\textwidth}
  \centering
  \includegraphics[width=.77\linewidth]{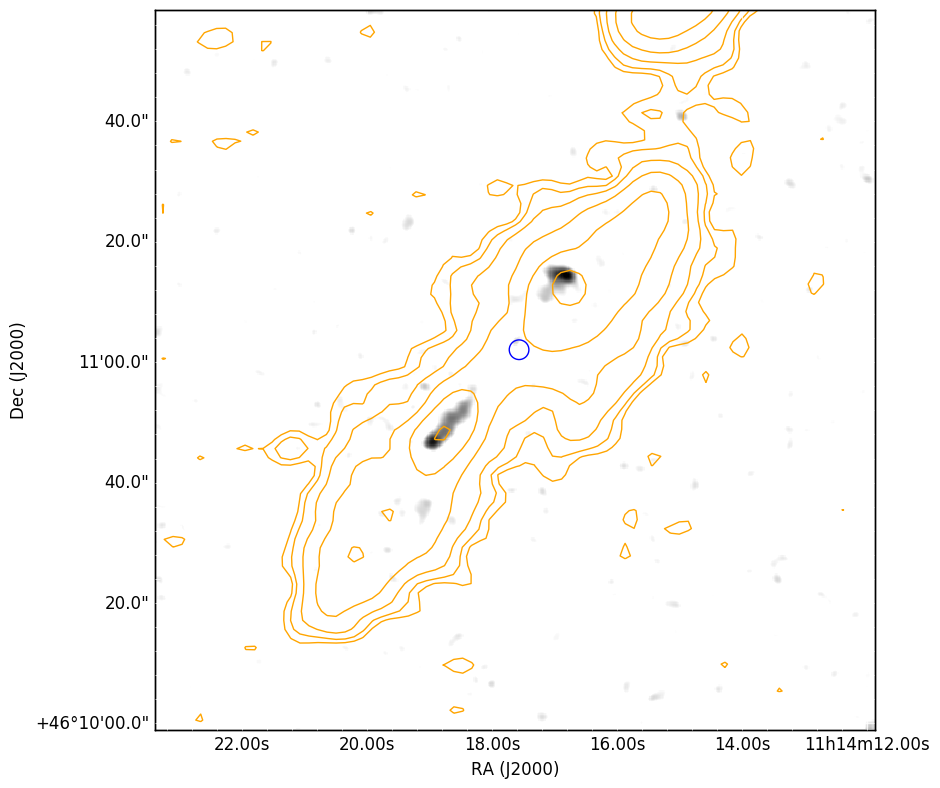}
  \caption{ILTJ111417.63+461058.9 \\
  \texttt{ROBUST: -0.5}}
  \label{fig:sfig2}
\end{subfigure}
\begin{subfigure}{.5\textwidth}
  \centering
  \includegraphics[width=.77\linewidth]{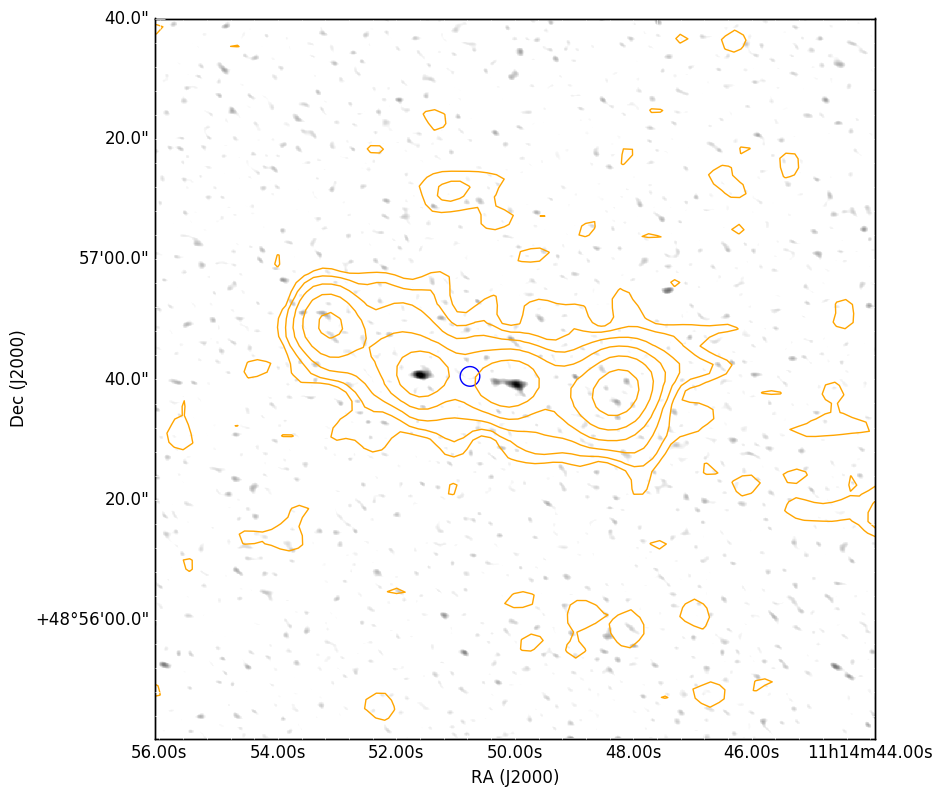}
  \caption{ILTJ111449.99+485640.2 \\
  \texttt{ROBUST: -0.5}}
  \label{fig:sfig2}
\end{subfigure}
\caption{1.4 GHz VLA images of the 40 candidate restarted sources shown in greyscale, overlaid with the 144 MHz LOFAR contours from the LoTSS DR1 in orange. The VLA images are scaled logarithmically and smoothed with a Gaussian function with FWHM of 3 times the beam size. The LOFAR contours denote the surface brightness levels starting at 3$\sigma$ and increasing at various powers of 3$\sigma$, where $\sigma$ denotes the local RMS noise. $^{\dagger}$Owing to dynamic range limitations for the brightest sources, $\sigma$ was instead chosen based on a particular value of the dynamic range in the LOFAR image, depending on the surface brightness of the source. The blue circles denote the optical ID. The source names in the sub-captions are the LOFAR source names presented in Table \ref{source_table}. The Briggs robust weighting parameters used for CLEANing the VLA images are labelled in the sub-captions.}
\label{overlays}
\end{figure*}
\begin{figure*}[!h]\ContinuedFloat
\centering
\begin{subfigure}{.5\textwidth}
  \centering
  \includegraphics[width=.84\linewidth]{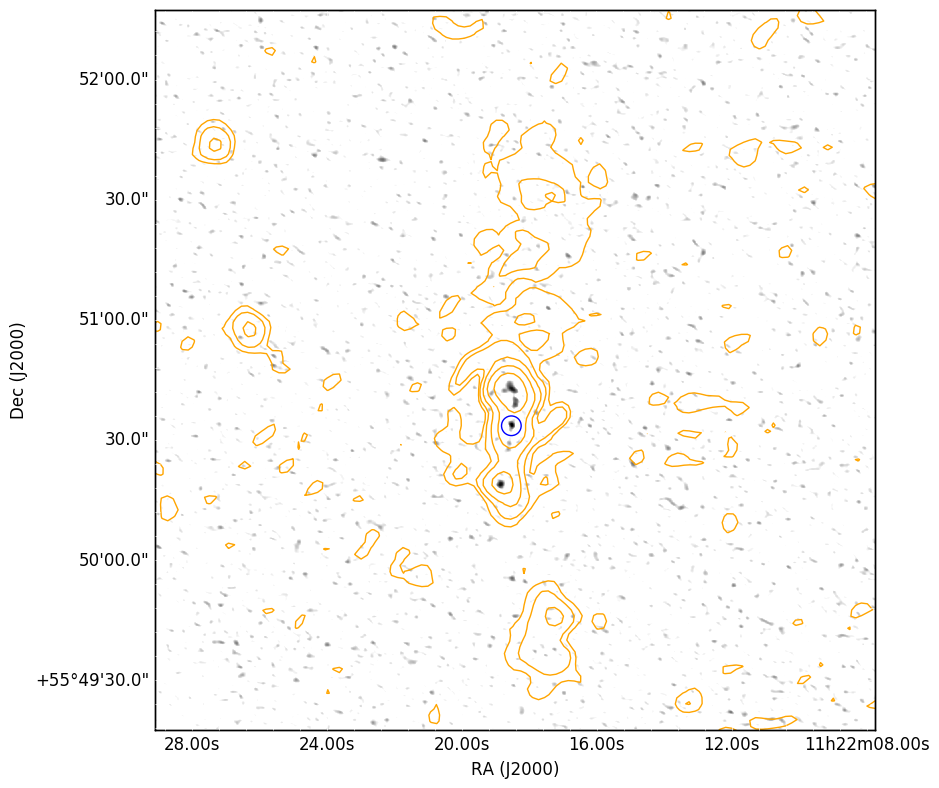}
  \caption{ILTJ112218.41+555047.7 \\
  \texttt{ROBUST: -0.5}}
  \label{fig:sfig1}
\end{subfigure}%
\hspace*{\fill}
\begin{subfigure}{.5\textwidth}
  \centering
  \includegraphics[width=.84\linewidth]{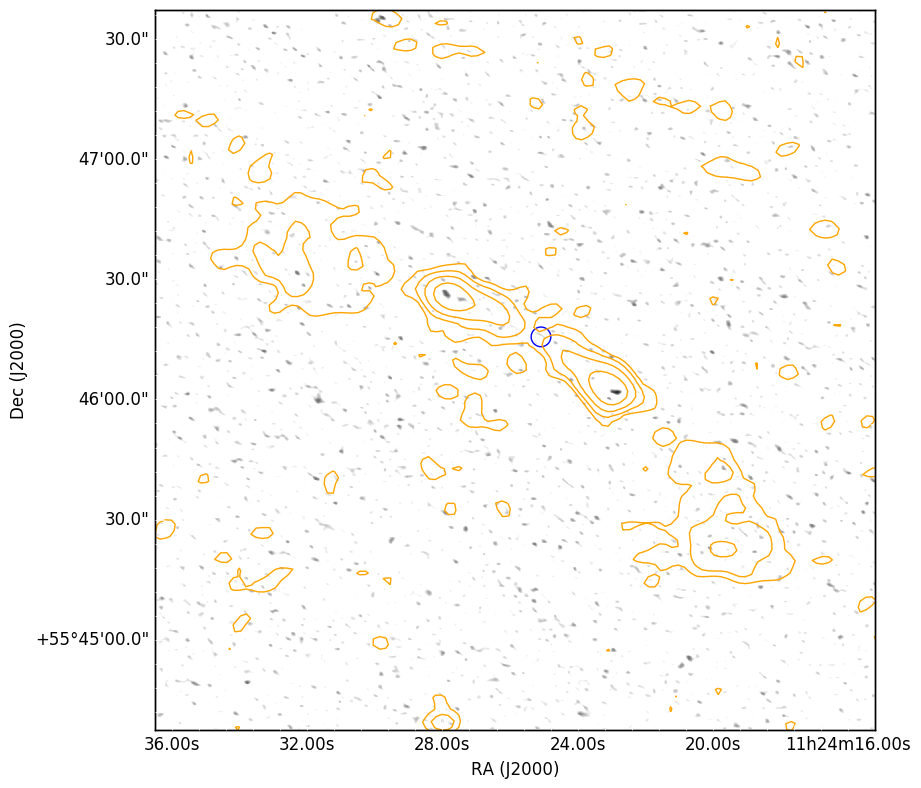}
  \caption{ILTJ112425.85+554607.6 \\
  \texttt{ROBUST: -0.5}}
  \label{fig:sfig1}
\end{subfigure}%
\vskip\baselineskip
\begin{subfigure}{.5\textwidth}
  \centering
  \includegraphics[width=.8\linewidth]{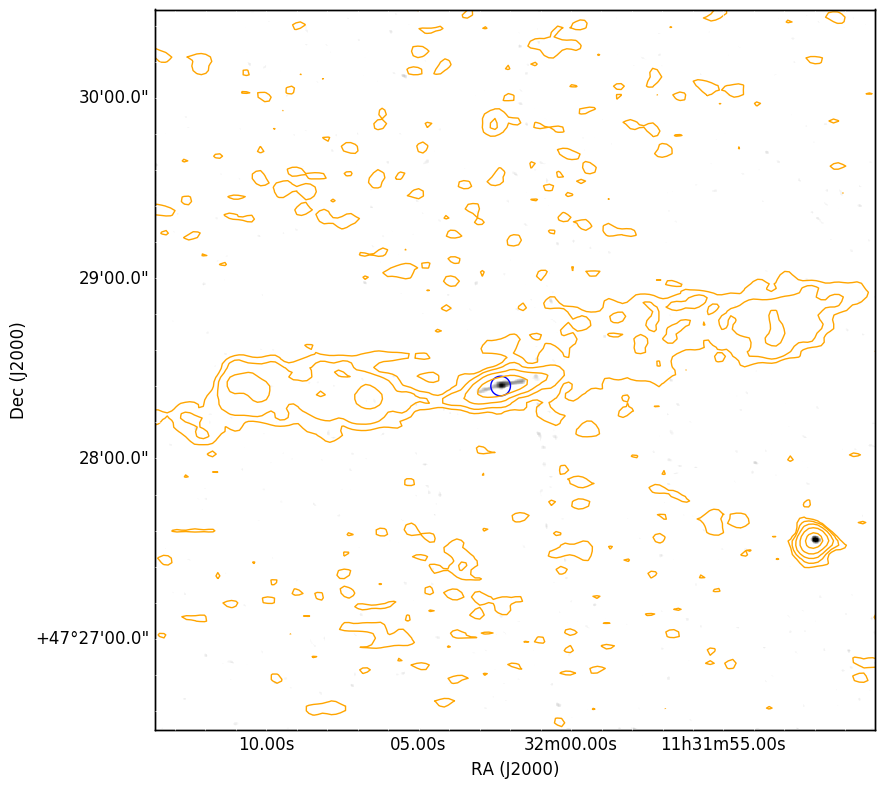}
  \caption{ILTJ113201.82+472829.9 (removed from sample) \\
  \texttt{ROBUST: 0.0}}
  \label{fig:sfig1}
\end{subfigure}%
\begin{subfigure}{.5\textwidth}
  \centering
  \includegraphics[width=.8\linewidth]{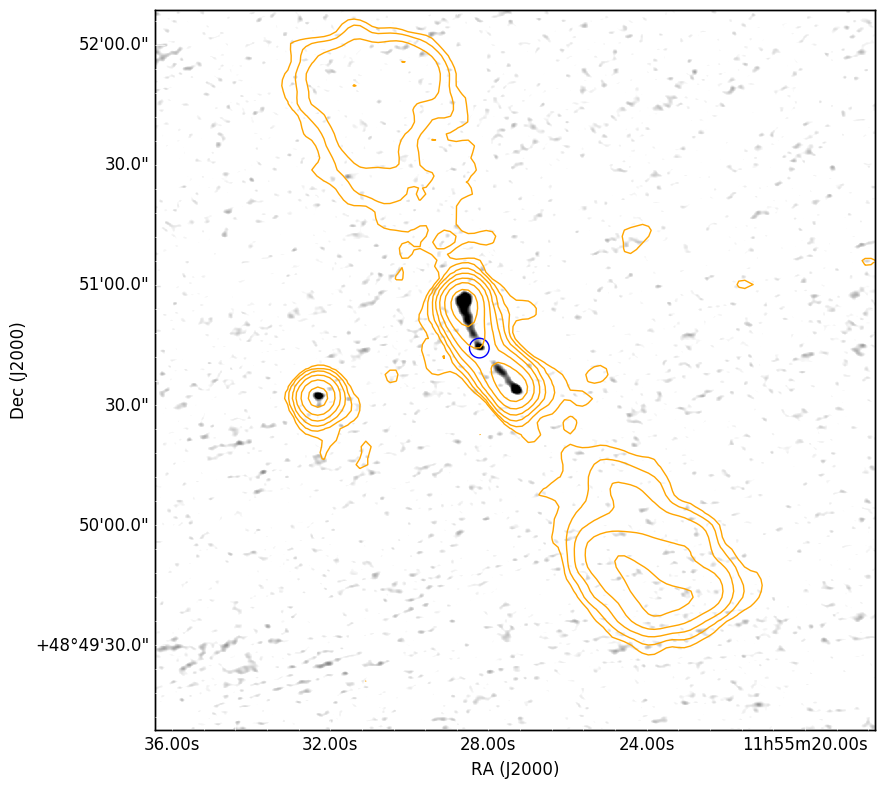}
  \caption{ILTJ115527.32+485039.0 \\
  \texttt{ROBUST: -0.5}}
  \label{fig:sfig1}
\end{subfigure}%
\vskip\baselineskip
\begin{subfigure}{.5\textwidth}
  \centering
  \includegraphics[width=.8\linewidth]{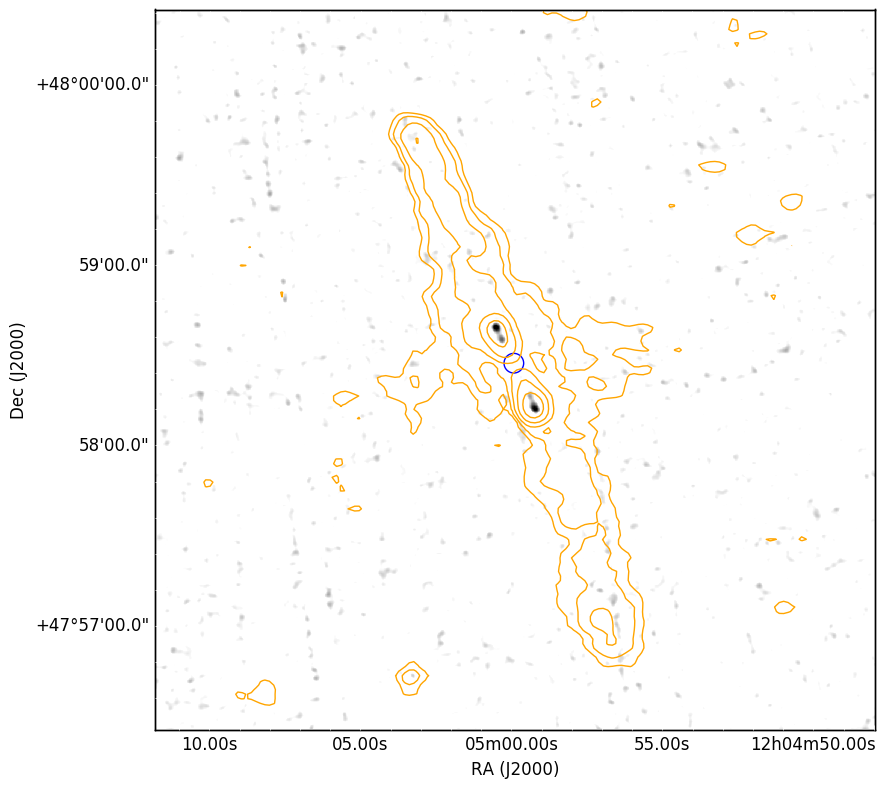}
  \caption{ILTJ120459.87+475825.4$^{\dagger}$ \\
  \texttt{ROBUST: -0.5}}
  \label{fig:sfig1}
\end{subfigure}%
\begin{subfigure}{.5\textwidth}
  \centering
  \includegraphics[width=.8\linewidth]{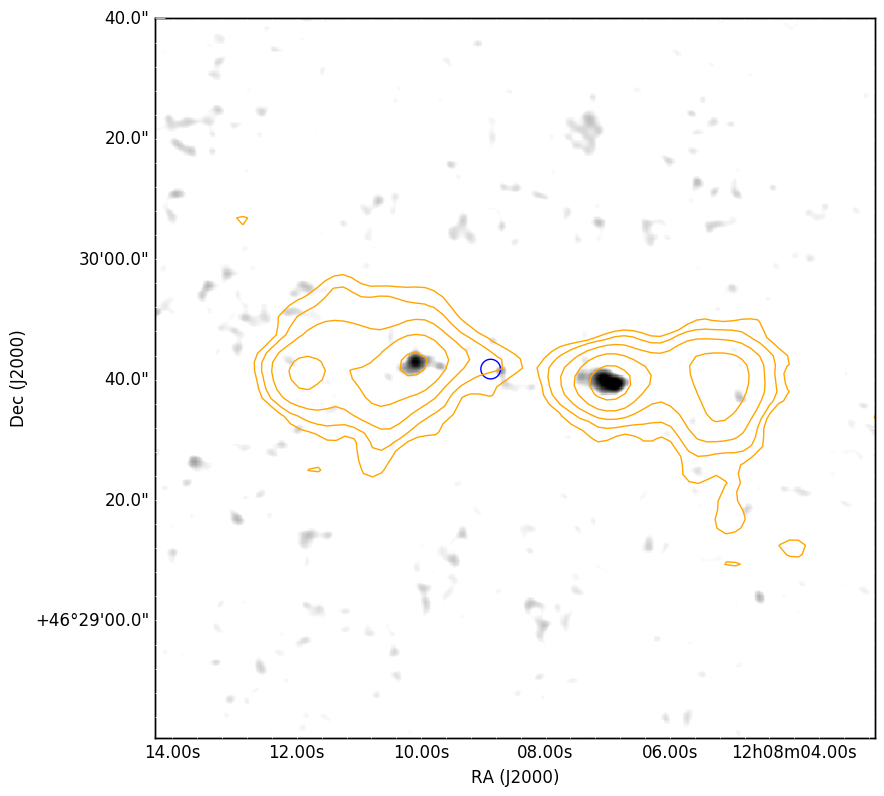}
  \caption{ILTJ120808.48+462940.6 \\
  \texttt{ROBUST: 0.5}}
  \label{fig:sfig1}
\end{subfigure}%
\caption{Continued}
\end{figure*}
\begin{figure*}\ContinuedFloat
\centering
\begin{subfigure}{.5\textwidth}
  \centering
  \includegraphics[width=.8\linewidth]{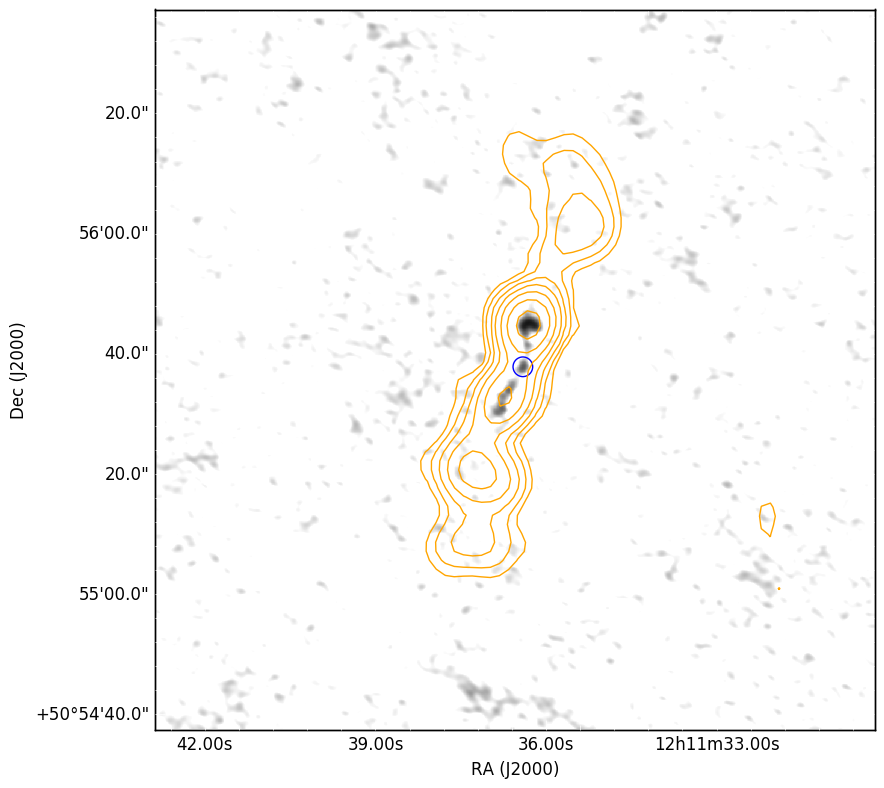}
  \caption{ILTJ121136.54+505537.5 \\
  \texttt{ROBUST: 0.5}}
  \label{fig:sfig1}
\end{subfigure}%
\begin{subfigure}{.5\textwidth}
  \centering
  \includegraphics[width=.8\linewidth]{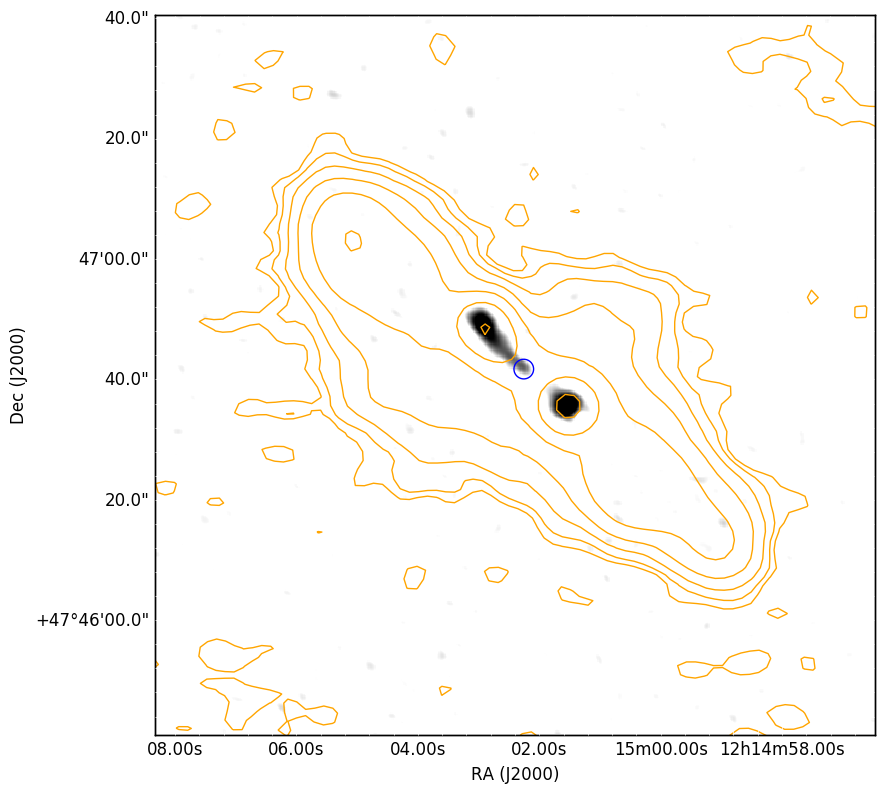}
  \caption{ILTJ121502.39+474641.1 \\
  \texttt{ROBUST: -0.5}}
  \label{fig:sfig1}
\end{subfigure}%
\vskip\baselineskip
\begin{subfigure}{.5\textwidth}
  \centering
  \includegraphics[width=.83\linewidth]{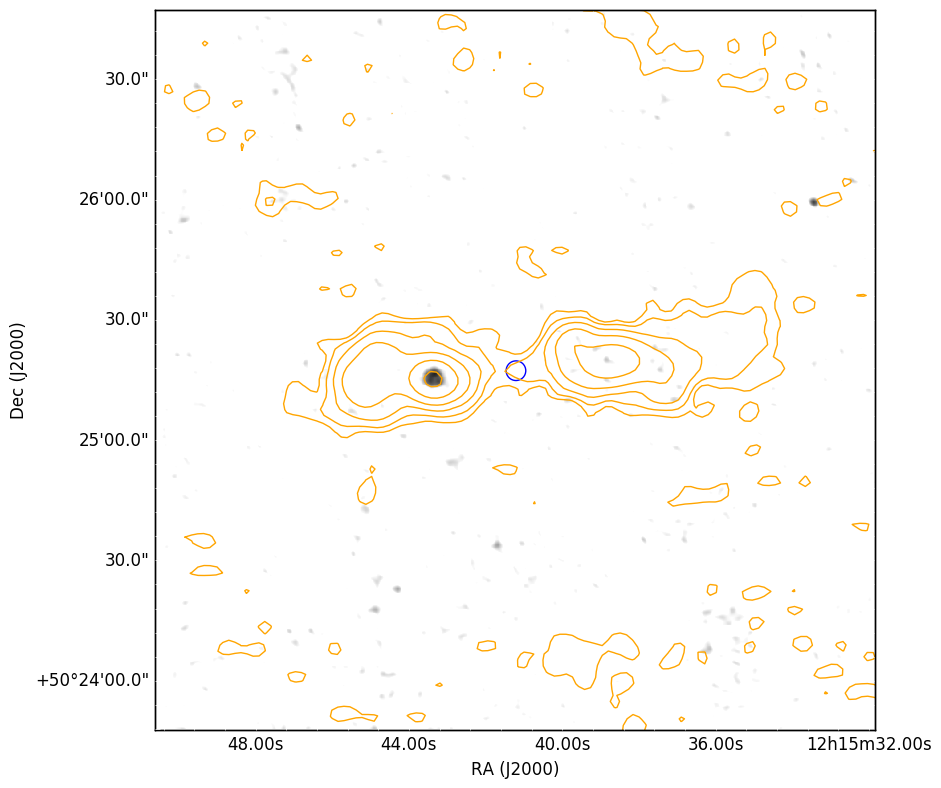}
  \caption{ILTJ121541.21+502517.9 \\
  \texttt{ROBUST: 1.0}}
  \label{fig:sfig1}
\end{subfigure}%
\begin{subfigure}{.5\textwidth}
  \centering
  \includegraphics[width=.83\linewidth]{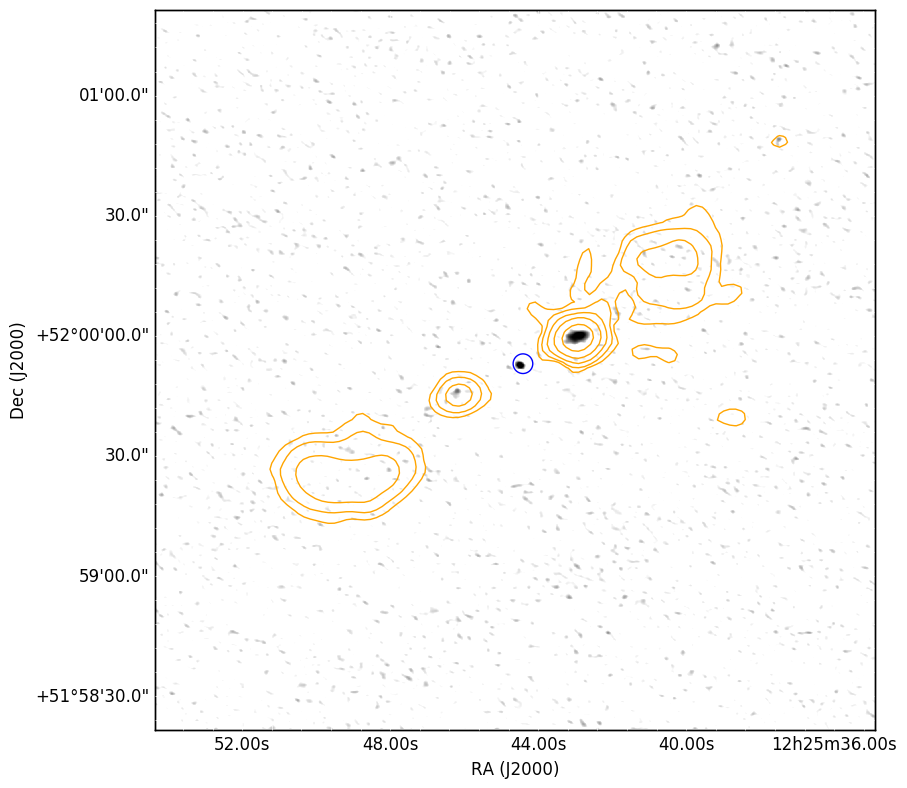}
  \caption{ILTJ122544.63+515951.7$^{\dagger}$ \\
  \texttt{ROBUST: -0.5}}
  \label{fig:sfig1}
\end{subfigure}%
\vskip\baselineskip
\begin{subfigure}{.5\textwidth}
  \centering
  \includegraphics[width=.8\linewidth]{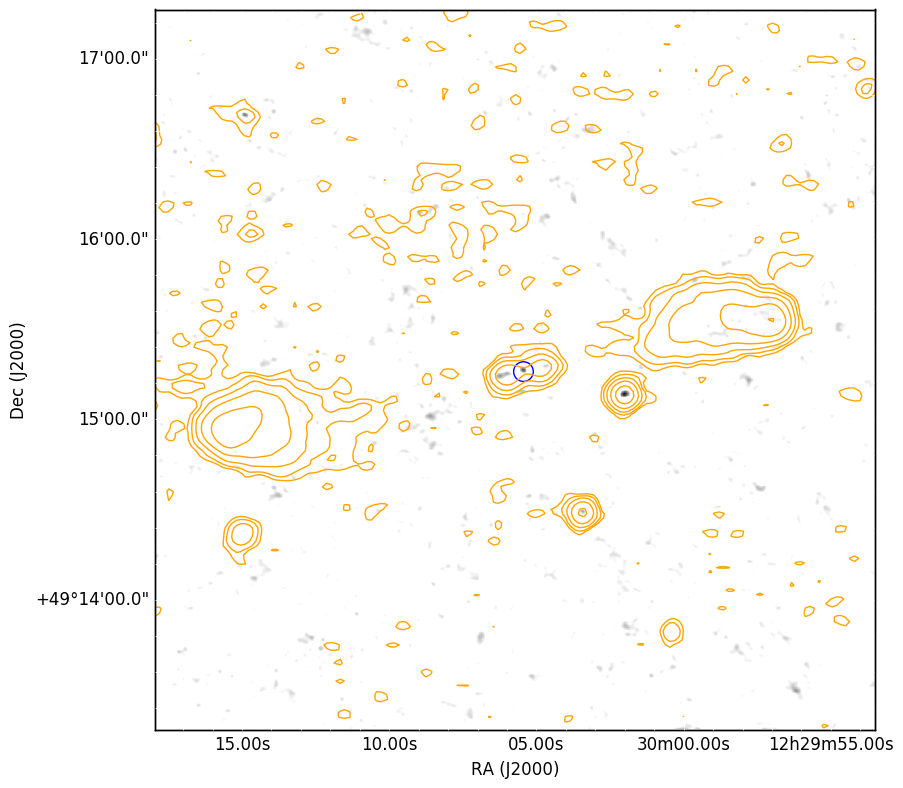}
  \caption{ILTJ123005.72+491516.8 \\
  \texttt{ROBUST: -0.5}}
  \label{fig:sfig1}
\end{subfigure}%
\begin{subfigure}{.5\textwidth}
  \centering
  \includegraphics[width=.83\linewidth]{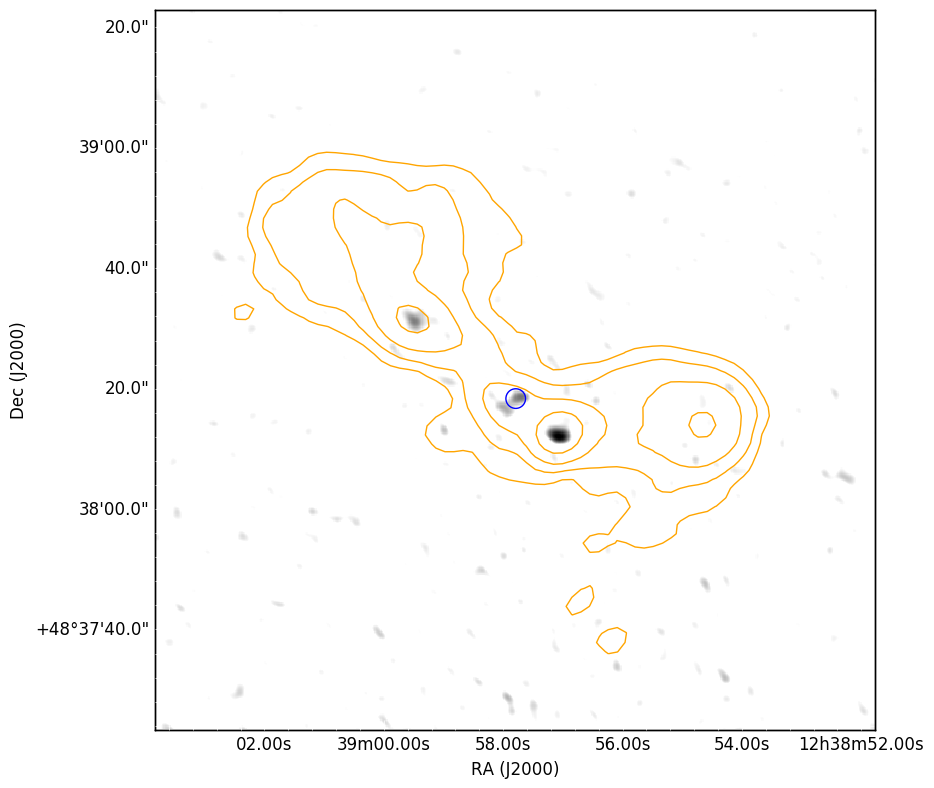}
  \caption{ILTJ123857.80+483823.5 \\
  \texttt{ROBUST: 0.5}}
  \label{fig:sfig1}
\end{subfigure}%
\caption{Continued}
\end{figure*}
\begin{figure*}\ContinuedFloat
\centering
\begin{subfigure}{.5\textwidth}
  \centering
  \includegraphics[width=.8\linewidth]{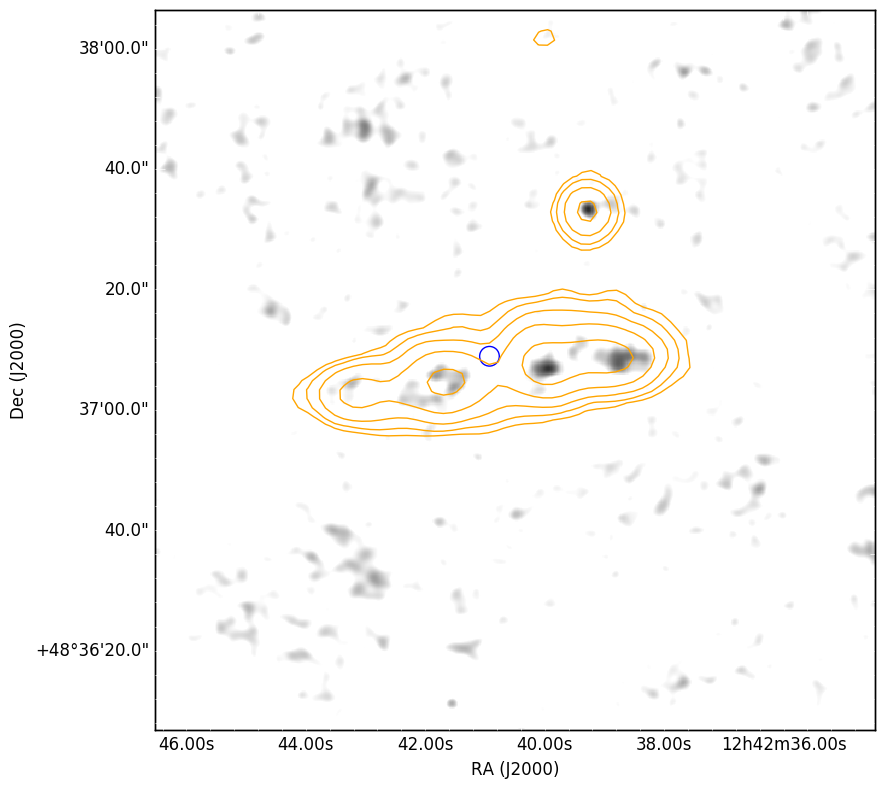}
  \caption{ILTJ124240.48+483706.8 (removed from sample) \\
  \texttt{ROBUST: 1.5}}
  \label{fig:sfig1}
\end{subfigure}%
\begin{subfigure}{.5\textwidth}
  \centering
  \includegraphics[width=.8\linewidth]{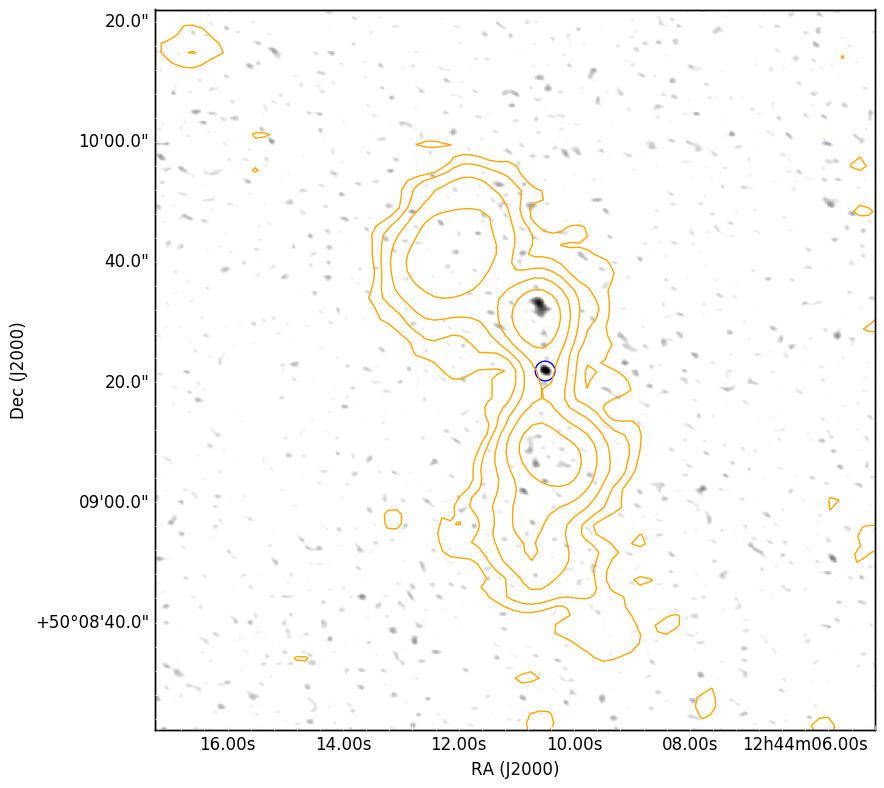}
  \caption{ILTJ124411.02+500922.1 \\
  \texttt{ROBUST: -0.5}}
  \label{fig:sfig1}
\end{subfigure}%
\vskip\baselineskip
\begin{subfigure}{.5\textwidth}
  \centering
  \includegraphics[width=.83\linewidth]{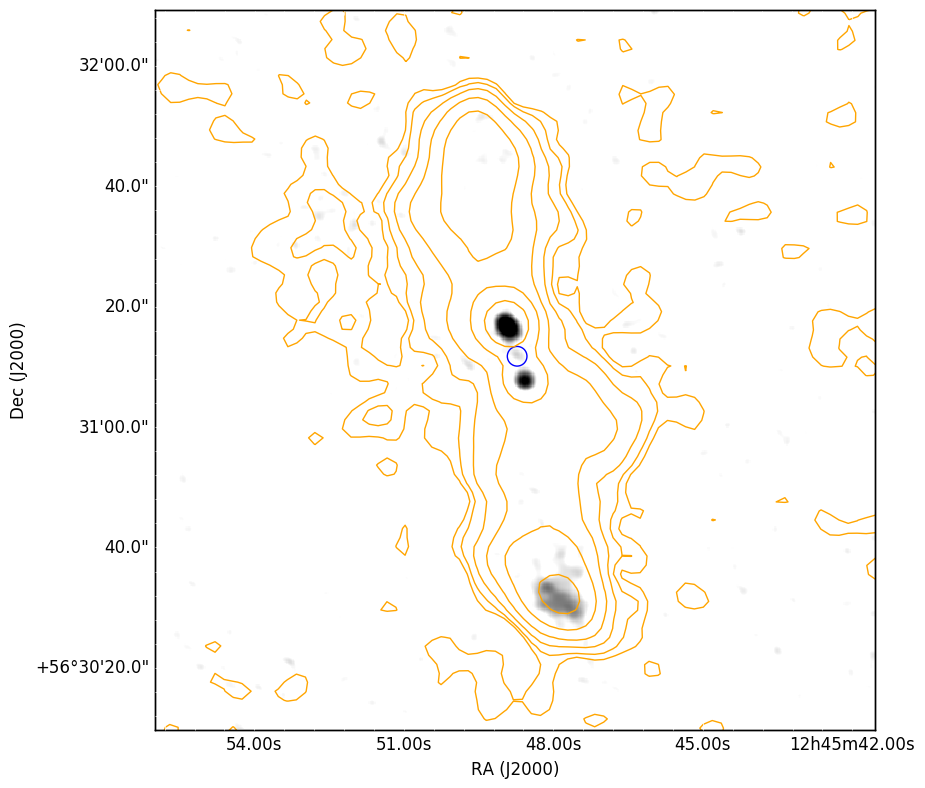}
  \caption{ILTJ124548.75+563109.7 \\
  \texttt{ROBUST: -0.5}}
  \label{fig:sfig1}
\end{subfigure}%
\begin{subfigure}{.5\textwidth}
  \centering
  \includegraphics[width=.8\linewidth]{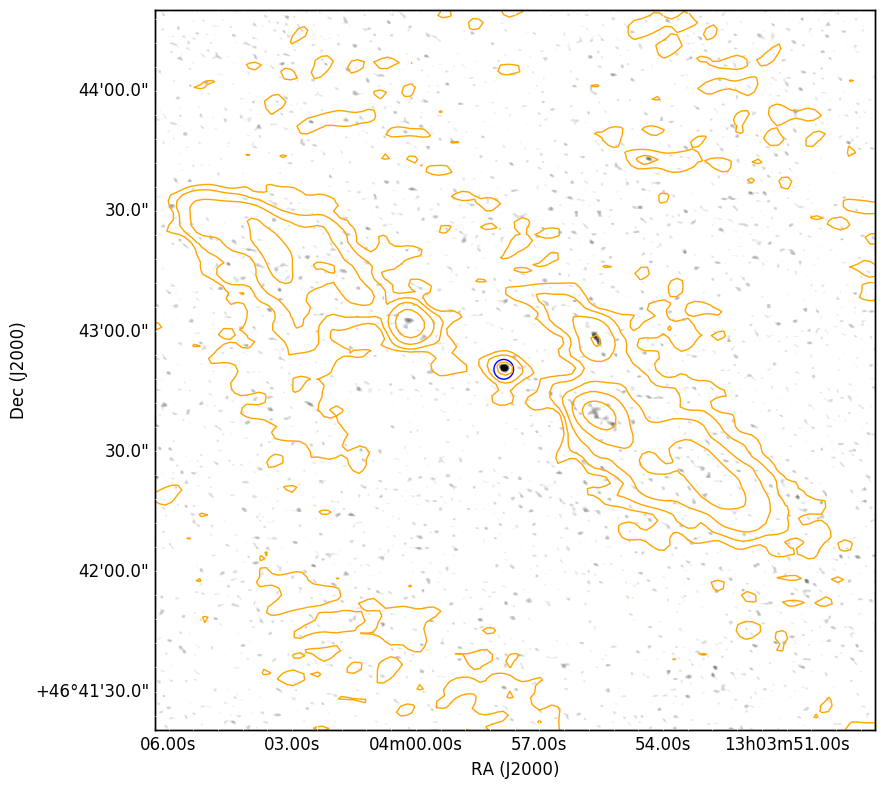}
  \caption{ILTJ130357.58+464250.4 \\
  \texttt{ROBUST: -0.5}}
  \label{fig:sfig1}
\end{subfigure}%
\vskip\baselineskip
\begin{subfigure}{.5\textwidth}
  \centering
  \includegraphics[width=.83\linewidth]{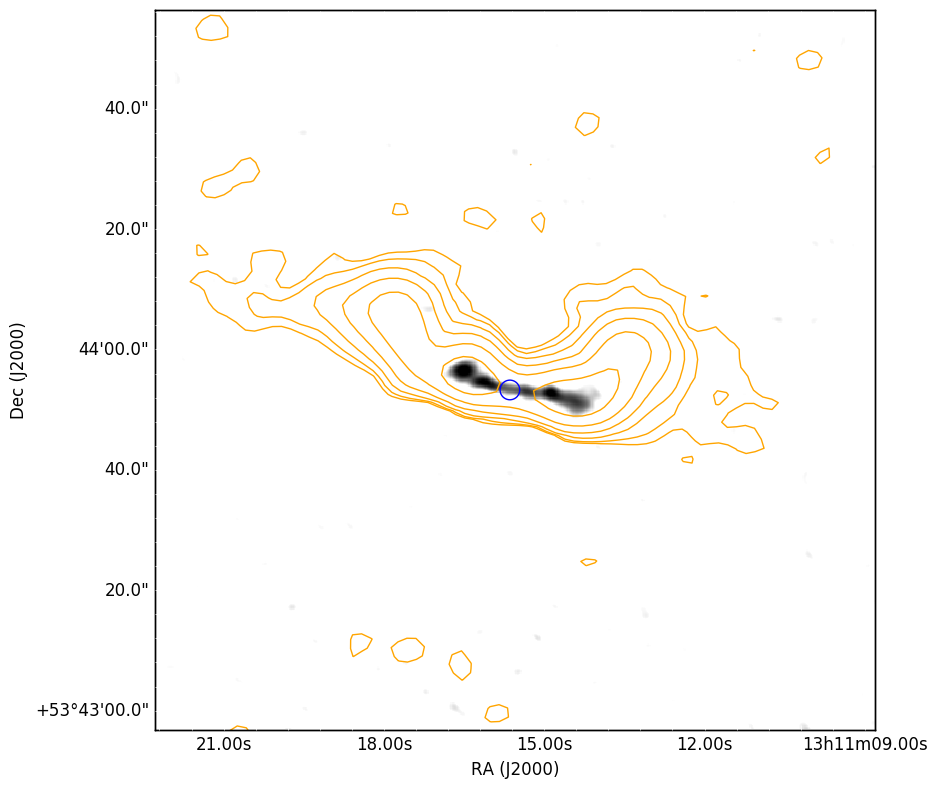}
  \caption{ILTJ131115.53+534356.8 (removed from sample) \\
  \texttt{ROBUST: -0.5}}
  \label{fig:sfig1}
\end{subfigure}%
\begin{subfigure}{.5\textwidth}
  \centering
  \includegraphics[width=.8\linewidth]{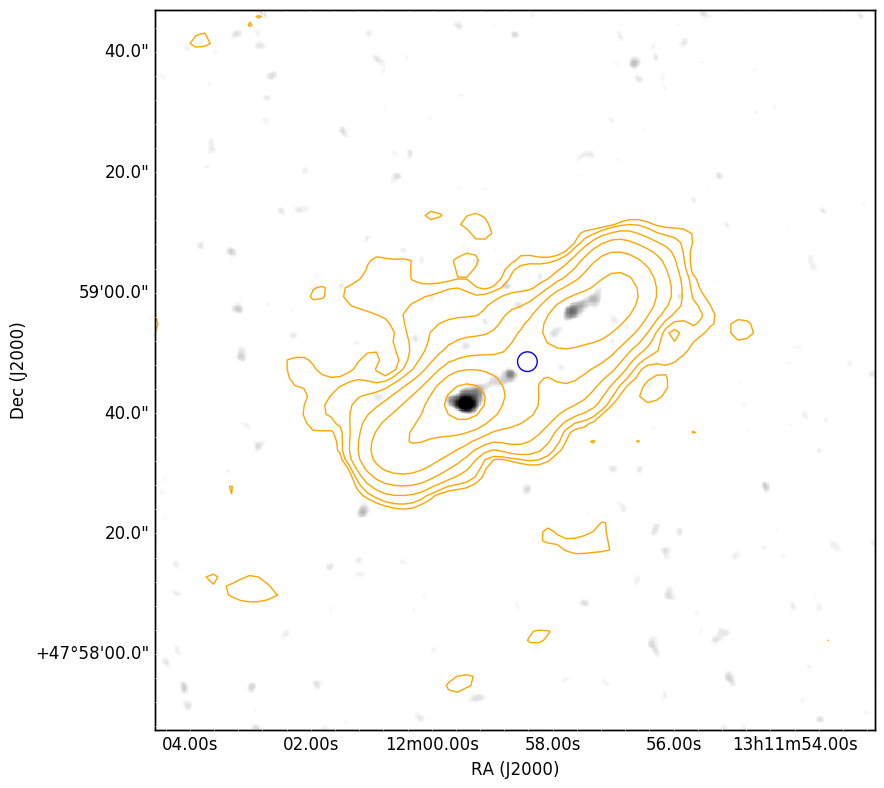}
  \caption{ILTJ131158.61+475847.5 \\
  \texttt{ROBUST: -0.5}}
  \label{fig:sfig1}
\end{subfigure}%
\caption{Continued}
\end{figure*}
\begin{figure*}\ContinuedFloat
\centering
\begin{subfigure}{.5\textwidth}
  \centering
  \includegraphics[width=.8\linewidth]{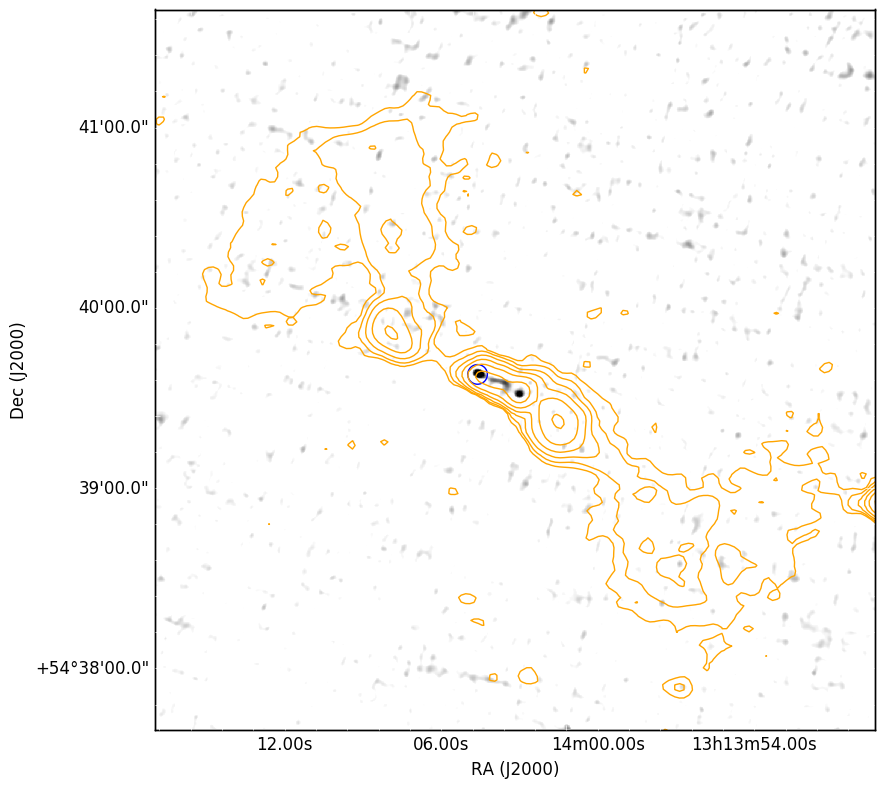}
  \caption{ILTJ131403.17+543939.6 \\
  \texttt{ROBUST: -0.5}}
  \label{fig:sfig1}
\end{subfigure}%
\begin{subfigure}{.5\textwidth}
  \centering
  \includegraphics[width=.8\linewidth]{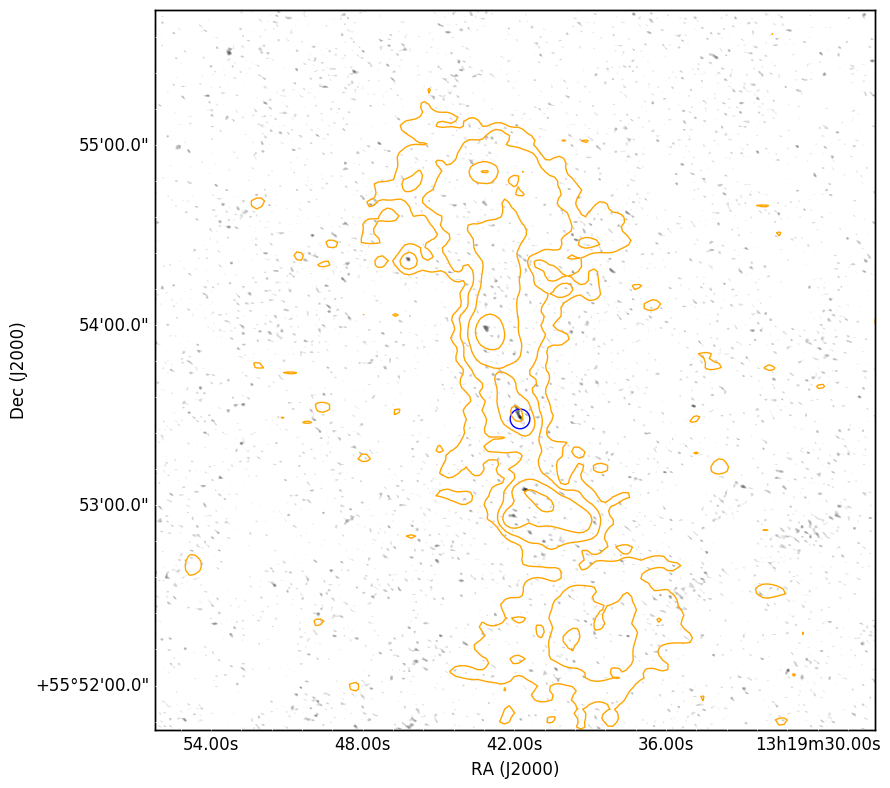}
  \caption{ILTJ131941.97+555345.3 \\
  \texttt{ROBUST: -0.5}}
  \label{fig:sfig1}
\end{subfigure}%
\vskip\baselineskip
\begin{subfigure}{.5\textwidth}
  \centering
  \includegraphics[width=.82\linewidth]{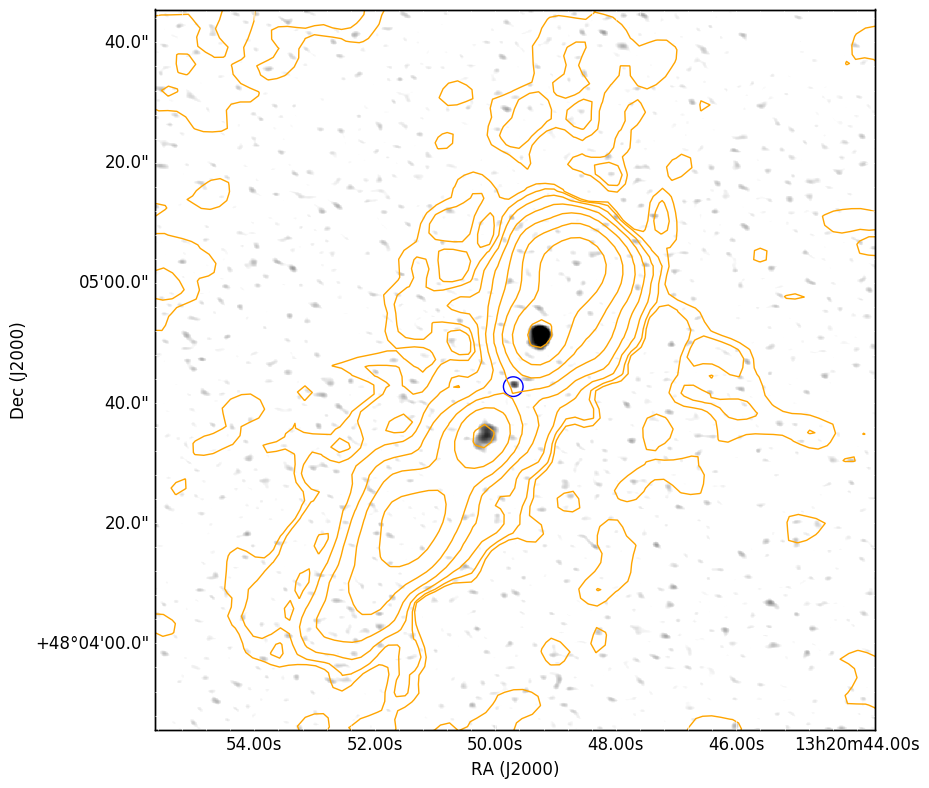}
  \caption{ILTJ132049.67+480445.6 \\
  \texttt{ROBUST: -0.5}}
  \label{fig:sfig1}
\end{subfigure}%
\begin{subfigure}{.5\textwidth}
  \centering
  \includegraphics[width=.8\linewidth]{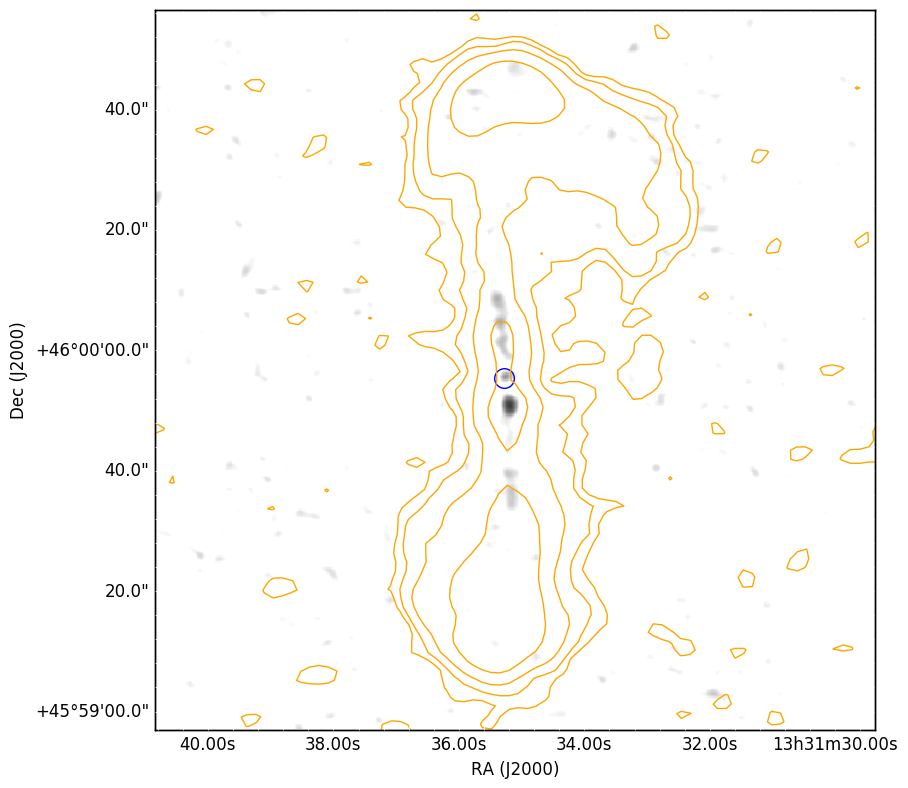}
  \caption{ILTJ133135.09+455957.0 (removed from sample) \\
  \texttt{ROBUST: 0.5}}
  \label{fig:sfig1}
\end{subfigure}%
\vskip\baselineskip
\begin{subfigure}{.5\textwidth}
  \centering
  \includegraphics[width=.8\linewidth]{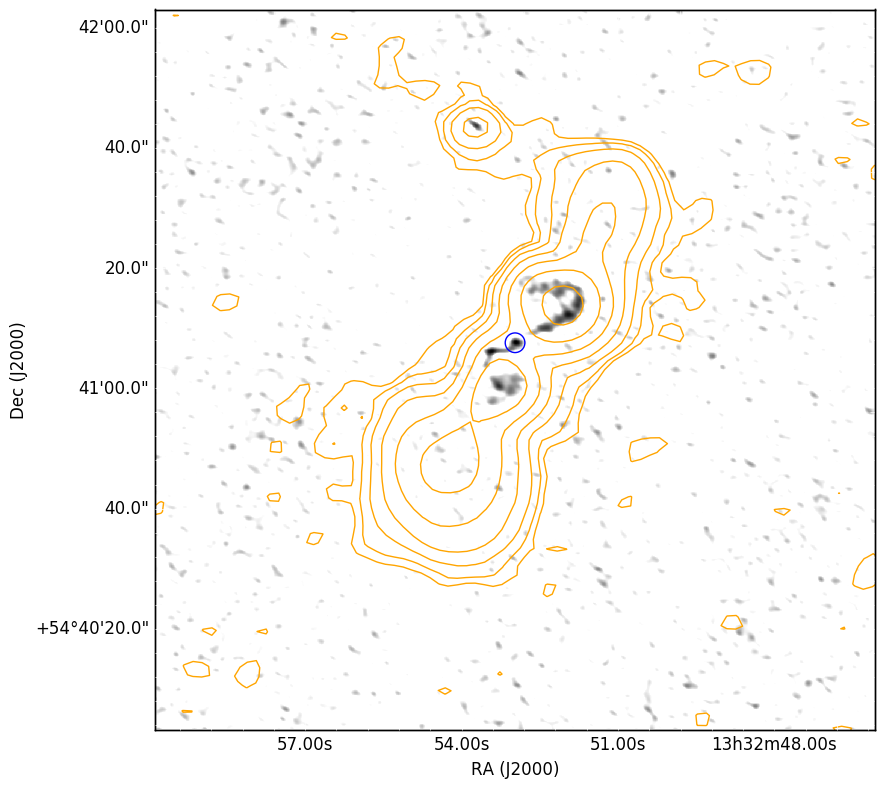}
  \caption{ILTJ133252.97+544103.2 (removed from sample) \\
  \texttt{ROBUST: -0.5}}
  \label{fig:sfig1}
\end{subfigure}%
\begin{subfigure}{.5\textwidth}
  \centering
  \includegraphics[width=.8\linewidth]{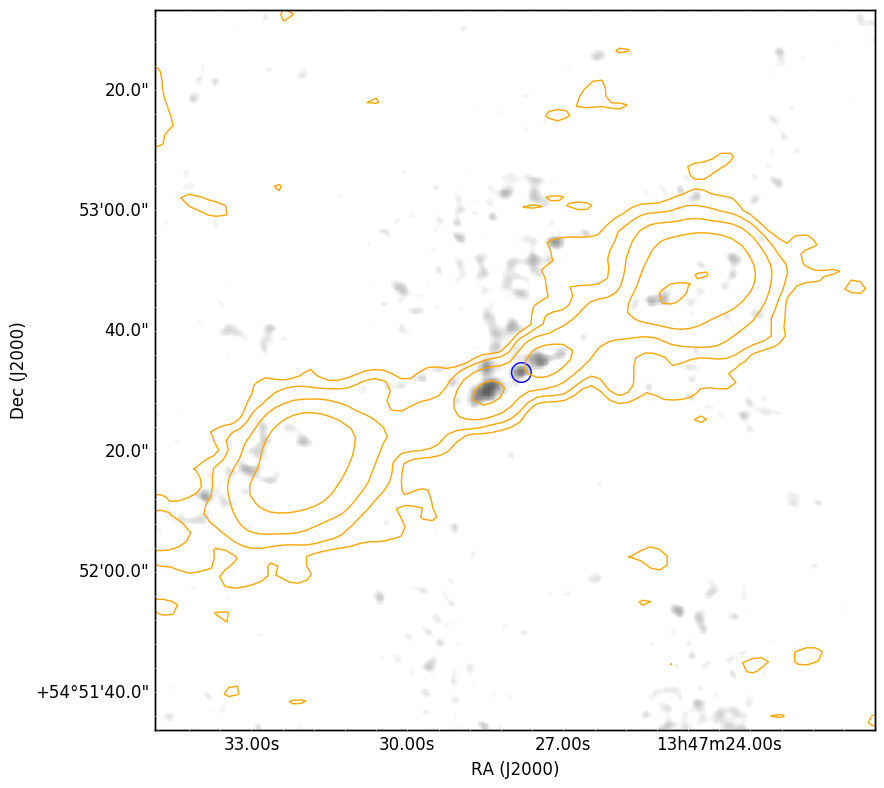}
  \caption{ILTJ134727.92+545233.7 \\
  \texttt{ROBUST: 1.0}}
  \label{fig:sfig1}
\end{subfigure}%
\caption{Continued}
\end{figure*}
\begin{figure*}\ContinuedFloat
\centering
\begin{subfigure}{.5\textwidth}
  \centering
  \includegraphics[width=.8\linewidth]{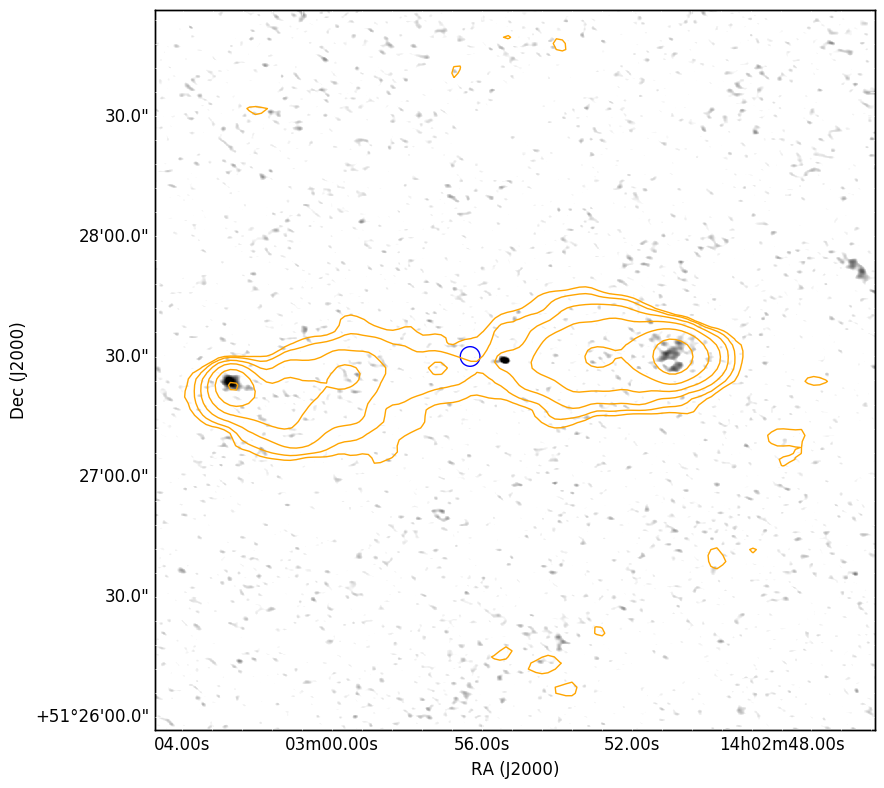}
  \caption{ILTJ140255.12+512726.8 (removed from sample -- misidentified) \\
  \texttt{ROBUST: -0.5}}
  \label{fig:sfig1}
\end{subfigure}%
\begin{subfigure}{.5\textwidth}
  \centering
  \includegraphics[width=.8\linewidth]{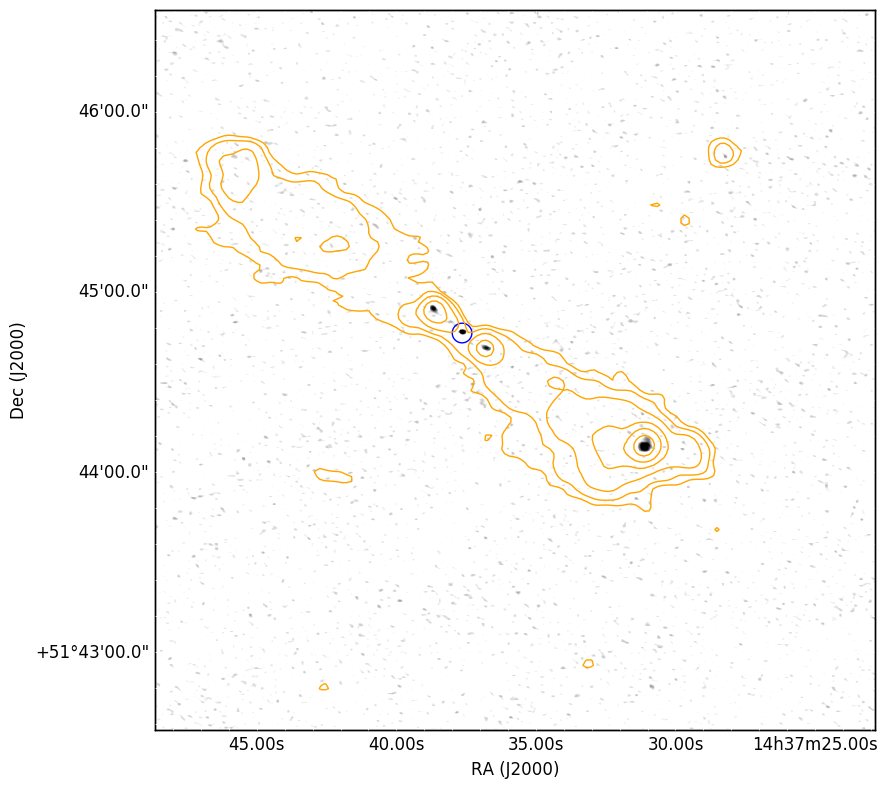}
  \caption{ILTJ143735.74+514434.3$^{\dagger}$ \\
  \texttt{ROBUST: -0.5}}
  \label{fig:sfig1}
\end{subfigure}%
\vskip\baselineskip
\begin{subfigure}{.5\textwidth}
  \centering
  \includegraphics[width=.83\linewidth]{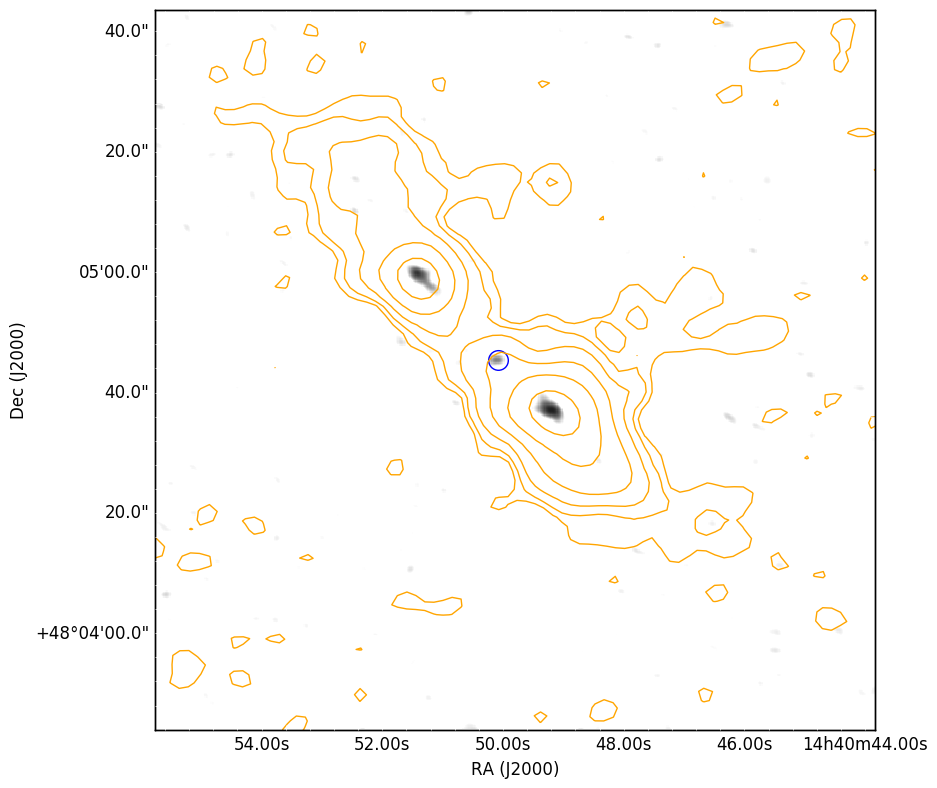}
  \caption{ILTJ144049.79+480444.0 \\
  \texttt{ROBUST: -0.5}}
  \label{fig:sfig1}
\end{subfigure}%
\begin{subfigure}{.5\textwidth}
  \centering
  \includegraphics[width=.8\linewidth]{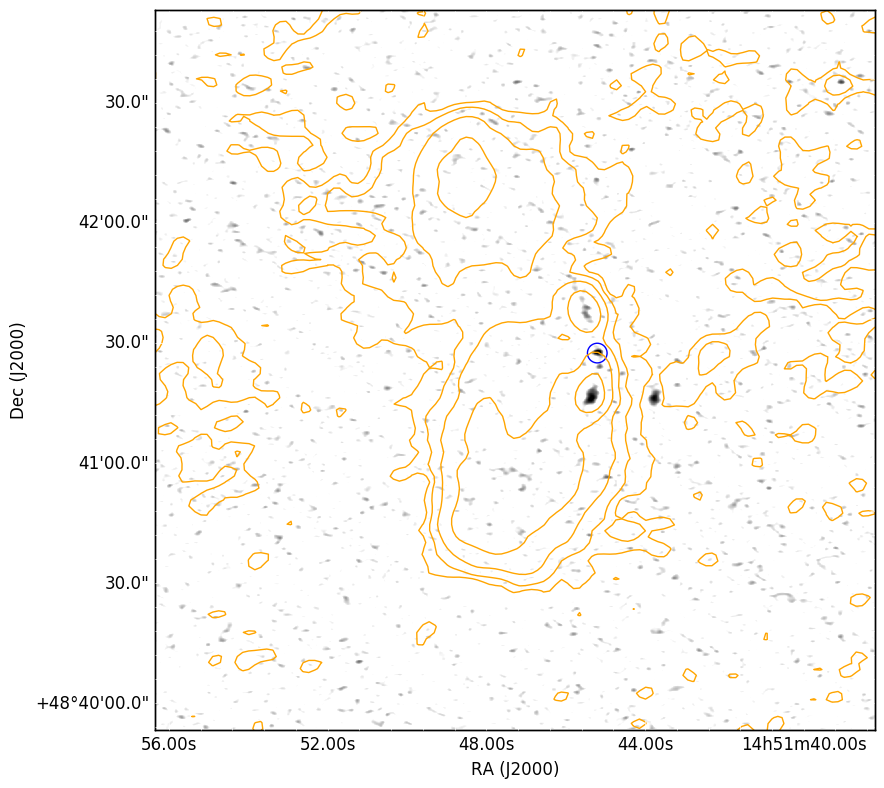}
  \caption{ILTJ145147.28+484123.5 \\
  \texttt{ROBUST: -0.5}}
  \label{fig:sfig1}
\end{subfigure}%
\vskip\baselineskip
\begin{subfigure}{.5\textwidth}
  \centering
  \includegraphics[width=.8\linewidth]{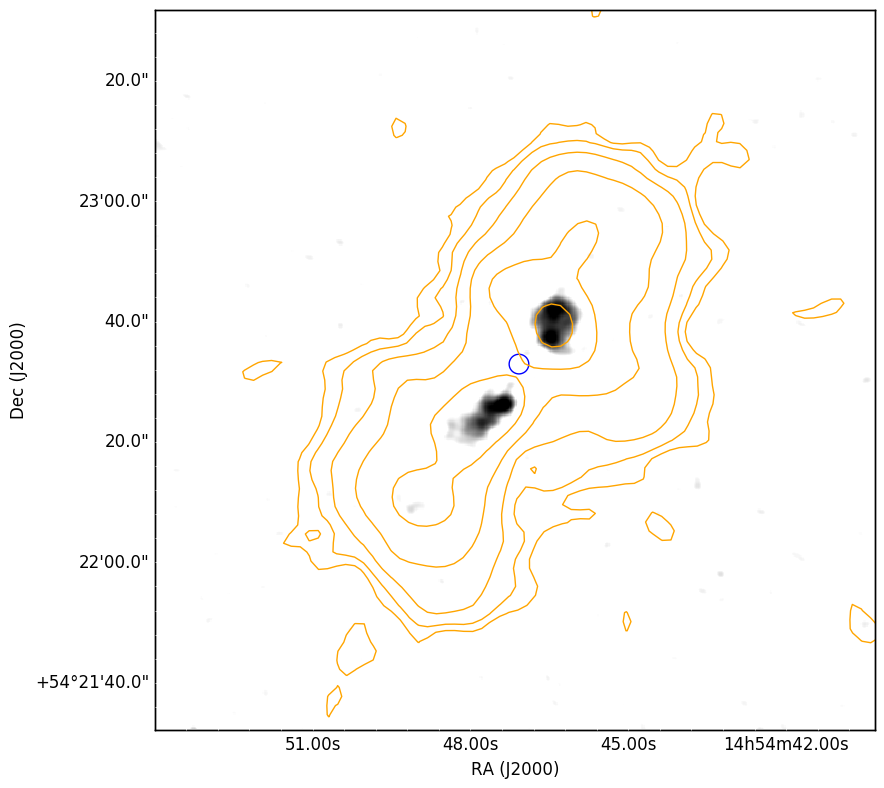}
  \caption{ILTJ145447.14+542232.2$^{\dagger}$ \\
  \texttt{ROBUST: -0.5}}
  \label{fig:ILTJ145447.14+542232.2}
\end{subfigure}%
\begin{subfigure}{.5\textwidth}
  \centering
  \includegraphics[width=.8\linewidth]{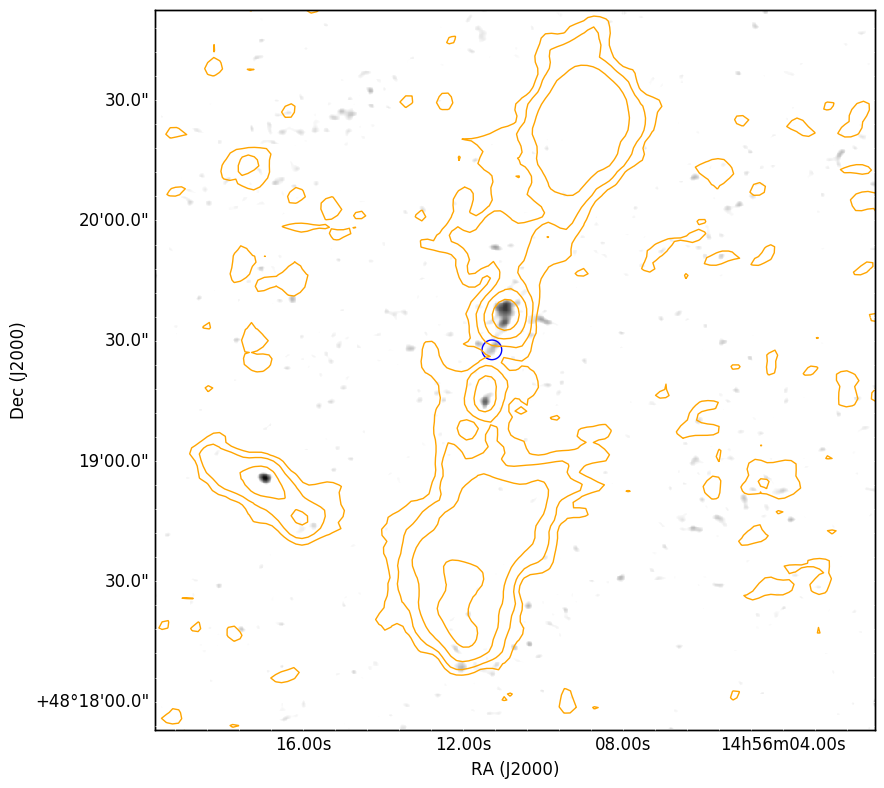}
  \caption{ILTJ145610.69+481923.0 \\
  \texttt{ROBUST: 0.5}}
  \label{fig:sfig1}
\end{subfigure}%
\caption{Continued}
\end{figure*}
\begin{figure*}\ContinuedFloat
\centering
\begin{subfigure}{.5\textwidth}
  \centering
  \includegraphics[width=.8\linewidth]{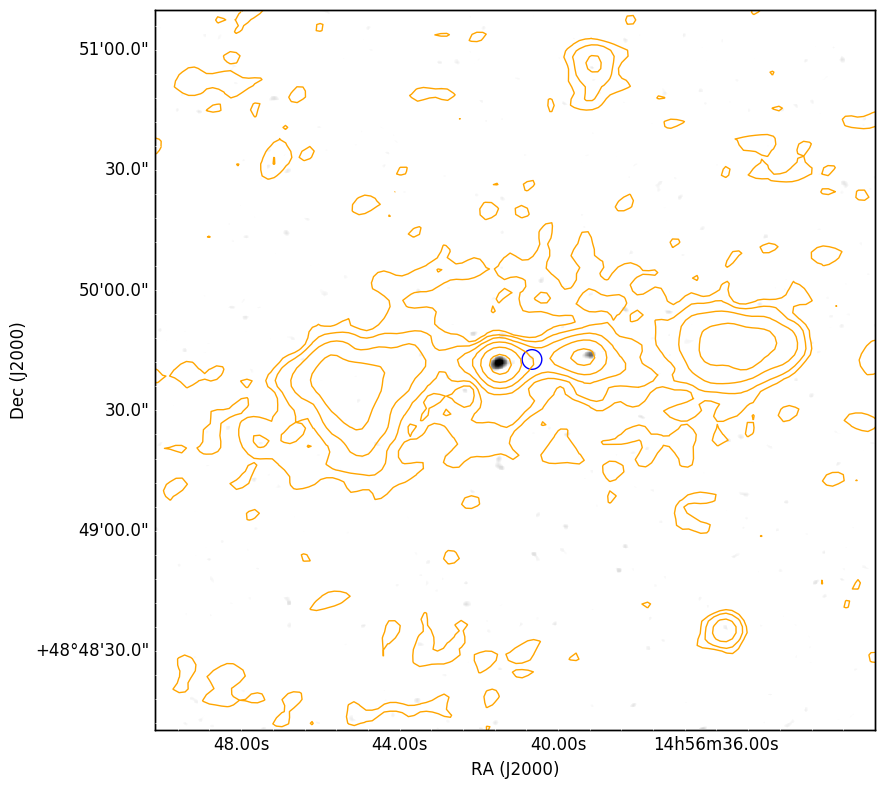}
  \caption{ILTJ145641.07+484940.5 \\
  \texttt{ROBUST: -0.5}}
  \label{fig:sfig1}
\end{subfigure}%
\begin{subfigure}{.5\textwidth}
  \centering
  \includegraphics[width=.8\linewidth]{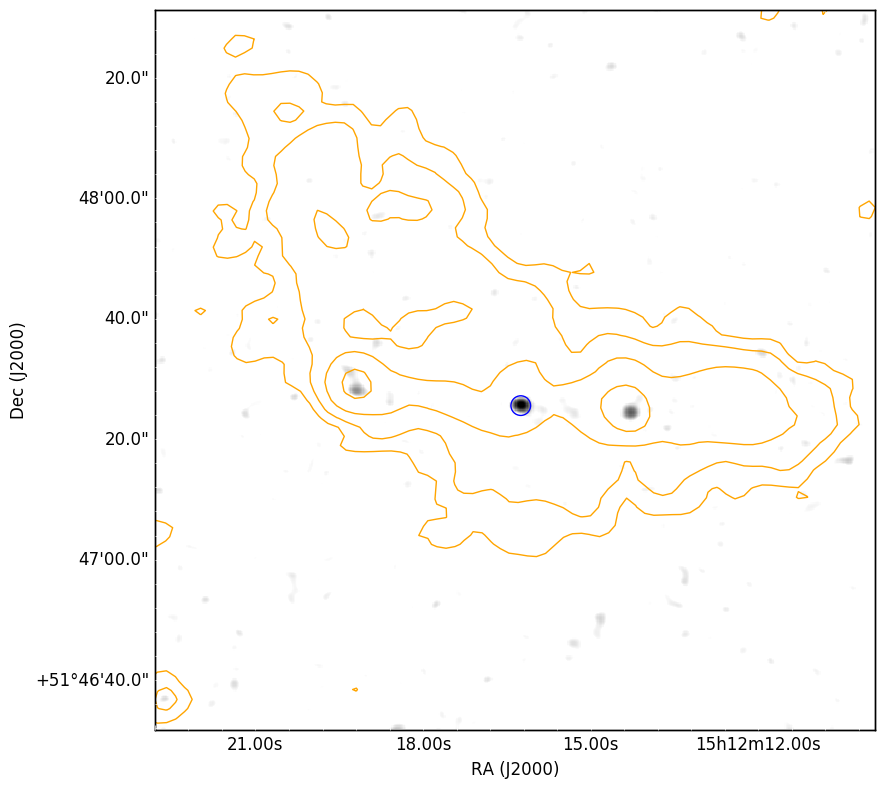}
  \caption{ILTJ151216.35+514731.8$^{\dagger}$ \\
  \texttt{ROBUST: 1.0}}
  \label{fig:sfig1}
\end{subfigure}%
\vskip\baselineskip
\begin{subfigure}{.5\textwidth}
  \centering
  \includegraphics[width=.83\linewidth]{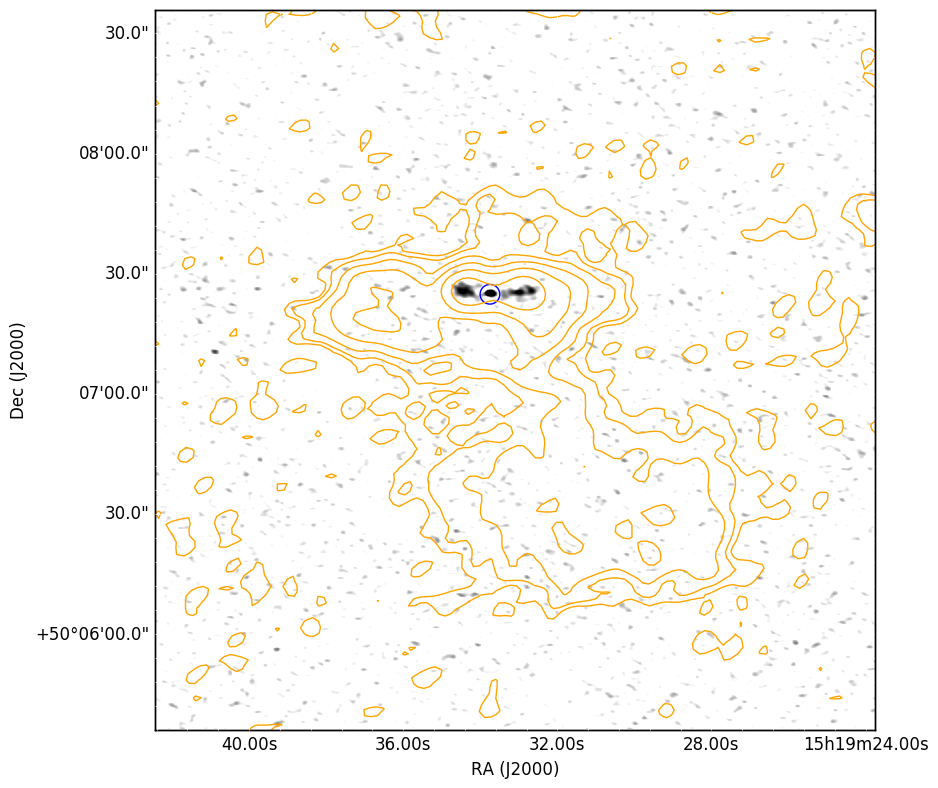}
  \caption{ILTJ151933.09+500706.2 \\
  \texttt{ROBUST: -0.5}}
  \label{fig:sfig1}
\end{subfigure}%
\begin{subfigure}{.5\textwidth}
  \centering
  \includegraphics[width=.8\linewidth]{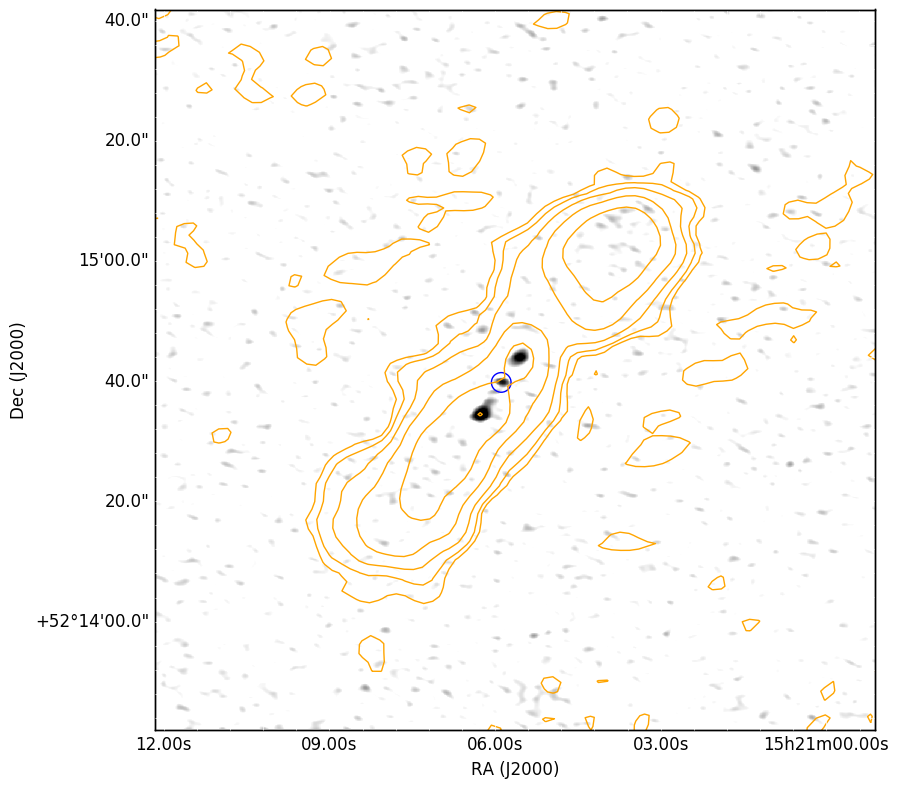}
  \caption{ILTJ152105.64+521442.0 \\
  \texttt{ROBUST: -0.5}}
  \label{fig:sfig1}
\end{subfigure}%
\caption{Continued}
\end{figure*}

\end{document}